\documentclass[
 reprint,
nofootinbib,
pra,
 amsmath,amssymb,
 aps,
]{revtex4-2}

\usepackage{graphicx}
\usepackage{dcolumn}
\usepackage{bm}
\usepackage[colorlinks]{hyperref}

\usepackage[justification=justified,
   format=plain]{caption} 
\usepackage{subcaption}
\usepackage{ragged2e}

\usepackage{physics}
\usepackage{dsfont}
\usepackage{mathtools}
\usepackage{amsthm}

\usepackage{dsfont}
\usepackage{faktor}

\usepackage{xcolor}

\usepackage[T1]{fontenc}
\usepackage[utf8]{inputenc}

\usepackage{empheq}
\usepackage[most]{tcolorbox}
\newtcbox{\mymath}[1][]{%
    nobeforeafter, math upper, tcbox raise base,
    enhanced, colframe=blue!30!black,
    colback=blue!30, boxrule=1pt,
    #1}

\graphicspath{ {./images/} }

\begin{document}

\preprint{hYp3R-Qu451}

\title{Spacetime Quasicrystals}


\renewcommand{\andname}{\ignorespaces}

\author{Latham Boyle}
\affiliation{Higgs Centre for Theoretical Physics, University of Edinburgh, UK}
\affiliation{Perimeter Institute for Theoretical Physics, Waterloo, Ontario, Canada}
\author{Sotirios Mygdalas}
\affiliation{Perimeter Institute for Theoretical Physics, Waterloo, Ontario, Canada}
\affiliation{Department of Physics \& Astronomy, University of Waterloo, Waterloo, Ontario, Canada}
\date{\today}

\begin{abstract}

Self-similar quasicrystals (like the famous Penrose and Ammann-Beenker tilings) are exceptional geometric structures in which long-range order, quasi-periodicity, non-crystallographic orientational symmetry, and discrete scale invariance are tightly interwoven in a beautiful way.  In this paper, we show how such structures may be generalized from Euclidean space to Minkowski spacetime.  We construct the first examples of such {\it Lorentzian quasicrystals} (the spacetime analogues of the Penrose or Ammann-Beenker tilings), and point out key novel features of these structures (compared to their Euclidean cousins).  We end with some (speculative) ideas about how such spacetime quasicrystals might relate to reality.  This includes an intriguing scenario in which our infinite $(3+1)$D universe is embedded (like one of our spacetime quasicrystal examples) in a particularly symmetric $(9+1)$D torus $T^{9,1}$ (which was previously found to yield the most symmetric toroidal compactification of the superstring). We suggest how this picture might help explain the mysterious seesaw relationship $M_{\rm Pl}M_{\rm vac}\approx M_{\rm EW}^{2}$ between the Planck, vacuum energy, and electroweak scales ($M_{\rm Pl}$, $M_{\rm vac}$, $M_{\rm EW}$).

\end{abstract}

\maketitle

\tableofcontents

\section{Introduction}
\label{sec:Introduction}

\begin{figure*}
    \begin{minipage}{0.55\textwidth}
        \includegraphics[width=.95\linewidth]{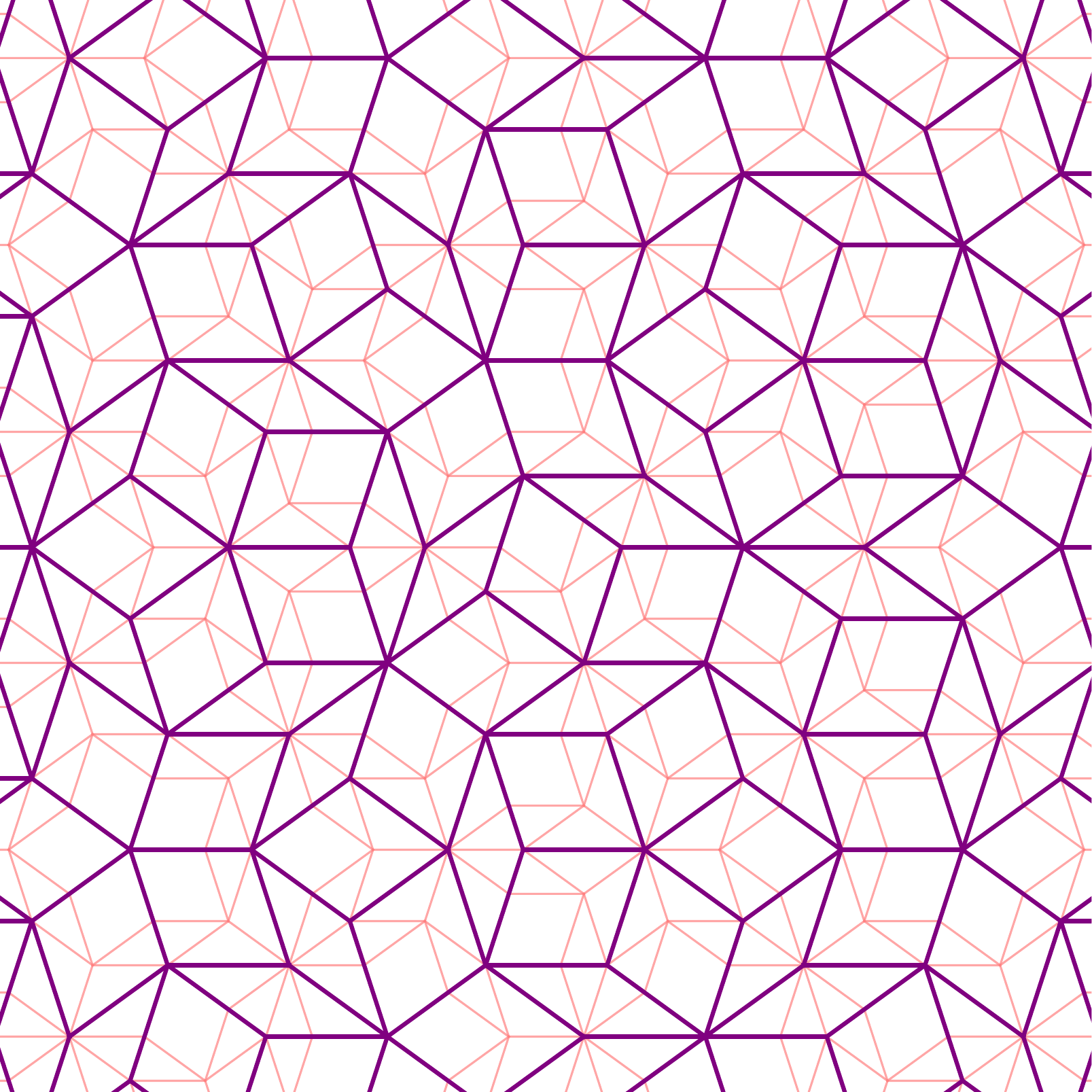}
	\end{minipage}
    \begin{minipage}{0.35\textwidth}
        \includegraphics[width=.95\linewidth]{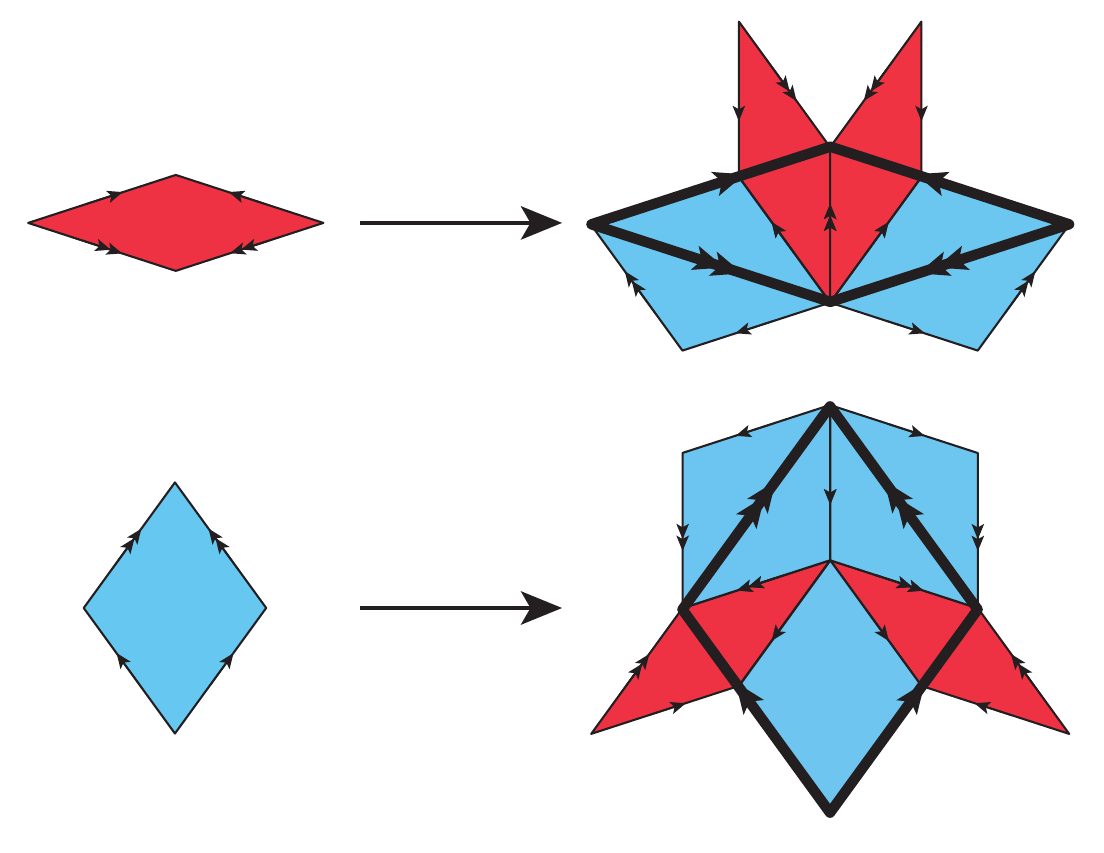}
        \includegraphics[width=.95\linewidth]{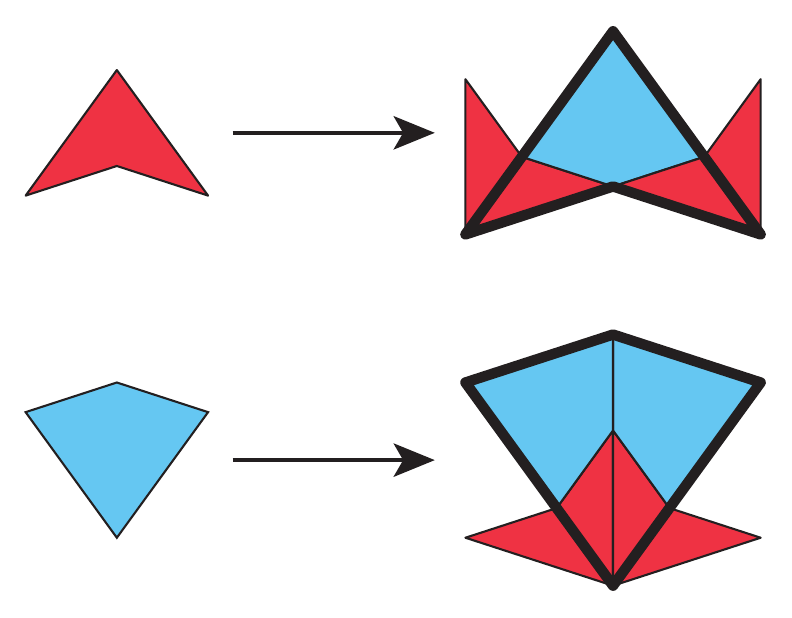}
	\end{minipage}
	\caption{\justifying Left: a patch of the Penrose tiling \cite{penrose1974role} given by rhombs in purple, underlaid by its ``inflation'' in pink. Right top: the thin (red) and thick (blue) Penrose rhombs, along with their edge-matching rules and inflation rules. Right bottom: Conway's version of the tiles: ``darts'' (red) and ``kites'' (blue) are given with their inflation rules. The two versions are equivalent since tilings built from kites and darts are mutually locally derivable from tilings by rhombs. For an introduction to the Penrose tiling, see  Ref.~\cite{gardner1977extraordinary}.}
	\label{fig:penroseTiling}
\end{figure*}

Self-similar quasicrystals (like the Penrose tiling \cite{penrose1974role, gardner1977extraordinary} in Fig.~\ref{fig:penroseTiling}) are mathematical structures whose beautiful and unexpected properties have fascinated mathematicians since their discovery in the 1970s \cite{gardner1977extraordinary, grunbaum1987tilings, baake1999guidemathematicalquasicrystals, baake2013aperiodic}.  Among many other fascinating features, they:~(i)~are intrinsically non-periodic, yet perfectly long-range ordered (with a pure-point diffraction spectrum, like a crystal);~(ii) have a ``non-crystallographic'' orientational symmetry -- too large to be compatible with any periodic pattern; and (iii) embody a kind of discrete scale-invariance or self-similarity that is also impossible in a periodic pattern.  Moreover, these three features are tightly intertwined -- a manifestation of the deep connections between the geometry of lattices \cite{conway2013sphere}, Coxeter's theory of reflection symmetry \cite{coxeterSimplices, coxeterDiscreteGroups, coxeterCompleteEnumeration, humphreys_1990}, and algebraic number theory \cite{serre2012arithmetic, lang1994algebraic, cassels1967algebraic}.

Such patterns were initially of interest to mathematicians, but beautiful mathematical ideas often manage to find a place in the real world!  In the 1980s, these patterns turned out to be the blueprints for a new class of materials (quasicrystals), first discovered in the lab \cite{Schechtman1984}, and later found to also occur naturally \cite{bindi2009natural} (forming {\it e.g.}\ in the birth of the Solar System \cite{hollister2014impact}, in lightning strikes \cite{bindi2023electrical}, and in the first atomic bomb test \cite{bindi2021accidental}).  More recently, they have also appeared in other, completely different situations: as the blueprints for quantum error correcting codes \cite{li2024penrosetilingquantumerrorcorrecting}, and in the context of discretizing scale-invariant/conformally-invariant systems and ``discrete holography'' \cite{Boyle:2018uiv, Boyle:2024qzn}, all of which suggests a possible connection to quantum gravity.

Motivated both by these physical connections, and also by pure mathematical interest, in this paper we show how self-similar quasicrystals may be naturally generalized from Euclidean {\it space} to Minkowski {\it spacetime}; and we construct the first examples of such {\it Lorentzian quasicrystals} (essentially the spacetime analogues of Euclidean quasicrystals like the Penrose tiling or the Ammann-Beenker tiling).\footnote{We draw the reader's attention to previous references \cite{Flicker:2018tkr, vacaru2018space, friedland2024spacetime, he2025experimental} that have studied dynamical/physical models that they dub ``time quasicrystals'' or ``spacetime quasicrystals.''  These are something different from the spacetime patterns constructed in present paper, in which quasi-periodicity, non-crystallographic orientational symmetry, and discrete self-similarity are intrinsically knit together as in the Penrose or Ammann-Beenker tilings.}  We point out key novel features of these objects (compared to their Euclidean cousins), and end with some (speculative) ideas about how such spacetime quasicrystals might relate to reality.

A summary of the paper is as follows:  

In Sec.~\ref{sec:LatticesAndCoxeter}, we introduce some background about lattices and their symmetries, Coxeter's general theory of reflection groups, and algebraic number theory.  

In Sec.~\ref{sec:CNPSchemes}, as a warm-up for spacetime, we first develop the theory of quasicrystals in ordinary Euclidean space.  After reviewing the standard ``cut-and-project'' (C\&P) scheme (for constructing a quasicrystal by taking an irrationally-sloped ``\textit{cut}'' through a higher-dimensional lattice, and \textit{projecting} nearby lattice points onto the cut), we describe a variant of particular interest, which we call the ``\textit{symmetric} cut-and-project'' (sC\&P) scheme, in which quasi-periodicity, non-crystallographic symmetry, and discrete scale-invariance are intrinsically interwoven in the desired fashion (as in the Penrose or Ammann-Beenker tilings).  Whereas the most standard C\&P scheme uses a ``window'' to choose lattice points near the cut, we also discuss a more general C\&P scheme based on a ``weighting function'' (which weights different points differently, according to their distance from the cut), and this allows us to define ``self-dual'' and ``globally-scale-invariant'' quasicrystals (this will be important when we turn to Lorentzian quasicrystals.)  To illustrate all of this, we use sC\&P to construct the standard Ammann-Beenker tiling, as well as its self-dual variant (see Fig.~\ref{fig:AmmannBeenker}).  We end by discussing the types of (quasi)symmetries/dualities that sC\&P quasicrystals have, and delineating two distinct types of scale invariance (``local'' and ``global'').  

This section is mainly a review, although the reader may find certain aspects of our formulation novel or helpful, as it has been designed (i) to clarify the logical connection between quasi-periodicity, non-crystallographic symmetry and discrete self-similarity or scale-invariance; and (ii) relatedly, to make the generalization from Euclidean to Lorentzian spacetime simple and natural (and, to our knowledge, our notions of self-dual and globally-scale-invariant quasicrystals are also new.)

In Sec.~\ref{sec:TowardsLorentzianQuasicrystal}, we are finally ready to extend our discussion to spacetime.  First, in Sec.~\ref{subsec:MathematicalPreliminaries}, we introduce Lorentzian lattices ({\it i.e.}\ lattices in Minkowski spacetime) and their reflection symmetries, focusing on the key novel features compared to lattices in Euclidean space. Then in Sec.~\ref{subsec:SpacetimeCNP}, we explain how the sC\&P algorithm may be naturally extended to spacetime.  Crucially, the quasicrystal's non-crystallographic orientational symmetry is now a dramatically larger (in fact infinite!) group, and consequently, the sC\&P window (or weighting function) must also be much more symmetric.  In Sec.~\ref{subsec:implications}, we discuss further consequences of this fact: spacetime quasicrystals are a symmetric point set without an associated symmetric tiling; and they do not have local scale invariance, but instead can have global scale invariance or self-duality.  Then, in Sec.~\ref{subsec:ExamplesSpacetimeQuasicrystals}, we construct the first two examples of such spacetime quasicrystals: (i) the $1+1$ dimensional (self-dual or globally-scale-invariant) spacetime quasicrystal obtained from the $(3+1)$ dimensional odd self-dual Lorentzian lattice $\textrm{I}_{3,1}$, and (ii) the four particularly interesting $3+1$ dimensional (self-dual or globally-scale-invariant) spacetime quasicrystals obtained from the remarkable $(9+1)$ dimensional even self-dual Lorentzian lattice $\textrm{II}_{9,1}$.

In Sec.~\ref{sec:PhysicalSpeculations}, we present some (speculative) ideas about how such spacetime quasicrystals might relate to reality.  

Sec.~\ref{subsec:FittingUniverseNutshell} is devoted to one such speculation: In Ref.~\cite{moore1993finitedirections}, Moore pointed out that the most symmetric toroidal compactification of the superstring was obtained by curling all 10 dimensions of spacetime (including time) into a particularly symmetric torus $T^{9,1}$ obtained by modding out $(9+1)$D Minkowski space by the remarkable $(9+1)$D even self-dual lattice $\mathrm{II}_{9,1}$: $T^{9,1}=\mathbb{R}^{9,1}/\mathrm{II}_{9,1}$.  At the time, it seemed that this 10D torus couldn't have anything to do with the real world and was of purely academic interest (as it seemed to have closed time like curves, and lack our $3+1$ macroscopic dimensions).  However, we explain how the quasicrystal construction in Sec.~\ref{subsec:4DLorentzianQuasicrystal} suggests a potential way around this conclusion: the most symmetric embeddings of $(3+1)$D Minkowski spacetime $\mathbb{R}^{3,1}$ in the 10D torus $T^{9,1}$ are {\it irrationally-sloped} with respect to the torus, so that $\mathbb{R}^{3,1}$ wraps around and around the torus, densely filling it, without ever intersecting itself.  An observer confined to this $\mathbb{R}^{3,1}$ would regard themselves as living in infinite 4D Minkowski spacetime, despite the fact that this spacetime is embedded in the completely compactified 10D spacetime torus $T^{9,1}$ (the universe in a nutshell!).  Moreover, we explain some (very tentative) arguments suggesting that this scenario might even be a helpful starting point for explaining the mysteriously large hierarchies between the Planck scale $M_{\rm Pl}$, the electroweak scale $M_{\rm EW}$ and the vacuum energy scale $M_{\rm vac}$ ($M_{\rm Pl}\gg M_{\rm EW}\gg M_{\rm vac}$), as well as the seesaw relationship they are observed to obey ($M_{\rm Pl}M_{\rm vac}\approx M_{\rm EW}^{2}$).

Sec.~\ref{subsec:AdditionalRemarks} points out various other speculations and suggestions for why spacetime quasicrystals may be useful as a tool for discretizing/studying a variety of (nearly) scale-invariant physical systems of interest; and how they may connect to (or be useful in) other approaches to studying quantum gravity.

Finally, two appendices contain more technical results:  

Appendix \ref{app:SymmetriesLorentzianLattices} summarizes Vinberg's analysis of the reflection symmetries of self-dual Lorentzian lattices (since some of Vinberg's relevant papers are hard to obtain, have not been translated into English, and also contain potentially-confusing typos in their Coxeter-Dynkin diagrams, which we correct).  

Appendix \ref{app:inflation} explains how the scale factors characterizing the discrete scale invariance of a quasicrystal are related to the units of an associated algebraic number field, and tabulates these number fields for the self-dual Lorentzian lattices up to dimension 10.

\section{Background Material}
\label{sec:LatticesAndCoxeter}

We begin in this section by reviewing some useful background material about mathematical lattices, Coxeter's theory of reflection groups, and algebraic number theory.

\subsection{Lattices}
\label{subsec:Lattices}

In this subsection we give a brief summary of some of the key properties of mathematical lattices we will need in this paper.  For more details see \cite{conway2013sphere}.

Let $\{{\bf b}_i\}$ ($i=1,\ldots,d$) be a basis for $V=\mathbb{R}^{d}$.  The $\mathbb{Z}$-span of $\{{\bf b}_i\}$ is a $d$-dimensional {\bf lattice} $\Lambda\in V$, {\it i.e.}\ $\Lambda$ is the set of points in $\mathbb{R}^{d}$ given by
\begin{equation}
  \Lambda=\left\{\sum_{i=1}^{d}n_{i}{\bf b}_{i}\;(n_i\in \mathbb{Z})\right\}.
\end{equation}
Note that the choice of lattice basis $\{{\bf b}_i\}$ is not unique: another basis $\{{\bf b}_i'\}$ generates the same lattice $\Lambda$ if and only if it is related to $\{{\bf b}_i\}$ by ${\bf b}_i'=M_{ij}{\bf b}_j$, where $M_{ij}\in{\rm GL}(d,\mathbb{Z})$ is an integer $d\times d$ matrix with determinant $\pm1$.

We let $\Lambda^{{\bf t}}$ denote the same lattice $\Lambda$ translated (relative to the origin of $\mathbb{R}^{d}$) by the vector ${\bf t}\in\mathbb{R}^{d}$:
\begin{equation}
  \Lambda^{{\bf t}}=\left\{\sum_{i=1}^{d}{\bf t}+n_{i}{\bf b}_{i}\;(n_i\in \mathbb{Z})\right\}.
\end{equation}

The $d$-dimensional parallelotope $\tau$ with edges ${\bf b}_i$
\begin{equation}
  \tau=\left\{\sum_{i=1}^{d}x_i {\bf b}_i,\;0\leq x_i<1\right\}
\end{equation}
is called a {\bf unit cell} for the lattice (since every point in the lattice is related to a unique point in the fundamental domain by translation by an integer linear combination of the basis vectors).  In other words, $\tau=\mathbb{R}^{d}/\Lambda$.

We will regard $\mathbb{R}^{d}$ as a $d$-dimensional vector space $V$ equipped with a real-valued inner product $\langle\cdot,\cdot\rangle$.\footnote{In Sec.~\ref{sec:TowardsLorentzianQuasicrystal}, in order to pass from space to spacetime, we will generalize $\langle\cdot,\cdot\rangle$ from a (positive definite) inner product to a (indefinite) non-degenerate bilinear form.} Since the inner product is positive definite, we can choose an orthonormal basis (or ``Cartesian basis'') $\{{\bf e}_{\alpha}\}$ for $V$ with
\begin{equation}
  \langle {\bf v},{\bf w}\rangle=
  \langle v^{\alpha}{\bf e}_{\alpha},
  w^{\beta}{\bf e}_{\beta}\rangle=
  v^{\alpha}\delta_{\alpha\beta}w^{\beta}
\end{equation}
where $\delta_{\alpha\beta}$ is the Kronecker delta.  

Given any $\Lambda$, we can define its {\bf dual lattice} $\Lambda^{\ast}$ as the set of all vectors that have integer inner product with any point in $\Lambda$:
\begin{equation}
  \Lambda_{\ast}=\{{\bf v}_{\ast}\in V|\langle {\bf v}_{\ast},{\bf v}\rangle\in\mathbb{Z}, \forall\,{\bf v}\in \Lambda\}.
\end{equation}
We say $\Lambda$ is an {\bf integer lattice} if the inner product of any two vectors in the lattice is an integer
\begin{equation}
  \langle {\bf v},{\bf v}'\rangle\in\mathbb{Z}\quad(\forall\,{\bf v},{\bf v}'\in\Lambda).
\end{equation}
Note that if $\Lambda$ is an integer lattice it is a subset of its dual lattice ($\Lambda\subseteq\Lambda_{\ast}$). 

We say $\Lambda$ is a {\bf unimodular lattice} if its unit cell $\tau$ has unit volume; or, in other words, if the $d\times d$ matrix whose columns are the basis vectors ${\bf b}_{i}$ has:
\begin{equation}
  {\rm det}\{{\bf b}_i\}=\pm1.
\end{equation}

We say $\Lambda$ is {\bf self-dual} if it is the same as its dual lattice
\begin{equation}
  \Lambda=\Lambda_{\ast}.
\end{equation}
$\Lambda$ is self-dual if and only if it is integer and unimodular.

A self-dual lattice $\Lambda$ is {\bf even} or {\bf Type II} if the inner product of any two points in $\Lambda$ is an {\it even} integer; otherwise, it is {\bf odd} or {\bf Type I}.

Self-dual lattices are relatively rare: Euclidean self-dual lattices have been classified up to dimension $d=25$ and are summarized in Table 2.2 in \cite{conway2013sphere}.  Type I lattices exist in all dimensions: up to $d=8$, the only Type I lattice is the hypercubic lattice $\mathbb{Z}^{d}$; for $d\geq 9$ a second lattice $E_{8}\otimes\mathbb{Z}^{d}$ appears; in $d=12$ there is a third lattice $D_{12}^{+}$; and the number increases in higher dimensions.  

Type II lattices have especially remarkable properties and are even rarer: they only occur when the dimension $d$ is a multiple of 8.  In the lowest dimension ($d=8$) the unique Type II lattice the $E_8$ root lattice.  Next, in $d=16$, there are two such lattices: $E_{8}\oplus E_{8}$ and the $SO(16)$ root lattice $D_{16}^{+}$.  Then, in $d=24$, there are 24 such lattices, including one particularly remarkable lattice with no roots (the Leech lattice).  And in higher dimensions ($d=32,40,\ldots$) the number grows rapidly.

In
Sec.~\ref{sec:TowardsLorentzianQuasicrystal}, we will see there is a much simpler classification of self-dual lattices in spacetime and, more generally, in spaces with indefinite signature $(p,q)$.

\begin{figure*}
    \captionsetup{justification=justified}
	\begin{minipage}{0.48\textwidth}
    \includegraphics[width=.95\linewidth]{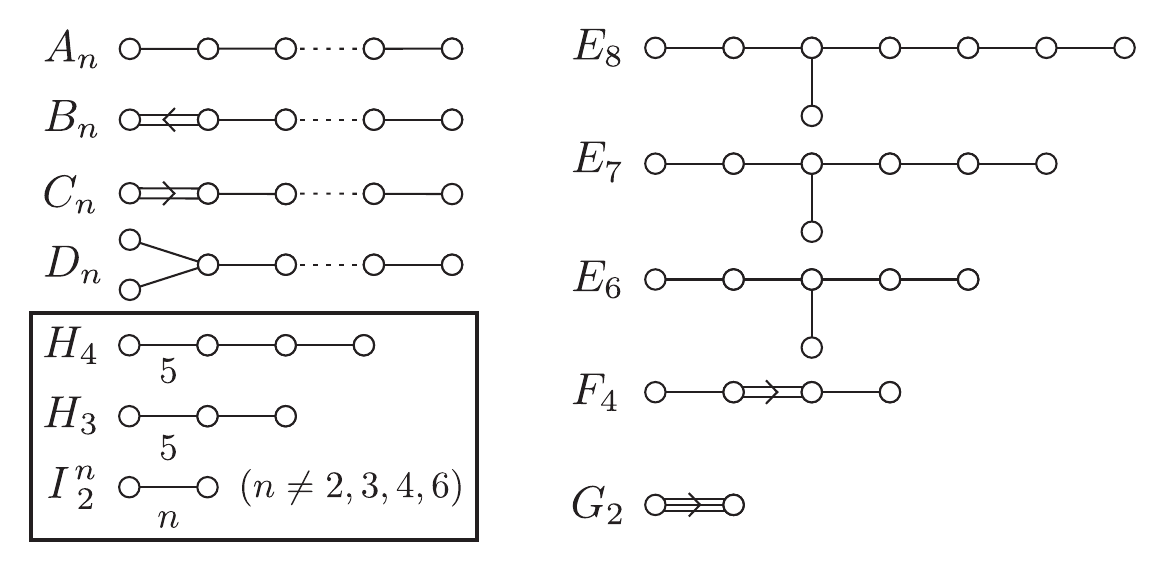}
	\end{minipage}
    \begin{minipage}{0.48\textwidth}
        \includegraphics[width=.95\linewidth]{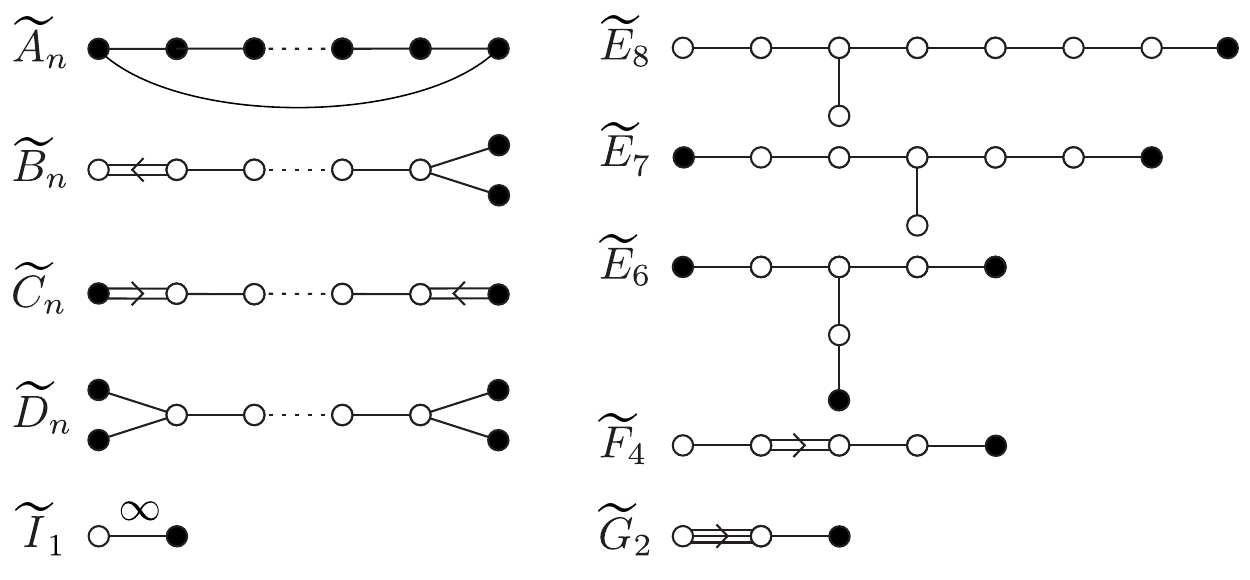}
	\end{minipage}
	\caption{\justifying The Coxeter-Dynkin diagrams for the irreducible Coxeter groups of finite/spherical type (left panel) and affine/planar type (right panel). The non-crystallographic cases ($H_{4}$, $H_{3}$ and $I_{2}^{n}$) are grouped in the box at lower left.  Each affine diagram (at right) may be obtained by adding a single ``extending root'' to the corresponding finite crystallographic diagram (at left); or, alternatively, by removing a single extending node (any of the black nodes) from any affine diagram (at right) we obtain the corresponding finite crystallographic diagram (at left).}
	\label{fig:Irreducible_Root_Systems}
\end{figure*}

\subsection{Coxeter Theory}
\label{subsec:Coxeter_Theory}

In this subsection, we briefly summarize some basic notions of Coxeter's theory of reflection groups \cite{coxeterSimplices, coxeterDiscreteGroups, coxeterCompleteEnumeration, humphreys_1990, mcmullen2002coxeter}.

Consider the vector space $V=\mathbb{R}^{d}$ equipped with the (positive-definite) inner product $\langle\cdot,\cdot\rangle$.  In the basis $\{{\bf v}_{\mu}\}$ ($\mu=1,\ldots,d$), the metric is a symmetric, positive-definite matrix $g_{\mu\nu}=\langle {\bf v}_{\mu}|{\bf v}_{\nu}\rangle$. 

Now consider a codimension-one hyperplane through the origin in $V$ (a mirror $M$), with normal vector ${\bf r}=r^{\mu}{\bf v}_{\mu}$ (its root).  Reflection through $M$ corresponds to the linear map
\begin{equation}
  {\bf x}\to{\bf x}-2
  \frac{\langle {\bf x},{\bf r}\rangle}
  {\langle {\bf r}, {\bf r}\rangle}{\bf r}.
\end{equation}
In the basis $\{{\bf v}_{\mu}\}$, this is described by $x^{\mu}\to R^{\mu}_{\;\;\nu}x^{\nu}$ with the reflection matrix
\begin{equation}
  \label{R_eq}
  R^{\mu}_{\;\;\nu}=\delta^{\mu}_{\;\;\nu}-2\frac{r^{\mu}r_{\nu}}{r^{2}}
\end{equation} 
where, throughout this paper, we use the standard physics notation where Greek indices are lowered and raised using the metric $g_{\mu\nu}$ and its inverse $g^{\mu\nu}$, respectively, the ``Einstein summation convention'' in which a pair of repeated indices (one up and one down) are implicitly summed from $1$ to $d$, and 
$r^{2}\equiv \langle{\bf r},{\bf r}\rangle=r^{\mu}r_{\mu}$.  Note that the matrix $R^{\mu}_{\;\;\nu}$ squares to the identity matrix
\begin{equation}
  R^{2}=1.
\end{equation}
If two different mirrors $M_i$ and $M_j$ meet at an angle $\pi/m_{ij}$ ($m_{ij}\in\mathbb{Z}^{+})$, so that their roots $\mathbf{r}_{i}$ and $\mathbf{r}_{j}$ satisfy 
\begin{equation}
  \frac{\langle {\bf r}_i,{\bf r}_j\rangle}{(r_i^2 r_j^2)^{1/2}}
  =-{\rm cos}\frac{\pi}{m_{ij}},
\end{equation}
the corresponding reflection matrices $R_i$ and $R_j$ satisfy
\begin{equation}
  (R_i R_j)^{m_{ij}}=1.
\end{equation}
A group that is generated by successive reflections in such mirrors is a {\bf reflection group}.

A {\bf Coxeter group} $G$ is an abstraction and generalization of a reflection group: it is a group generated by $n$ generators $S=\{R_1,\ldots,R_n\}$, subject to the relations $(R_i R_j)^{m_{ij}}=1$, where $m_{ij}$ is a symmetric integer $n\times n$ matrix $(m_{ij}=m_{ji},m_{ij}\in \mathbb{Z})$, with $m_{ii}=1$ along the diagonal, and $m_{ij}>1$ off the diagonal ({\it i.e.}\ when $i\neq j$).  The pair $(G,S)$ is called a {\bf Coxeter system}.  The matrix $m_{ij}$ is called the {\bf Coxeter matrix}, and the corresponding reflections $R_{i}$, mirrors $M_i$, and roots ${\bf r}_i$ ($i=1,\ldots,n)$ are called the {\bf fundamental reflections}, {\bf fundamental mirrors}, and {\bf fundamental roots}, respectively. The number $n$ of fundamental roots is called the {\bf rank} of $G$.

So a Coxeter group is specified entirely by its Coxeter matrix $m_{ij}$, which is conveniently summarized by a {\bf Coxeter diagram} (or {\bf Coxeter-Dynkin diagram}) in which we draw $n$ dots -- one for each fundamental reflection (or mirror, or root) -- and then draw a line between each pair of distinct dots $i$ and $j$, labelled by the corresponding integer $m_{ij}$.  For convenience, we then adopt the further standard conventions: if $m_{ij}=2$ we omit both the line and its integer label; if $m_{ij}=3$ we draw a single line and suppress the label; if $m_{ij}=4$ we draw a double line and suppress the label; and if the line connects a long root to a short root, we indicate this by adding an arrow to the line pointing to the short root. 

A Coxeter group is {\bf irreducible} (resp.\ {\bf reducible}) if its Coxeter diagram is {\it connected} (resp.\ {\it disconnected}).  Any reducible Coxeter group naturally decomposes into a product of irreducible Coxeter groups (corresponding to the connected pieces of its Coxeter diagram).   

\begin{figure*}
\captionsetup{justification=justified}
\begin{minipage}{0.32\textwidth}
        \centering
    \includegraphics[width=.95\linewidth]{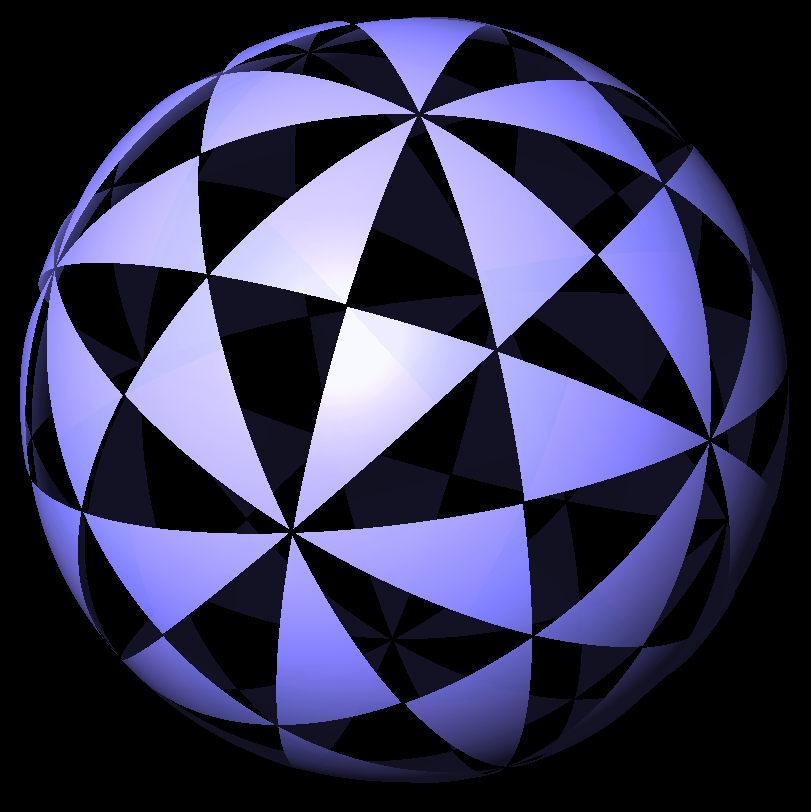}
\end{minipage}
    \begin{minipage}{0.32\textwidth}
        \centering
    \includegraphics[width=.95\linewidth]{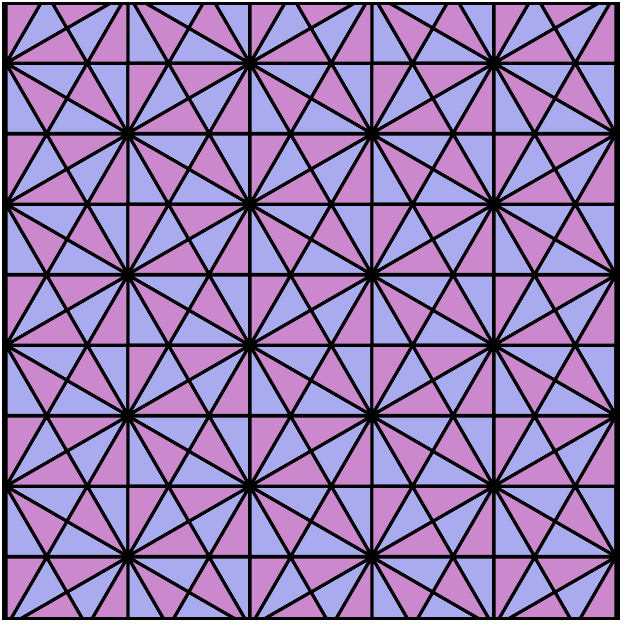}
\end{minipage}
\begin{minipage}{0.32\textwidth}
        \centering
    \includegraphics[width=1\linewidth]{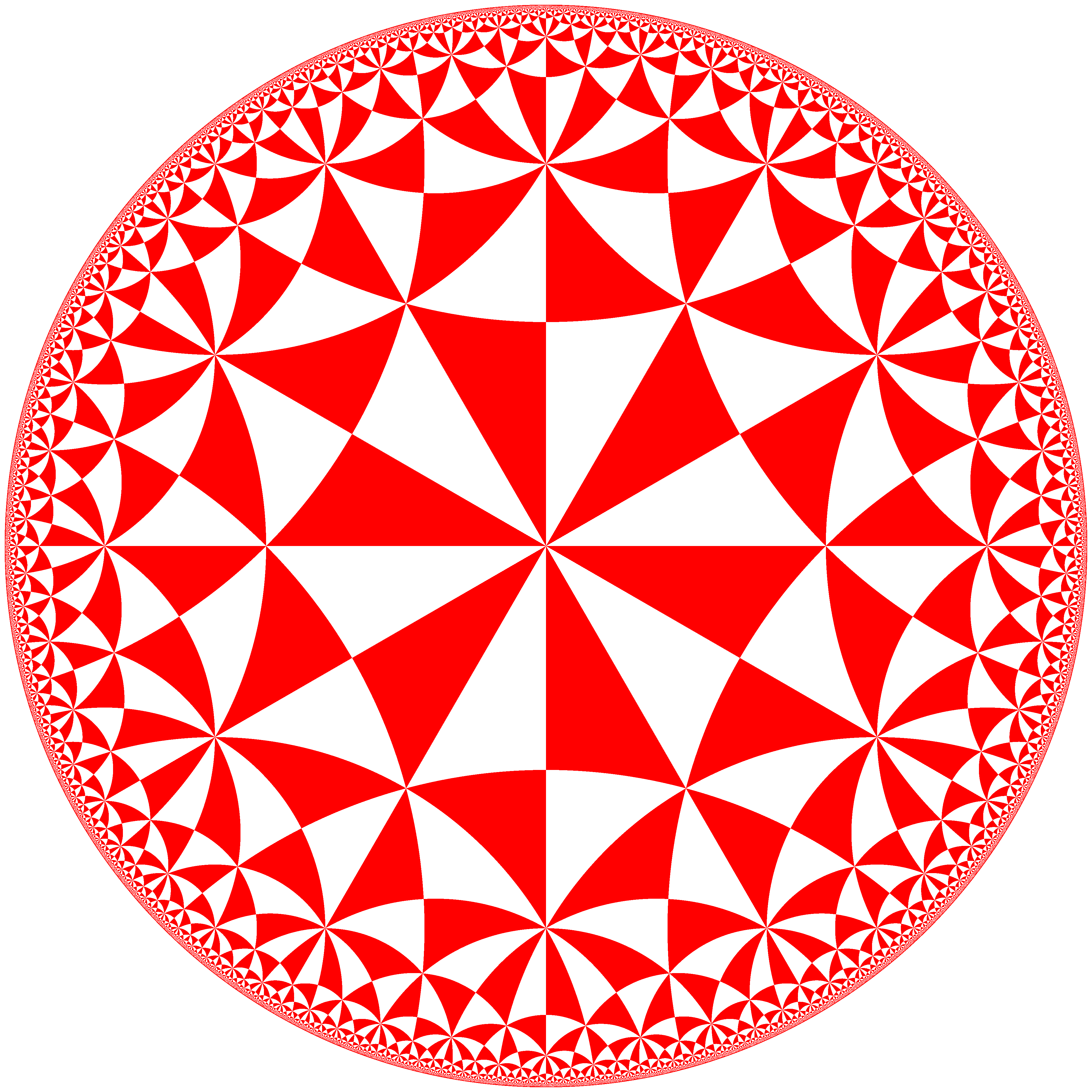}
\end{minipage}
\caption{\justifying The ``Coxeter complex'' showing mirrors (geodesic curves) and images of the fundamental domain (triangular regions) for various Coxeter groups: \href{https://commons.wikimedia.org/wiki/File:Icosahedral_reflection_domains.png}{left}: the spherical (finite) $[5,3]$ (icosahedral) group; \href{https://commons.wikimedia.org/wiki/File:Tiling_Dual_Semiregular_V4-6-12_Bisected_Hexagonal.svg}{center}: the affine (planar) $[6,3]$ group; \href{https://en.wikipedia.org/wiki/File:Hyperbolic_domains_642.png}{right}: the hyperbolic $[6,4]$ group.}
\label{fig:SphEucHyp}
\end{figure*}

A Coxeter group has an associated {\bf geometric representation} in the vector space $V=\mathbb{R}^{n}$ spanned by the fundamental roots $\{{\bf r}_1,\ldots,{\bf r}_n\}$: if we take $V$ to have a symmetric bilinear form $B({\bf v},{\bf w})$ such that
\begin{equation}
  B({\bf r}_i,{\bf r}_j)=-{\rm cos}\frac{\pi}{m_{ij}},
\end{equation}
then the generator $R_{i}$ is represented by the reflection 
\begin{equation}
  R_i({\bf v})={\bf v}-2 B({\bf v},{\bf r}_i){\bf r}_i.
\end{equation}
(Note that vector ${\bf r}_{i}$ is reversed by $R_i$, while the hyperplane orthogonal to ${\bf r}_{i}$ is fixed pointwise by $R_i$.)  The $n\times n$ {\bf Schl\"{a}fli matrix} $S_{ij}$ is twice the bilinear form $B$, expressed in the root basis:
\begin{equation}
  S_{ij}=-2\,{\rm cos}\frac{\pi}{m_{ij}}.
\end{equation}
Let $S_{ij}$ have $p$ positive and $q$ negative eigenvalues: we say it has signature $\{p,q\}$.~(Note that $p+q\leq n$ since $S_{ij}$ may also have some zero eigenvalues.)  

Using the terminology conventions in \cite{mcmullen2002coxeter}, irreducible Coxeter groups may be classified into one of four types according to the signature of $S_{ij}$
\begin{itemize}
  \item {\bf Spherical}: if $\{p,q\}=\{n,0\}$ ({\it i.e.}\ if all eigenvalues of $S_{ij}$ are positive);
  \item {\bf Affine}: if $\{p,q\}=\{n-1,0\}$ ({\it i.e.}\ if all eigenvalues of $S_{ij}$ are positive, except one which vanishes);
  \item {\bf Hyperbolic}: if $\{p,q\}=\{p,1\}$ ({\it i.e.}\ if exactly one eigenvalue of $S_{ij}$ is negative);
  \item {\bf Higher-rank}: if $q\geq2$ ({\it i.e.}\ if at least two eigenvalues of $S_{ij}$ are negative).
\end{itemize}  

From a geometrical perspective, we can think about the spherical, affine, and hyperbolic Coxeter groups as reflection groups in the  $k$-dimensional maximally-symmetric space $Y^k$ of positive, zero or negative curvature, respectively -- {\it i.e.}\ the $k$-dimensional sphere $S^k$, $k$-dimensional Euclidean space $E^k$, or $k$-dimensional hyperbolic space $H^k$, respectively (see Fig.~\ref{fig:SphEucHyp}: left, center and right panels, respectively).  For spherical and affine Coxeter groups, we have $k=n-1$.  In the spherical case, we can think of the $k$-dimensional sphere $S^{k}$ as embedded as the surface $(X^0)^{2}+(X^1)^{2}+\ldots+(X^k)^2=1$ in Euclidean space $E^{k+1}$; and the mirrors as $k$-dimensional hyperplanes passing through the origin of $E^{k+1}$ and intersecting the $S^k$ along ``great'' $(k-1)$-dimensional spheres $S^{k-1}$, so that the reflections across these hyperplanes in $E^{k+1}$ naturally induce corresponding reflections across the $(k-1)$-spheres $S^{k-1}$ in $S^{k}$. For hyperbolic Coxeter groups, we have $k=p$.  We can think of the $k$-dimensional hyperbolic space $H^k$ as the upper ($X^0>0$) sheet of the hyperboloid $-(X^0)^{2}+(X^1)^{2}+\ldots+(X^k)^2=-1$ in Minkowski space $M^{k+1}$; and the mirrors as $n$-dimensional hyperplanes passing through the origin of $M^{k+1}$ and intersecting the $H^k$ along $(k-1)$-dimensional hyperbolic spaces $H^{k-1}$, so that the reflections across these hyperplanes in $M^{k+1}$ naturally induce corresponding reflections across the $H^{k-1}$ in $H^{k}$ (see Sec.~\ref{sec:TowardsLorentzianQuasicrystal}).

If we draw all the mirrors in a given reflection group $G$, they divide up the space $Y^k$ into a set of congruent convex {\bf fundamental domains}, which together form a tessellation $T$ of $Y^k$.  If we pick one such fundamental domain (or {\bf Coxeter polytope}) $\Gamma$, the fundamental roots are the outward pointing normal vectors to its bounding mirror ``faces'' (or the hyperplanes defining them).  So the corresponding Coxeter diagram specifies the shape of the fundamental domain $\Gamma$: the number of its faces, and the angle at which each pair of faces meets (see Fig.~\ref{fig:SphEucHyp}).  And the symmetries of the Coxeter diagram reflect symmetries of the Coxeter polytope itself: {\it e.g.}\ the $\mathbb{Z}_{2}$ symmetry of the $A_{n}$, $D_{n}$ and $E_{6}$ graphs, or the larger $S_{3}$ symmetry of the $D_{4}$ graph which gives rise to the famous {\bf triality symmetry} of the $\mathfrak{so}(8)$ Lie algebra.
Given a choice of fundamental domain $\Gamma$, successive reflections through its faces can map it to any other copy of $\Gamma$ in the tessellation $T$ of $Y^k$, so that the ``tiles'' in $T$ are in one-to-one correspondence with the elements in the group $G$.

The spherical Coxeter groups have been completely classified (see the left panel of Fig.~\ref{fig:Irreducible_Root_Systems}); and, moreover, Coxeter showed that these are precisely the {\it finite} Coxeter groups.  The affine Coxeter groups have also been completely classified (see the right panel of Fig.~\ref{fig:Irreducible_Root_Systems}).  The hyperbolic Coxeter groups have not been classified, although the hyperbolic reflection groups whose fundamental domains are simplices {\it have} been classified (see Secs.~6.8 \& 6.9 and Figs.~2 \& 3 in \cite{humphreys_1990}).  Vinberg found additional hyperbolic reflection groups (with non-simplicial fundamental domains) by studying the reflection symmetries of Lorentzian lattices (see Sec.~\ref{sec:TowardsLorentzianQuasicrystal} and App.~\ref{app:SymmetriesLorentzianLattices}).

A Coxeter group is {\bf crystallographic} if its geometric representation in $V=\mathbb{R}^{n}$ (the space spanned by the fundamental roots) can stabilize a lattice $\Lambda$ in $V$ ({\it i.e.}\ the $\mathbb{Z}$-span of some basis of $V$).  Otherwise it is {\bf non-crystallographic}. A necessary and sufficient condition for a Coxeter group  to be crystallographic is that: (i) all of the edges in its Coxeter diagram are labelled by $2$, $3$, $4$, $6$, or $\infty$; and (ii) for each circuit (closed loop) in the Coxeter diagram, the number of edges labelled 4 (resp. 6) around the circuit is even (see Sec.~6.6 in \cite{humphreys_1990}).

The {\it crystallographic} finite/spherical groups are closely related to affine groups: each affine diagram may be obtained by adding a single ``extending node'' to one of the crystallographic finite groups (see Fig.~\ref{fig:Irreducible_Root_Systems}).  The affine diagrams describe the full reflection symmetries of some periodic pattern in Euclidean space $\mathbb{R}^{n}$, while the {\it crystallographic} finite/spherical diagrams arise as the point subgroups of those periodic patterns that fix a point where $n-1$ of the mirrors cross.\footnote{The crystallographic finite  Coxeter groups (and their associated Coxeter-Dynkin diagrams) also play a fundamental role in the classification of the finite-dimensional semi-simple Lie algebras (and their associated Lie groups) \cite{Humphreys1973}.  Thus, the list of finite crystallographic Coxeter-Dynkin diagrams in Fig.~\ref{fig:Irreducible_Root_Systems} (namely $A_{n}$, $B_{n}$, $C_{n}$, $D_{n}$, $E_{6}$, $E_{7}$, $E_{8}$, $F_{4}$ and $G_{2}$) is also famous as the list of simple finite-dimensional Lie algebras. More general crystallographic Coxeter-Dynkin diagrams (including affine, hyperbolic and indefinite diagrams) play an analogous role in the classification of Kac-Moody algebras \cite{kac1990infinite}.}  

By contrast, the {\it non-crystallographic} finite/spherical groups are those that {\it do not} arise as the point groups within a larger periodic pattern.~From Fig.~\ref{fig:Irreducible_Root_Systems}, the only finite irreducible non-crystallographic Coxeter groups occur in two dimensions ($I_{2}^{n}$, the symmetries of a regular $n$-gon, with $n\neq 2,3,4,6$), three dimensions ($H_{3}$, the symmetries of the icosahedron) or four dimensions ($H_{4}$, the symmetries of the 600-cell, the four-dimensional analogue of the icosahedron).~These non-crystallographic groups play an important role in the theory of quasi-periodic tilings and quasicrystals, where they govern the orientational symmetries of such structures: {\it e.g.}\ the 2D Penrose tiling (with decagonal orientational symmetry $I_{2}^{10}$) \cite{penrose1974role, gardner1997penrose}, the 2D Ammann-Beenker tiling (with octagonal symmetry $I_{2}^{8}$) \cite{ammann1992aperiodic, beenker1982algebraic}, various 3D quasicrystals and tilings (with icosahedral symmetry $H_{3}$) \cite{baake2013aperiodic}, and the 4D Elser-Sloane quasicrystal (with symmetry $H_{4}$) \cite{ElserSloane4D}.

\begin{figure}
\includegraphics[width=0.95\linewidth]{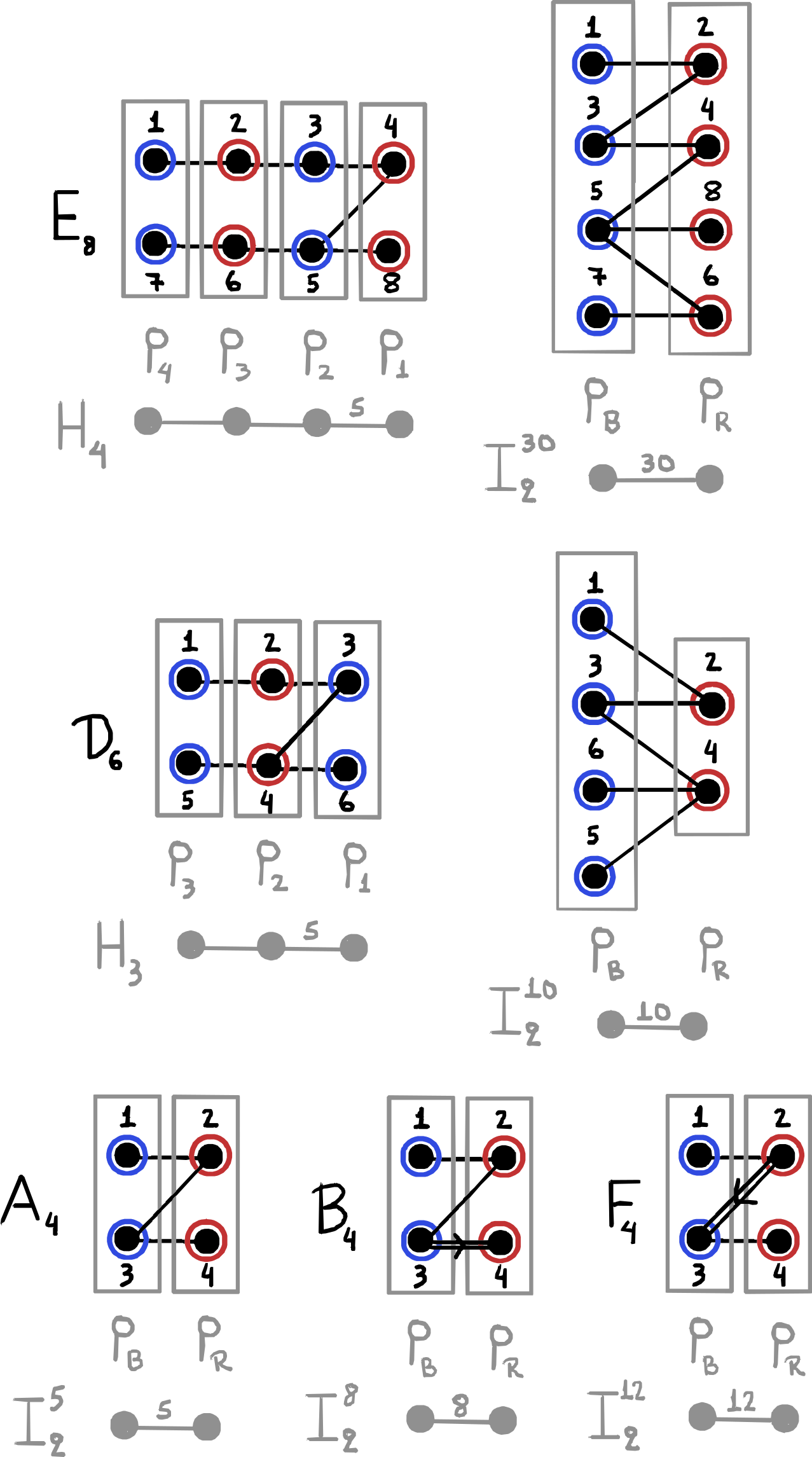}
\caption{\justifying Obtaining non-crystallographic subgroups of finite Coxeter groups by folding.  Top row: the foldings yielding the $H_{4}$ and $I_{2}^{30}$ subgroups of $E_{8}$.
Middle row: the foldings yielding the $H_{3}$ and $I_{2}^{10}$ subgroups of $D_{6}$. Bottom row: the foldings yielding the $I_{2}^{5}$, $I_{2}^{8}$ and $I_{2}^{12}$ subgroups of $A_{4}$, $B_{4}$ and $F_4$ respectively.}
\label{fig:non-crystallographicEuclideanFoldings}
\end{figure}

One way to obtain a non-crystallographic Coxeter subgroup $H$ of a Coxeter group $G$ is by {\bf folding} its Dynkin diagram. This is a process where we partition the fundamental generators of $G$ into subsets, where all the generators in a given subset commute with one another; and then, for each subset, we take the product of all the generators in that subset, and regard that product as a new fundamental generator (see also \cite{Dechant2016}). For example (see the upper left of Fig.~\ref{fig:non-crystallographicEuclideanFoldings}), if we partition the 8 generators $\{R_{1},\ldots,R_{8}\}$ of $E_{8}$ into four subsets; $\{R_{4},R_{8}\}$, $\{R_{3},R_{5}\}$, $\{R_{2},R_{6}\}$, $\{R_{1},R_{7}\}$, and use these subsets to define the new generators $\rho_{1}=R_{4}R_{8}$, $\rho_{2}=R_{3}R_{5}$, $\rho_{3}=R_{2}R_{6}$ and $\rho_{4}=R_{1}R_{7}$, we find that the $\rho_{i}$'s obey the Coxeter relations of non-crystallographic Coxeter subgroup $H_{4}\in E_{8}$.~Fig.~\ref{fig:non-crystallographicEuclideanFoldings} illustrates some important examples in which finite Coxeter groups are folded to obtain their non-crystallographic subgroups. These are the most important foldings from the standpoint of constructing ordinary Euclidean quasicrystals.~All foldings that lead to the 2D non-crystallographic $I_2^n$ arise from grouping all blue and red generators together to form $\rho_B$ and $\rho_R$, and correspond geometrically to a projection onto the 2D Coxeter planes (see Subsections \ref{subsec:Coxeter_Elements} and \ref{subsubsec:Hyperbolic_Coxeter_Elements}).

\subsection{Symmetries of Lattices}
\label{subsec:Symmetries_of_Lattices}

Given a lattice $\Lambda$ in vector space $V$ with inner product $\langle\,,\,\rangle$, the {\bf automorphism group} of $\Lambda$, ${\rm Aut}(\Lambda)$,
is the group of linear transformations from $V\to V$ that map $\Lambda$ to itself and preserve the inner product between vectors. It is a discrete subgroup of the orthogonal group $O(V)$.

A {\bf root} of the lattice $\Lambda$ is a vector ${\bf r}$ of positive norm $N({\bf r})=\langle {\bf r},{\bf r}\rangle > 0$ for which the associated reflection is a symmetry of $\Lambda$. If $\Lambda$ is self-dual, its roots are just the lattice vectors ${\bf r}\in \Lambda$ whose norm $N({\bf r})$ equals $1$ or $2$. In particular, the lattice vectors with $N({\bf r})=1$ or $N({\bf r})=2$ are called the {\bf short roots} and {\bf long roots}, respectively.  

The group generated by reflections in the mirrors corresponding to these roots is the {\bf reflection subgroup} $\Gamma$ of ${\rm Aut}(\Lambda)$.  An important subclass of lattices are {\bf reflective lattices} -- these are lattices for which the reflection subgroup $\Gamma$ has finite index in ${\rm Aut}(\Lambda)$. In other words, in a reflective lattice, nearly all of the automorphisms are generated by reflections in mirrors.  

\subsection{Coxeter Elements}
\label{subsec:Coxeter_Elements}

The product of all of a Coxeter group's fundamental reflections (in any order) gives a special type of element in the group called a {\bf Coxeter element} $C$:
\begin{equation}
  C=R_1R_2\ldots R_n.
\end{equation}
If the Coxeter-Dynkin graph is a tree or a forest of trees (i.e. no closed loops, as is always the case for finite and affine Coxeter groups) -- the different Coxeter elements (obtained by multiplying the fundamental roots in different orders) are all equivalent up to symmetry ({\it i.e.}\ in the same conjugacy class in the Coxeter group \cite{humphreys_1990}), so we may speak of {\it the} Coxeter element.  However, we will later see that for some hyperbolic Coxeter groups, the diagrams do have closed loops, and the various Coxeter elements are {\it not} equivalent to one another ({\it e.g.}\ they have different eigenvalues).

The order $h$ of a Coxeter element, {\it i.e.} the smallest positive integer $h$ such that
\begin{equation}
  C^{h}=1,
\end{equation}
is called the {\bf Coxeter number}, and will be finite (resp. infinite) for a finite (resp. infinite) Coxeter group.  

For finite $G$, the eigenvalues of $C$ are the {\bf primitive $h$-th roots of unity} $\zeta_h^{m_i}$, where $\zeta_h\equiv{\rm exp}(2\pi i/h)$, and $m_i$ runs over the {\bf totatives} of $h$ (i.e. the positive integers less than $h$ and coprime to $h$).  These eigenvalues occur in complex conjugate pairs $\zeta_{h}^{m_i}$ and $(\zeta_{h}^{m_i})^{\ast}=\zeta_{h}^{h-m_i}$. Letting $\mathbf{z}_{m_i},\mathbf{z}_{m_i}^*$ be the corresponding pair of complex-conjugate eigenvectors of $C$, combinations of the form $\alpha \mathbf{z}_{m_i} + \alpha^*\mathbf{z}_{m_i}^*$ span a real, two-dimensional eigenspace called a \textbf{Coxeter plane}, and the Coxeter element rotates this plane by an angle equal to $2\pi m/h$. 

Since the Coxeter elements $C$ are the highest-order elements in $G$, these Coxeter planes are particularly symmetric ({\it i.e.}\ they are the planes preserved by the largest subgroup of $G$), and hence can be used to construct particularly symmetrical quasicrystals. 

\subsection{Algebraic Number Theory}
\label{subsec:AlgebraicNumberTheory}

Here we briefly introduce a few basic definitions, ideas and results in {\bf algebraic number theory}, which will be useful below. For more details, see Refs.~\cite{serre2012arithmetic, lang1994algebraic, cassels1967algebraic} for standard introductions to algebraic number theory.

A {\bf field} is a commutative ring in which the additive identity $0$ and the multiplicative identity $1$ are distinct, and any non-zero element has a multiplicative inverse.  Classic examples are the rational numbers $\mathbb{Q}$, the real numbers $\mathbb{R}$, and the complex numbers $\mathbb{C}$.

An {\bf algebraic number field} $K$ is a field that is also a finite dimensional vector space over $\mathbb{Q}$.  In particular, if the $K$ is a $d$-dimensional vector space over $\mathbb{Q}$, we say that the {\bf field extension} $K/\mathbb{Q}$ has {\bf degree} $d$.  For example, the field $\mathbb{Q}[\sqrt{D}]$ (of all numbers of the form $a+b\sqrt{D}$, where $a$ and $b$ are rational numbers, and $D$ is some fixed square-free integer) may be regarded as a two-dimensional vector space over $\mathbb{Q}$ (with basis vectors $1$ and $\sqrt{D}$), and hence is an algebraic number field of degree two. Algebraic number fields of degree two are particularly simple, and are also known as {\bf quadratic fields}.  $\mathbb{Q}[\sqrt{D}]$ is a {\bf real quadratic field} when $D>0$ and an {\bf imaginary quadratic field} when $D<0$.

A {\bf monic polynomial} $P(x)$ is a polynomial in $x$ whose leading coefficient is $1$.  To any element $\alpha$ in the algebraic number field $K$ is associated a {\bf minimal polynomial} $P_{\alpha,\mathbb{Q}}(x)$, which is the lowest-order monic polynomial with coefficients in $\mathbb{Q}$, having $\alpha$ as a root.  The roots of $P_{\alpha,\mathbb{Q}}(x)$ are called the {\bf algebraic conjugates} (or just {\bf conjugates}) of $\alpha$.\footnote{If the extension $K/\mathbb{Q}$ is {\it normal}, one can equivalently say the conjugates of $\alpha$ are the images of $\alpha$ under the field automorphisms of $K$ that fix the elements of $\mathbb{Q}$ -- a key idea in Galois theory.}

An {\bf algebraic integer} is a complex number that is the root of a monic polynomial $P(x)$ with integer coefficients.  Let ${\cal O}_{K}$ denote the {\bf ring of algebraic integers} in $K$.  An element of ${\cal O}_{K}$ whose multiplicative inverse is also in ${\cal O}_{K}$ is called a {\bf unit} of $K$.  The units of $K$ form a group under multiplication: the {\bf unit group} ${\cal O}_{K}^{\times}$. 

If we regard $K$ as a finite-dimensional vector space over $\mathbb{Q}$, then multiplication by any element $\alpha\in K$ defines a linear transformation $m_{\alpha}:K\to K$, $m_{\alpha}(x)=\alpha x$.  The determinant of this linear transformation is called the {\bf field norm} (or just {\bf norm}) of $\alpha$, denoted $N_{K/\mathbb{Q}}(\alpha)$.  The element $\alpha\in K$ is a unit of $K$ if and only if its norm is a unit in $\mathbb{Q}$, {\it i.e.}\ iff $N_{K/\mathbb{Q}}(\alpha)=\pm1$.  

An algebraic number field $K$ of degree $d = [K:\mathbb{Q})]$ may be embedded in $\mathbb{C}$ in $d$ different ways: $r_{1}$ of these are real embeddings (of $K$ in $\mathbb{R}$), while the remaining embeddings come in $r_{2}$ complex-conjugate pairs, so 
\begin{equation}
    d = r_1 + 2r_2.
\end{equation}
One way to determine the numbers $r_1$ and $r_2$ is as follows: if $K$ is $\mathbb{Q}[\alpha]$ (defined as the smallest field containing both $\mathbb{Q}$ and $\alpha$), and $P_{\alpha,\mathbb{Q}}(x)$ is the minimal polynomial of $\alpha$ over $\mathbb{Q}$, then $r_1$ is the number of its roots that are real, and $2r_2$ is the number that are complex.

{\bf Dirichlet's Unit Theorem} then tells us that the unit group $\mathcal{O}_{K}^\times$ of the number field $K$ contains exactly $r = r_1 + r_2 - 1$ generators $\{\varepsilon_i\}_{i=1,\dots,r}$, such that any unit can be written uniquely in the form
\begin{equation}
  \zeta \varepsilon_1^{m_1} \dots \varepsilon_{r}^{m_r}
\end{equation}
with $m_{i}\in\mathbb{Z}$, and $\zeta$ a root of unity in $\mathcal{O}_{K}^{\times}$. As an abstract group, $\mathcal{O}^\times_K \simeq \mu(\mathcal{O}) \times \mathbb{Z}^{r}$; $r$ is called the {\bf rank} of $\mathcal{O}_K$, $\mu(\mathcal{O})$ is called the {\bf torsion group} and the set $\{ \varepsilon_i\}_{i=1,\dots,r}$ are the {\bf fundamental units} of $K$.

\section{Quasicrystals in Euclidean Space}
\label{sec:CNPSchemes}

{\bf Quasicrystals} are aperiodic patterns that nevertheless have a pure-point Fourier transform (diffraction pattern), like a periodic crystal.  And for particularly interesting quasicrystals (like the Penrose tiling or Ammann-Beenker tiling), these properties (along with a type of discrete scale invariance or self-similarity) follow from the quasicrystal's underlying orientational symmetry.  

In this section we introduce the cut-and project (C\&P) scheme for constructing quasicrystals -- first the ordinary  C\&P scheme in Sec.~\ref{subsec:CNP}, and then the symmetric C\&P (sC\&P) scheme in Sec.~\ref{subsec:sCNP} -- before discussing the symmetries of these quasicrystals in Sec.~\ref{subsec:CNPsymmetries}.

First, let us introduce some terminology: Given a $d$-dimensional lattice $\Lambda\in \mathbb{R}^{d}$, and a $k$-dimensional subspace $W\subset V$, we say that $W$ is:
\begin{itemize}
\item {\bf rational} if $W\cap\Lambda$ has rank $\rho=k$.
\item {\bf partially irrational} if $W\cap\Lambda$ has rank $0<\rho<k$,
\item {\bf totally irrational} if $W\cap\Lambda$ has rank $0$.
\end{itemize}
In other words, the subspace $W$ is totally irrational if it intersects the origin of $\Lambda$, but no other points in $\Lambda$.

\subsection{The Cut and Project (``C\&P'') Scheme}
\label{subsec:CNP}

We first review the non-symmetric cut-and-project (C\&P) algorithm.  (Note that C\&P quasicrystals are also referred to as {\bf model sets} in the mathematical literature \cite{baake2013aperiodic}).  For more details, see Ch.~7 in \cite{baake2013aperiodic}.

\subsubsection{Construction}
\label{subsec:CNPconstruction}

\begin{figure*}[t]

\subfloat[C\&P Method: Pattern $\Lambda_{{\rm ph}}^{{\bf t},{\cal W}}$]{%
        \includegraphics[width=0.48\linewidth]{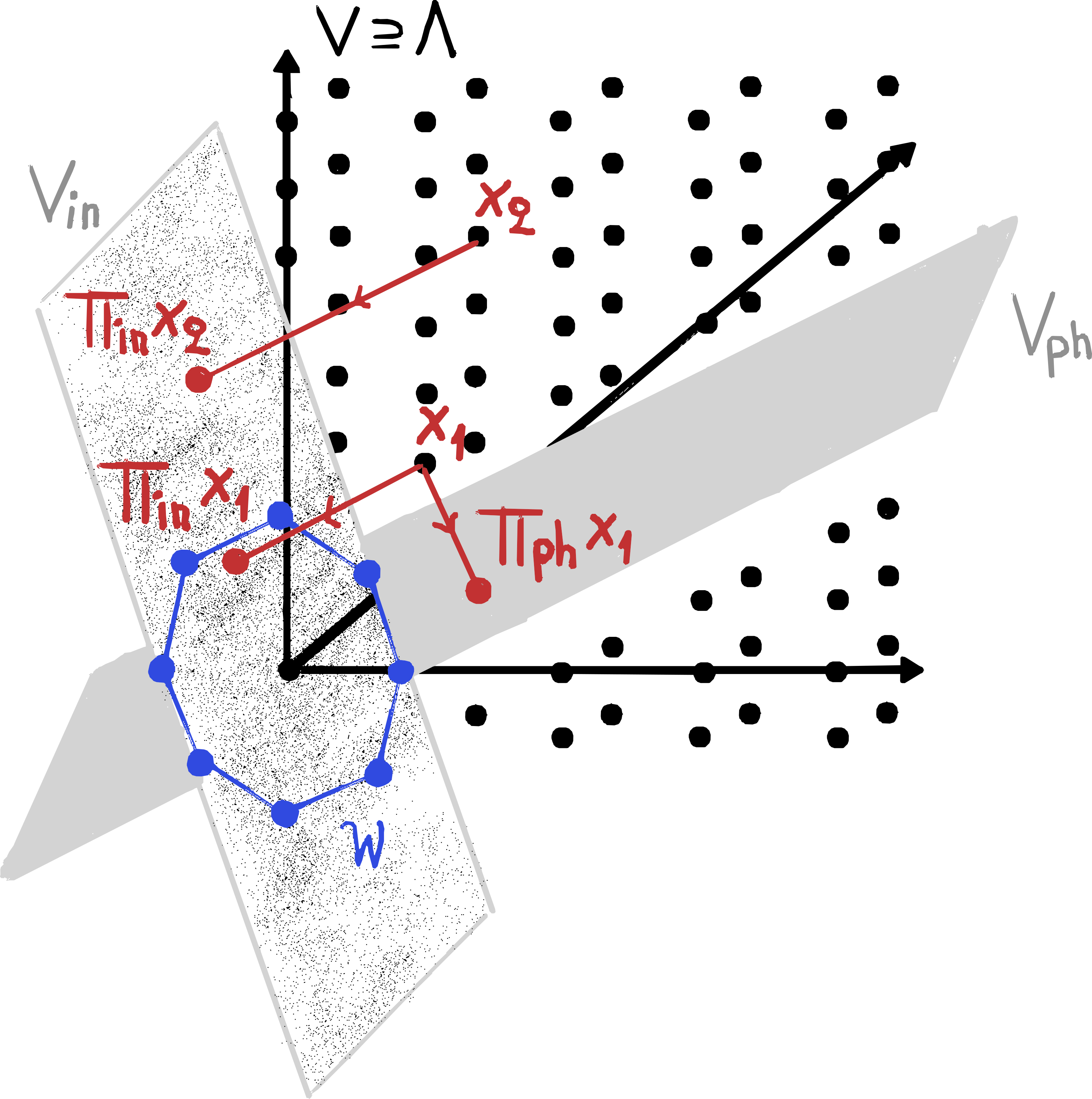}%
        \label{fig:CNPSchemeStandard}%
    }
    \hspace{+1ex}
    \subfloat[C\&P Method: Amplitude Pattern $\widehat{\Lambda}_{{\rm ph}}^{\,{\bf t},W}\!({\bf p}_{{\rm ph}})$]{%
        \includegraphics[width=0.48\linewidth]{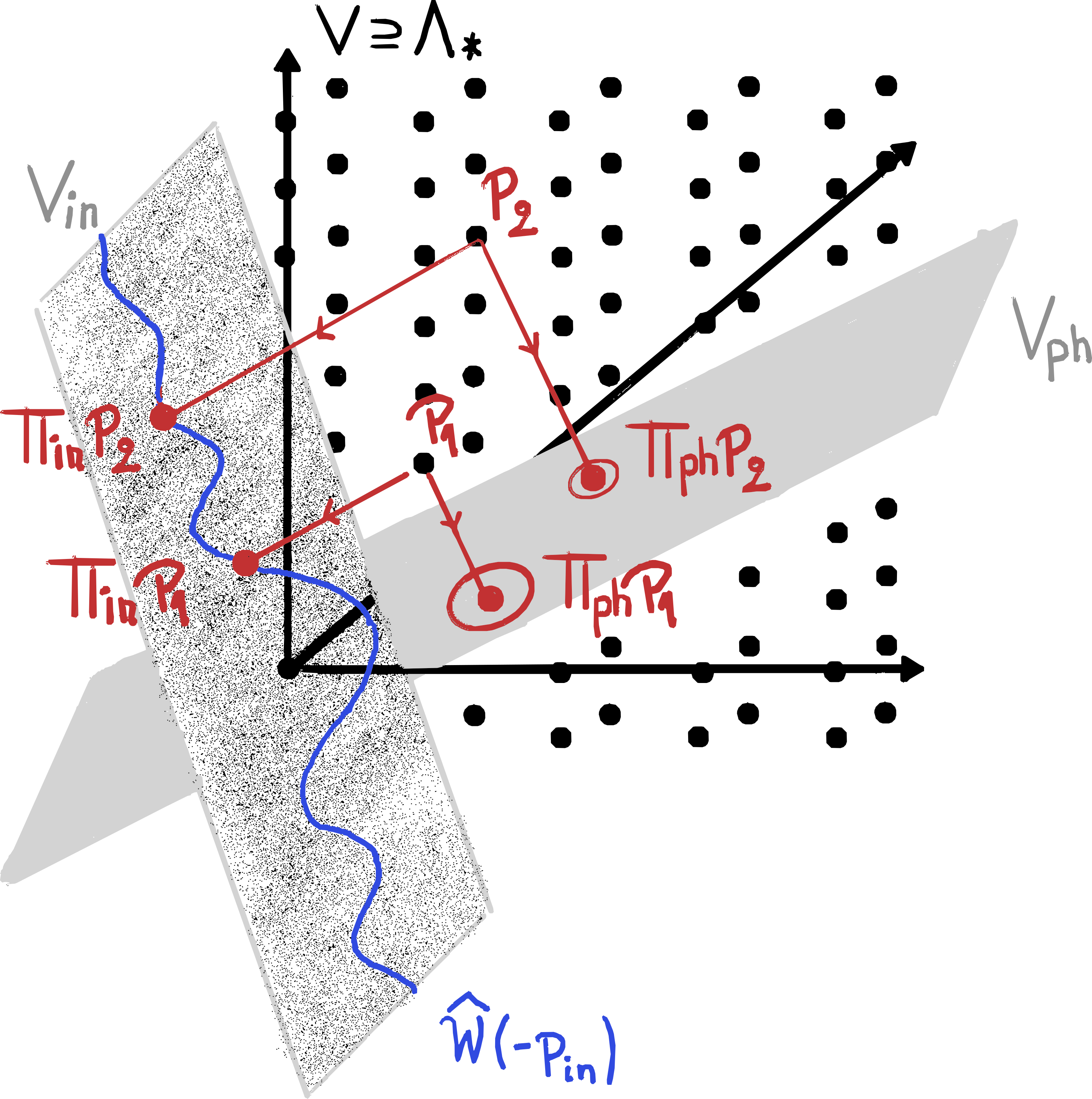}%
        \label{fig:CNPSchemeDiffraction}%
    }
    \caption{\justifying An illustration of the C\&P scheme explained in Sec.~\ref{subsec:CNP}. Left panel: constructing the quasicrystal $\Lambda_{{\rm ph}}^{{\bf t},W}$.  For each point ${\bf x}\in\Lambda^{{\bf t}}$, if the ``internal'' point ${\bf x}_{\rm in}=\Pi_{\rm in}{\bf x}$ lies inside the window ${\cal W}$, we include the ``physical'' point ${\bf x}_{\rm ph}=\Pi_{\rm ph}{\bf x}$ in the quasicrystal $\Lambda_{{\rm ph}}^{{\bf t},{\cal W}}$ (see Sec.~Sec.~\ref{subsec:CNPconstruction} for details, included the weighted variant of the C\&P scheme).  Right panel: The corresponding Fourier transform $\widehat{\Lambda}_{{\rm ph}}^{{\bf t},W}$ is produced by a closely analogous C\&P scheme: for each point ${\bf p}\in\Lambda_{\ast}$ (the dual lattice), the Fourier transform includes a delta function located at the ``physical'' momentum ${\bf p}_{\rm ph}=\Pi_{\rm ph}{\bf p}$, with a coefficient given by $\widehat{W}(-{\bf p}_{in})$, where $\widehat{W}$ is the Fourier transform of the ``top-hat'' or ``indicator'' function in the shape of the window ${\cal W}$, which is evaluated at the ``internal'' momentum ${\bf p}_{\rm in}=\Pi_{\rm in}{\bf p}$. As a display convention, one can represent the ``height'' of the delta function at ${\bf p}_{\rm ph}$ by a circular disk centered at ${\bf p}_{\rm ph}$, with area proportional to the coefficient $\widehat{W}(-{\bf p}_{\rm in})$, as indicated by the red circles in the right panel.} 
    \label{fig:CNPScheme}
\end{figure*}

To construct a specific quasicrystal via C\&P, we first make three choices:
\begin{itemize}
\item {\bf Choice 1}: a $d$-dimensional lattice $\Lambda^{{\bf t}}$ in $V=\mathbb{R}^{d}$.
\item {\bf Choice 2}: a totally irrational subspace $V_{{\rm ph}}\subset V$ of dimension $0<d_{{\rm ph}}<d$, with totally irrational orthogonal complement $V_{{\rm in}}$, so that $V=V_{{\rm ph}}\oplus V_{{\rm in}}$.  We call $V_{{\rm ph}}$ and $V_{{\rm in}}$ the ``{\bf physical space}'' and ``{\bf internal space},'' respectively.  
\item {\bf Choice 3}: a subset ${\cal W}$ of $V_{{\rm in}}$ (called the ``{\bf window}'') or a function 
$W:V_{{\rm in}}\to\mathbb{C}$ (called the ``{\bf weighting function''}).
\end{itemize}  

It is convenient to let $\Pi_{{\rm ph}}$ and $\Pi_{{\rm in}}$ denote the projection operators onto the physical and internal subspaces, respectively ($V_{{\rm ph}}=\Pi_{{\rm ph}}V$ and $V_{{\rm in}}=\Pi_{{\rm in}}V$); so we have $\Pi_{{\rm ph}}^{2}=\Pi_{{\rm ph}}$, $\Pi_{{\rm in}}^{2}=\Pi_{{\rm in}}$, $\Pi_{{\rm ph}}\Pi_{{\rm in}}=\Pi_{{\rm in}}\Pi_{{\rm ph}}=0$ and $\Pi_{{\rm ph}}+\Pi_{{\rm in}}=1$.  More explicitly, we can always choose a basis $\{{\bf u}_i\}$ for $V$ that is orthogonal ($\langle {\bf u}_i,{\bf u}_j\rangle=0$ for $i\neq j$) and adapted to the decomposition $V=V_{{\rm ph}}\oplus V_{{\rm in}}$:
\begin{equation}
  \{
  \underbrace{{\bf u}_{1},\ldots,{\bf u}_{d_{{\rm ph}}}^{}}_{\text{spans}\;V_{{\rm ph}}},
  \underbrace{{\bf u}_{d_{{\rm ph}}+1}^{},\ldots,{\bf u}_{d}}_{\text{spans}\;V_{{\rm in}}}\}.
\end{equation}
In an arbitrary basis $\{{\bf v}_{\mu}\}$, where ${\bf u}_{i}=u_{i}^{\mu}{\bf v}_{\mu}$, $\Pi_{{\rm ph}}$ is then 
\begin{equation}
  \label{def_Pi}
  (\Pi_{{\rm ph}})^{\mu}_{\;\;\nu}=\sum_{i=1}^{d_{{\rm ph}}}\frac{u_{i}^{\mu}u_{i,\nu}}{u_{i}^{2}}.
\end{equation}
In general, we will use the subscripts ``{\rm ph}'' and ``{\rm in}'' to denote quantities that have been projected onto the physical or internal subspaces, respectively.

Given these three choices, we construct the corresponding quasicrystal using one of the following two schemes (as appropriate):
\begin{itemize}
\item {\bf The window scheme.} If (in Choice 3) we choose the window ${\cal W}\subset V_{{\rm in}}$, then the corresponding quasicrystal $\Lambda_{{\rm ph}}^{{\bf t},{\cal W}}$ is a {\it subset} of $V_{{\rm ph}}$, given by
\begin{equation}
  \label{eq:window_scheme}
  \Lambda_{{\rm ph}}^{{\bf t},{\cal W}}
  =\{{\bf x}_{{\rm ph}}|{\bf x}\in\Lambda^{{\bf t}}, {\bf x}_{{\rm in}}\in{\cal W}\}
\end{equation}
In other words, for each lattice point ${\bf x}\in\Lambda^{{\bf t}}$, if its {\it internal} position ${\bf x}_{{\rm in}}=\Pi_{{\rm in}}{\bf x}$ lies in the window $\mathcal{W}$, we include its {\it physical} position ${\bf x}_{{\rm ph}}=\Pi_{{\rm ph}}{\bf x}$ as a point in the quasicrystal $\Lambda_{{\rm ph}}$ (see Fig.~\ref{fig:CNPSchemeStandard}).

\item {\bf The weighting scheme.} If we instead choose the weighting function $W:V_{{\rm in}}\to\mathbb{C}$, the corresponding quasicrystal $\Lambda_{{\rm ph}}^{{\bf t},W}$ is a {\it distribution} on $V_{{\rm ph}}$ given by
\begin{equation}
  \label{eq:weighting_scheme}
  \Lambda_{{\rm ph}}^{{\bf t},W}({\bf v}_{{\rm ph}})=
  \sum_{{\bf x}\in\Lambda^{{\bf t}}}W({\bf x}_{{\rm in}})\delta({\bf v}_{{\rm ph}}-{\bf x}_{{\rm ph}}).
\end{equation}
In other words, for each lattice point ${\bf x}\in \Lambda^{{\bf t}}$, we add a delta function at the corresponding {\it physical} position ${\bf x}_{{\rm ph}}=\Pi_{{\rm ph}}{\bf x}$, with an amplitude given by the weighting function $W$ evaluated at the {\it internal} position ${\bf x}_{{\rm in}}=\Pi_{{\rm in}}{\bf x}$.
\end{itemize}

The window scheme is essentially the special case of the weighting scheme where the weighting function $W$ is the ``indicator function'' for region ${\cal W}$:
\begin{equation}
  \label{indicator_function}
  W({\bf v}_{{\rm in}})=\left\{\begin{array}{ll}
  1, & \text{ if } {\bf v}_{{\rm in}}\in \mathcal{W}, \\
  0, & \text{ if } {\bf v}_{{\rm in}}\notin \mathcal{W}.\end{array}\right.
\end{equation}

In the window scheme, the average {\it number density} $\bar{n}_{\rm ph}$ (in $V_{{\rm ph}}$) of points in $\Lambda_{{\rm ph}}^{{\bf t},{\cal W}}$ equals the average number density $\bar{n}$ (in $V$) of point in $\Lambda$, times the volume $V_{{\cal W}}$ (in $V_{{\rm in}}$) of ${\cal W}$:
\begin{equation}
  \label{number_density_relation}
  \bar{n}_{\rm ph}=\bar{n}V_{\cal W}.
\end{equation}
Hence, to obtain a quasicrystal with a finite density of points, we must have a window $\mathcal{W}$ of finite volume.

In the weighting scheme, if the weighting function $W$ is non-zero over the entirety of $V_{{\rm in}}$, the resulting distribution $\Lambda_{{\rm ph}}^{{\bf t},W}$
will have delta function ``peaks'' living over a dense set of points in $V_{{\rm ph}}$.  In such a situation, it is less useful to talk about the number density of such delta function peaks (which is infinite); but rather, we can think about each unit delta function as the mass density distribution of a unit point mass, and talk about the average {\it mass density} of the distribution $\Lambda_{{\rm ph}}^{{\bf t},W}$ as a whole.  In particular, the average mass density $\bar{\rho}_{\rm ph}$ (in $V_{{\rm ph}}$) of points in $\Lambda_{{\rm ph}}^{{\bf t},W}$ equals the average mass density $\bar{\rho}$ (in $V$) of point in $\Lambda$, times the integral of the weighting function $W({\rm v}_{\rm in})$ over $V_{\rm in}$:
\begin{eqnarray}
  \label{mass_density_relation}
  \bar{\rho}_{\rm ph}&=&\lim_{R\to\infty}
  \frac{\int_{R}d{\bf v}_{{\rm ph}}\Lambda_{{\rm ph}}^{{\bf t},W}({\bf v}_{{\rm ph}})}{\int_{R}d{\bf v}_{{\rm ph}}} \nonumber\\
  &=&\bar{\rho}\!\int\!d{\rm v}_{{\rm in}}
  W({\bf v}_{{\rm in}}).
\end{eqnarray}
Hence, to obtain a quasicrystal with finite average mass density, the integral of $W({\bf v}_{{\rm in}})$ over $V_{{\rm in}}$ must be finite.

\subsubsection{Diffraction Pattern}
\label{subsec:CNPdiffraction}

Let us now compute the diffraction pattern (Fourier transform) of the quasicrystal $\Lambda_{{\rm ph}}^{{\bf t},W}({\bf v}_{{\rm ph}})$. 

In our Fourier conventions, a function $f({\bf v})$ and its Fourier transform $\widehat{f}({\bf k})$ are related by
\begin{subequations}
  \begin{eqnarray}
    \widehat{f}\,({\bf k})&=&\int d{\bf v}\, {\rm e}^{-2\pi i\langle {\bf k},{\bf v}\rangle} f({\bf v}), \\
    f({\bf v})&=&\int d{\bf k}\, {\rm e}^{+2\pi i\langle{\bf k}, {\bf v}\rangle}\widehat{f}\,({\bf k}).
  \end{eqnarray}
\end{subequations}
The Fourier transform of $\Lambda_{{\rm ph}}^{{\bf t},W}({\bf v}_{{\rm ph}})$ is then given by 
\begin{subequations}
  \begin{eqnarray}
    \label{Fourier1}
    \widehat{\Lambda}_{{\rm ph}}^{\,{\bf t},W}\!({\bf k}_{{\rm ph}})
    &=&\int d{\bf v}_{{\rm ph}}{\rm e}^{-2\pi i\langle{\bf k}_{{\rm ph}},{\bf v}_{{\rm ph}}\rangle}\Lambda_{{\rm ph}}^{{\bf t},W}({\bf v}_{{\rm ph}}) \\
    \label{Fourier2}
    &=&\sum_{{\bf x}\in\Lambda^{{\bf t}}}\!{\rm e}^{-2\pi i\langle{\bf k}_{{\rm ph}},{\bf x}_{{\rm ph}}\rangle}W({\bf x}_{{\rm in}}) \\
    \label{Fourier3}
    &=&\frac{{\rm e}^{-2\pi i \langle {\bf k}_{{\rm ph}},{\bf t}\rangle}}{{\rm Vol}(V/\Lambda)}\sum_{{\bf p}\in\Lambda_{*}}
    {\rm e}^{-2\pi i\langle{\bf p}_{{\rm in}},{\bf t}\rangle}\widehat{W}(-{\bf p}_{{\rm in}})\qquad\nonumber\\
    &&
    \qquad\qquad\qquad\qquad\times\delta({\bf k}_{{\rm ph}}-{\bf p}_{{\rm ph}})\qquad
  \end{eqnarray}
\end{subequations}
where, in (\ref{Fourier2}) we have substituted (\ref{eq:weighting_scheme}) and used the delta function $\delta({\bf v}_{{\rm ph}}-{\bf x}_{{\rm ph}})$ to evaluate the ${\bf v}_{{\rm ph}}$ integral, and in (\ref{Fourier3}) we have used the Poisson summation formula $\sum_{{\bf x}\in\Lambda}f({\bf x})\!=\!\frac{1}{{\rm Vol}(V/\Lambda)}\sum_{{\bf p}\in\Lambda_{\ast}}\!\widehat{f}({\bf p})$ \cite{baake2013aperiodic} and then simplified using $d{\bf x}=d{\bf x}_{{\rm ph}}d{\bf x}_{{\rm in}}$ and $\langle{\bf k},{\bf x}\rangle = \langle{\bf k}_{{\rm ph}}, {\bf x}_{{\rm ph}}\rangle+\langle{\bf k}_{{\rm in}},{\bf x}_{{\rm in}}\rangle$.  

In other words, (\ref{Fourier3}) says that $\widehat{\Lambda}_{{\rm ph}}^{\,{\bf t},W}({\bf k}_{{\rm ph}})$ is a pure-point diffraction spectrum that, apart from the pre-factor $\frac{1}{{\rm Vol}(V/\Lambda)}{\rm e}^{-2\pi i\langle {\bf k}_{{\rm ph}},{\bf t}\rangle}$, is given by a C\&P procedure closely analogous to the one for $\Lambda_{{\rm ph}}^{{\bf t},W}({\bf v}_{{\rm ph}})$: for each dual lattice point ${\bf p}\in\Lambda_{*}$, we add a delta function at the corresponding {\it physical} momentum ${\bf p}_{{\rm ph}}=\Pi_{{\rm ph}}{\bf p}$, with an amplitude determined by the {\it internal} momentum ${\bf p}_{{\rm in}}$ and given, in particular, by ${\rm e}^{-2\pi i\langle {\bf p}_{{\rm in}},{\bf t}\rangle}\widehat{W}(-{\bf p}_{{\rm in}})$ (see Fig.~\ref{fig:CNPSchemeDiffraction}).

\subsubsection{Self-Dual Quasicrystals}
\label{subsec:SelfDualQuasicrystals}

First, consider periodic crystals.  By the Poisson summation formula, the distribution
\begin{equation}
  \gamma({\bf v})=\sum_{{\bf x}\in\Lambda}\delta({\bf v}-{\bf x})
\end{equation}
has Fourier transform
\begin{equation}
  \widehat{\gamma}({\bf k})=\sum_{{\bf p}\in\Lambda_{*}}
  \delta({\bf k}-{\bf p}).
\end{equation}
So if the lattice $\Lambda$ is self-dual ($\Lambda=\Lambda_{*}$), the distribution $\gamma$ is also self-dual in the sense of invariance under Fourier transform ($\gamma=\widehat{\gamma}$).  We say that such a distribution $\gamma$ is a {\bf self-dual crystal}.

Analogously, if the C\&P quasicrystal $\Lambda_{{\rm ph}}^{{\bf t},W}$ is invariant under Fourier transform ($\Lambda_{{\rm ph}}^{{\bf t},W}=\widehat{\Lambda}_{{\rm ph}}^{\,{\bf t},W}$) we say it is a {\bf self-dual quasicrystal}.  Comparing Eqs.~(\ref{eq:weighting_scheme}) and (\ref{Fourier3}), we see this happens when the higher dimensional lattice $\Lambda$ is self-dual ($\Lambda=\Lambda_{*}$), untranslated (${\bf t}=0$ or, more generally ${\bf t}\in\Lambda$), and the weighting function is invariant under Fourier transform combined with inversion through the origin ($W({\bf x})=\widehat{W}(-{\bf x})$).

Note that this {\it linear} definition of self-duality is stronger than the {\it quadratic} one defined in Section \S 9.5.2 of \cite{baake2013aperiodic}: {\it i.e.}\ whereas our linear condition (invariance of the distribution under Fourier transform) implies the  quadratic condition (equality of the autocorrelation and the diffraction measure, both of which are quadratic in the weighting function $W$), the converse is not true (since the quadratic condition discards all phase information).

\begin{figure}
\includegraphics[width=0.6\linewidth]{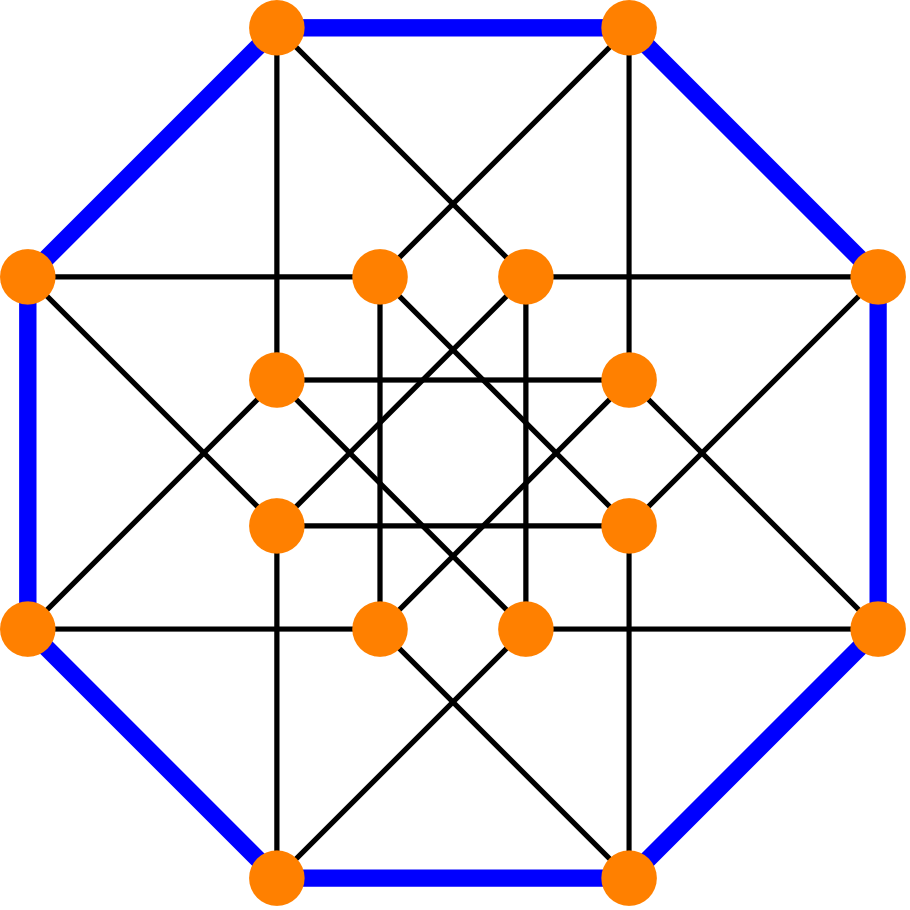}
\caption{\justifying The projection of the 4D unit hypercube $\gamma_{4}$ onto the internal space $V_{{\rm in}}$.  The convex hull (blue octagon) defines the window $\mathcal{W}_{AB}$ used to generate the 8-fold-symmetric Ammann Beenker tiling shown in Fig.~\ref{fig:AmmannBeenker}.} 
\label{fig:sCNPWindowABTiling} 
\end{figure}

\begin{figure*}
    \begin{tabular}{cc}
        \subfloat[\label{fig:sCNP-ABPattern}]{
            \includegraphics[width=0.3\linewidth]{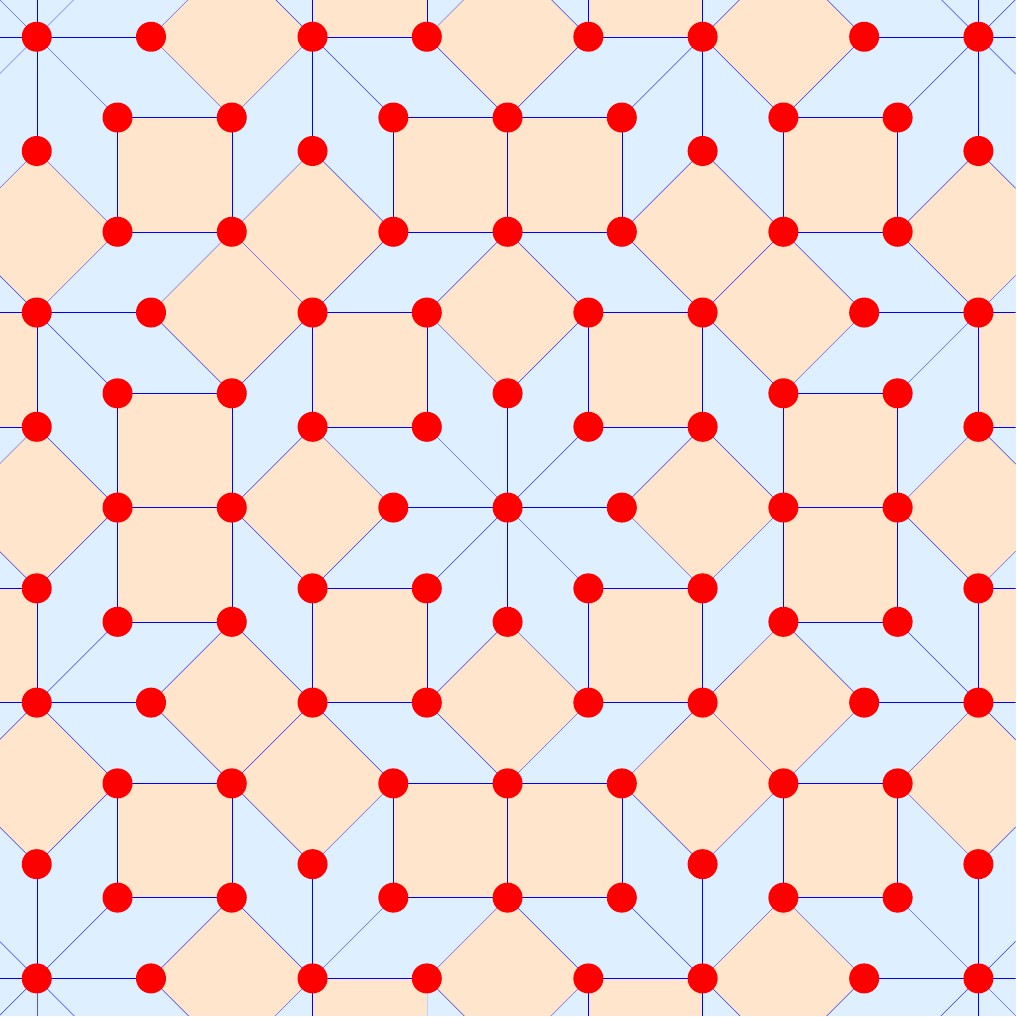}
            \hspace{1em}}
        \subfloat[\label{fig:sCNP-ABApproxAmplitudes}]{    
            \includegraphics[width=0.3\linewidth]{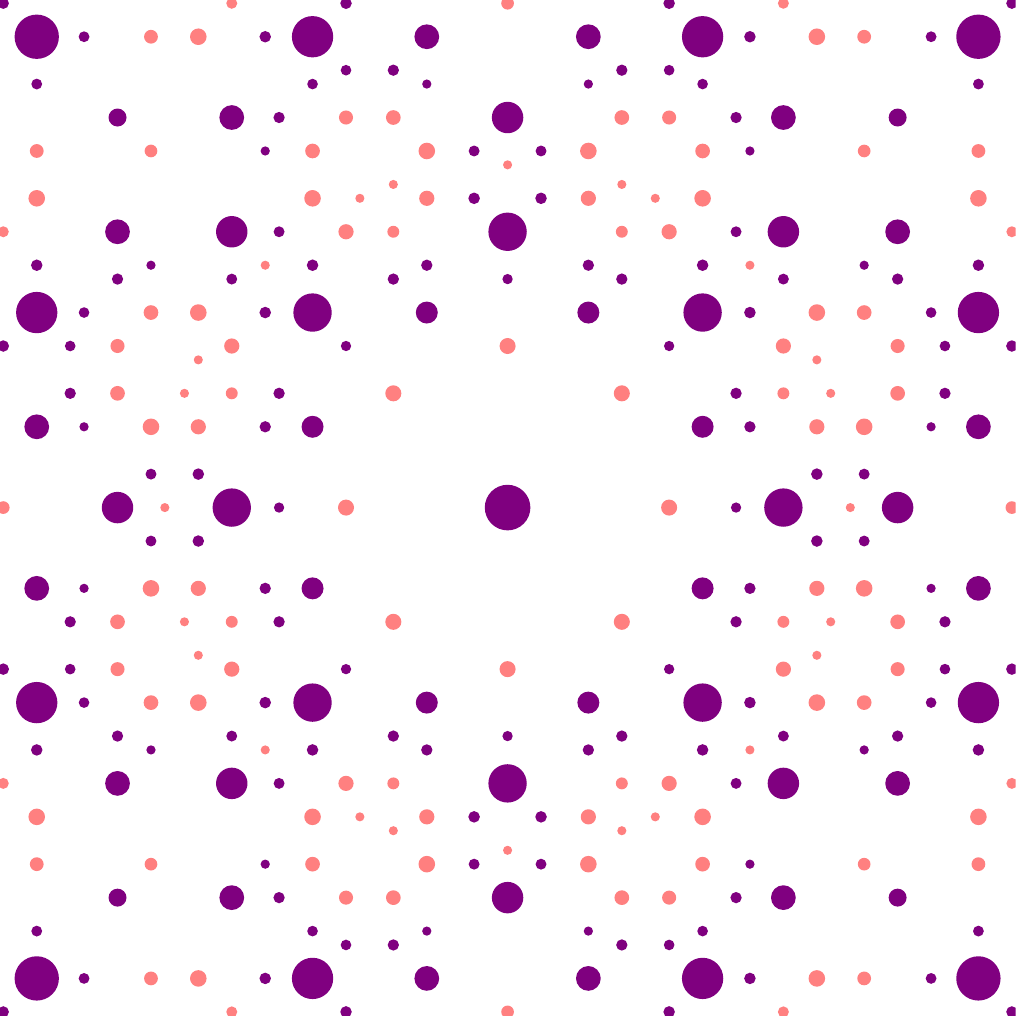}
        } \hspace{1em}
        \subfloat[\label{fig:sCNP-ABSelfDualAmplitudes}]{\includegraphics[width=0.3\linewidth]{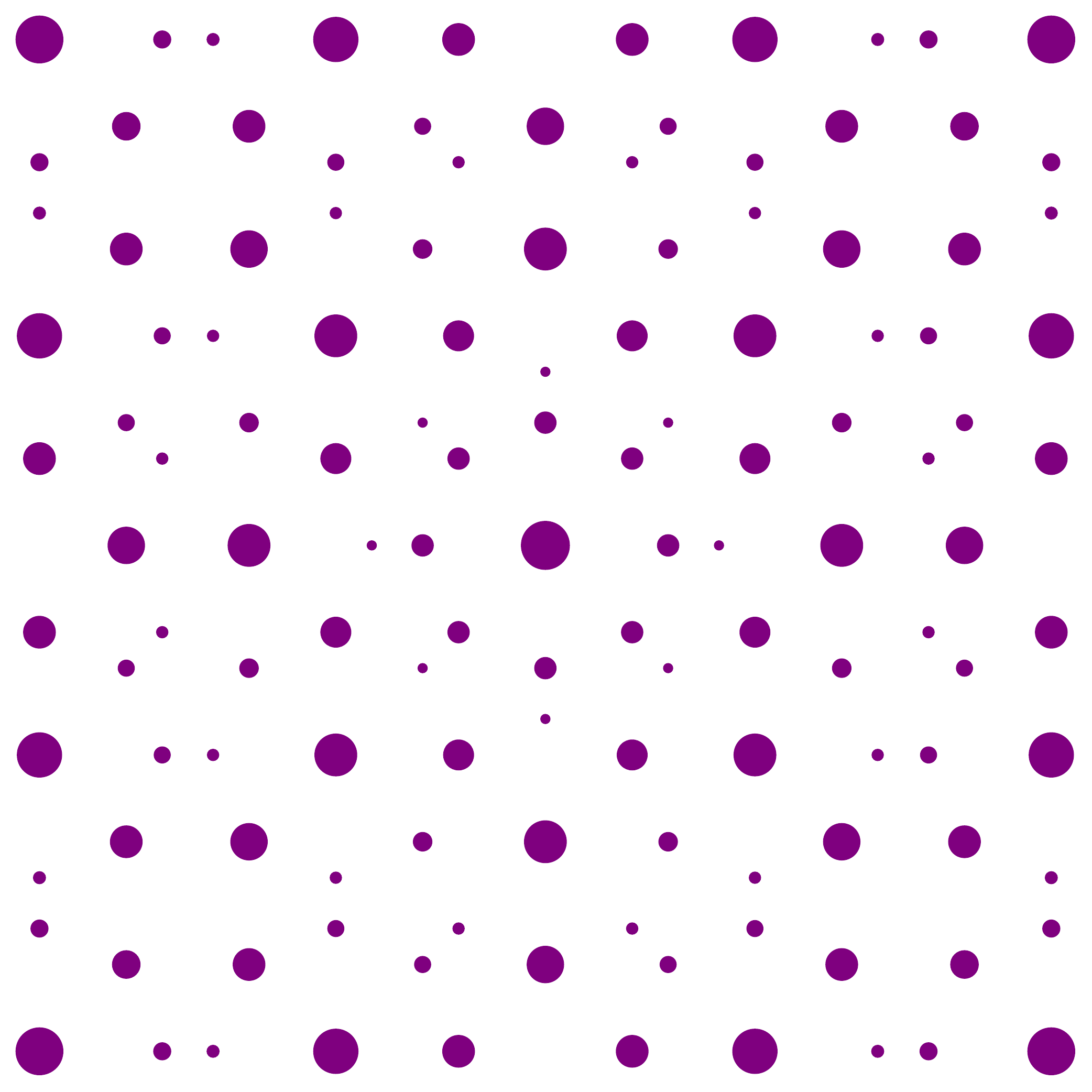}}
    \end{tabular}
\caption{\justifying Illustrating the symmetric cut-and-project (``sC\&P'') scheme (Sec.~\ref{subsec:sCNP}) for Euclidean quasicrystals with 8-fold orientational symmetry.  Left: a patch of the  Ammann-Beenker tiling, obtained via sC\&P as explained in  Sec.~\ref{subsec:AmmannBeenkerTilingsExample}. Middle: a corresponding patch of its Fourier transform. Right: the self-dual variant of the Ammann-Beenker quasicrystal (which is the same as its Fourier transform) -- see Sec.~\ref{self-dual_AB}.  In the middle and right panels, each circle's area represents the coefficient of the delta function located at its center; purple (pink) circles indicate positive (negative) amplitudes; and only amplitudes with absolute value $>10^{-1.4}$ of the central peak's amplitude are shown (since, if all peaks are included, they densely fill the plane, but most have tiny coefficients).  Note that all three patches/panels are 8-fold symmetric; and are "the same patch" (but with different weightings/coefficients assigned to the points in the patch).}
\label{fig:AmmannBeenker}
\end{figure*}

Also note that the {\it window} scheme, where $W$ is the indicator function (\ref{indicator_function}) for some compact window $\mathcal{W}$, does {\it not} yield a self-dual quasicrystal.  In particular, $W$ is non-vanishing on a compact region in $V_{{\rm in}}$, implying that $\Lambda_{{\rm ph}}^{{\bf t},W}$ has delta functions on a finite-density (non-dense) set of points in $V_{{\rm ph}}$; while $\widehat{W}$ is non-vanishing on a non-compact region of $V_{{\rm in}}$, implying that $\widehat{\Lambda}_{{\rm ph}}^{{\bf t},W}$ has delta functions on a dense set of points in $V_{{\rm ph}}$ ({\it i.e.}\ there are delta functions arbitrarily close to any point in $V_{{\rm ph}}$).  By contrast, in a self-dual quasicrystal, $W$ and $\widehat{W}$ are both non-vanishing on a non-compact region of $V_{{\rm in}}$, implying that $\Lambda_{{\rm phys}}^{{\bf t},W}$ and $\widehat{\Lambda}_{{\rm phys}}^{\,{\bf t},W}$ both have delta functions on a dense set of points in $V_{{\rm ph}}$.  For illustration, we present examples of both types (the 8-fold symmetric Ammann-Beenker tiling and its self-dual variant) in  Sec.~\ref{subsec:AmmannBeenkerTilingsExample} and Fig.~\ref{fig:AmmannBeenker}.

\subsection{The Symmetric C\&P (``sC\&P'') Scheme}
\label{subsec:sCNP}

In the general C\&P scheme of Sec.~\ref{subsec:CNP}, a particular quasicrystal is specified by making three choices.  But the most interesting quasicrystals ({\it e.g.}\ the Penrose tiling or the Ammann-Beenker tiling) are produced by a more refined version of this scheme (which we will call the ``symmetric cut-and-project'' or ``sC\&P'' scheme).  

\subsubsection{Construction}
\label{subsec:sCNPconstruction}

In the sC\&P scheme, we again make three choices:
\begin{itemize}
  \item {\bf Choice 1}: a $d$-dimensional lattice $\Lambda^{{\bf t}}$ in $V=\mathbb{R}^{d}$ with automorphism group ${\rm Aut}(\Lambda)$.
  \item {\bf Choice 2}: a non-crystallographic subgroup $G\subset {\rm Aut}(\Lambda)$, and a decomposition $V=V_{{\rm ph}}\oplus V_{{\rm in}}$ that is invariant under $G$: $\Pi_{{\rm ph}}=g\Pi_{{\rm ph}}g^{-1}$ (or equivalently $[\Pi_{{\rm ph}},g]=0$), $\forall g\in G$.
  \item {\bf Choice 3}: a window ${\cal W}\subset V_{{\rm in}}$ or weighting function $W:V_{{\rm in}}\to\mathbb{C}$ invariant under $G$: {\it i.e.}\ ${\cal W}=g{\cal W}$ or $W({\bf v}_{{\rm in}})=W(g\,{\bf v}_{{\rm in}})$, $\forall g\in G$. 
\end{itemize}
This produces a quasicrystal with orientational symmetry $G$ (in the sense discussed below, Sec.~\ref{subsec:CNPsymmetries}).  Note that the total irrationality of $V_{{\rm ph}}$ and $V_{{\rm in}}$ now follows from the fact that $G$ is non-crystallographic.  

\subsubsection{Example 1: Ammann-Beenker Tiling}
\label{subsec:AmmannBeenkerTilingsExample}

As a first example of this procedure, we construct the 8-fold-symmetric Ammann-Beenker tiling.

{\bf Choice 1}: As our lattice $\Lambda^{{\bf t}}$, we choose the standard integer lattice ($\Lambda=\mathbb{Z}^{4}$) with no translation (${\bf t}={\bf 0}$).  ${\rm Aut}(\Lambda)$ is the Coxeter group $B_{4}$, whose fundamental roots ${\bf r}_{i}$ are the columns of the matrix 
\begin{equation*}
B = \left(\begin{array}{cccc}
\,1\; & -1 & 0 & 0\\
0 & 1 & -1& 0 \\
0 & 0 & 1 & -1\\
0 & 0 & 0 & 1
\end{array}\right).
\end{equation*}
From each ${\bf r}_i$ we construct the corresponding reflection matrix $R_i$ from Eq.~(\ref{R_eq}), and hence the corresponding Coxeter element
\begin{equation*}
C = R_1 R_2 R_3 R_4 = 
\left(\begin{array}{cccc}
0 & 0 & 0 & -1\\
\,1\; & 0 & 0 & 0 \\
0 & \,1\; & 0 & 0 \\
0 & 0 & \,1\; & 0
\end{array}\right).
\end{equation*}
$B_{4}$ has Coxeter number $h=8$, so $C^{8}=1$, and $C$'s eigenvalues are primitive 8th roots of unity; in particular 
\begin{equation}
  \zeta_{8},\;\bar{\zeta}_{8},\;\zeta_{8}',\;\bar{\zeta}_{8}',
\end{equation}
where $\zeta_{8}={\rm e}^{2\pi i/8}=(1+i)/\sqrt{2}$, bar denotes complex conjugation $i\to-i$, and prime denote Galois conjugation
$\sqrt{2}\to-\sqrt{2}$.  The corresponding eigenvectors are
\begin{equation}
  {\bf v}, \bar{{\bf v}}, {\bf v}', \bar{{\bf v}}'.
\end{equation}

{\bf Choice 2}: For our decomposition $V=V_{{\rm ph}}\oplus V_{{\rm in}}$, we choose $V_{{\rm ph}}=\{z{\bf v}+\bar{z}\bar{{\bf v}},z\in\mathbb{C}\}$ (the real subspace spanned by ${\bf v}$ and $\bar{{\bf v}}$) and $V_{{\rm in}}=\{z'{\bf v}'+\bar{z}'\bar{{\bf v}}',z'\in\mathbb{C}\}$ (the real subspace spanned by ${\bf v}'$ and $\bar{{\bf v}}'$).  This decomposition is preserved by the cyclic subgroup $\mathbb{Z}_{8}$ generated by $C$ (which acts on both $V_{{\rm ph}}$ and $V_{{\rm in}}$ as the rotational symmetries of a regular octagon) and, more generally, by the non-crystallographic Coxeter group $I_{2}^{8}$ (which acts on both $V_{{\rm ph}}$ and $V_{{\rm in}}$ as the full symmetries of a regular octagon, including both rotations and reflections).  Then, from Eq.~(\ref{def_Pi}), we obtain the corresponding projectors
\begin{equation*}
\Pi_{\rm ph} = 
\frac{1}{2}
\begin{pmatrix}
1 & \frac{\sqrt{2}}{2} & 0 & -\frac{\sqrt{2}}{2}\\
\frac{\sqrt{2}}{2} & 1 & \frac{\sqrt{2}}{2} & 0 \\
0 & \frac{\sqrt{2}}{2} & 1 & \frac{\sqrt{2}}{2} \\
-\frac{\sqrt{2}}{2} & 0 & \frac{\sqrt{2}}{2} & 1
\end{pmatrix} \text{ and } \Pi_{\rm in} = \Pi_{\rm ph}'.
\end{equation*}

{\bf Choice 3}: To define the window ${\cal W}$, we start with the tesseract $\gamma_4$, {\it i.e.}\ the unit hypercube in four dimensions (centered at the origin) whose vertices are all 16 combinations of $\left( \pm \tfrac{1}{2}, \pm \tfrac{1}{2}, \pm \tfrac{1}{2}, \pm \tfrac{1}{2} \right)$. The polytope has 32 edges, 24 faces and 8 three-dimensional cells (or \textit{facets}).  The projection $\Pi_{\rm in}(\gamma_4)$ leads to the regular octagon $\mathcal{W}$ shown in Fig. \ref{fig:sCNPWindowABTiling}, which is invariant under the non-crystallographic subgroup $G=I_{2}^{8}$ from Choice 2.

We show the resulting 8-fold-symmetric ($I_{2}^{8}$-symmetric) quasicrystal $\Lambda_{{\rm ph}}^{{\bf 0},{\cal W}}$ (the vertices of the Ammann-Beenker tiling) in Fig.~\ref{fig:AmmannBeenker} (left panel), and its Fourier transform $\widehat{\Lambda}_{{\rm ph}}^{\,{\bf 0},W}$ in Fig.~\ref{fig:AmmannBeenker} (centre panel).

\subsubsection{Example 2: Self-Dual Variant}
\label{self-dual_AB}

Alternatively, we can replace our previous Choice 3 by:

{\bf Choice 3'}: For the weighting function, we now choose
\begin{equation}
  W({\bf v}_{{\rm in}})={\rm exp}(-\pi\, v_{{\rm in}}^{2}).
\end{equation}
Since this satisfies $W({\bf v}_{{\rm in}})=\widehat{W}(-{\bf v}_{{\rm in}})$, and we have also chosen a self-dual lattice $\Lambda=\mathbb{Z}^{4}=\Lambda_{\ast}$ with no translation (${\bf t}=0$), the resulting quasicrystal is self-dual (identical to its Fourier transform). This self-dual 8-fold-symmetric quasicrystal (the natural self-dual variant of the Ammann-Beenker quasicrystal) is given (for the first time, we believe) in Fig.~\ref{fig:AmmannBeenker} (right panel).

\subsection{Symmetries of Quasicrystals}
\label{subsec:CNPsymmetries}

In this section, we discuss the symmetries and transformations that naturally act on our quasicrystals.

\subsubsection{The Torus Parametrization}
\label{subsec:Torusparametrization}

In Sec.~\ref{subsec:CNPconstruction}, we explained that the C\&P quasicrystal $\Lambda_{{\rm ph}}^{{\bf t},W}$ is determined by three choices: the lattice $\Lambda^{{\bf t}}$, the decomposition $V=V_{{\rm ph}}\oplus V_{{\rm in}}$ ({\it i.e.}\ the projector $\Pi_{{\rm ph}}$), and the window ${\cal W}$ (or weighting function $W$). 

Let us describe three important relationships between quasicrystals described by different values of the parameter ${\bf t}$:
\begin{itemize}

\item {\bf Torus parametrization \cite{baake1997torus}.}  First note that if we shift ${\bf t}\to{\bf t}'={\bf t}+\bm{\lambda}$, with $\bm{\lambda}\in\Lambda$, the two quasicrystals are {\it identical}:
\begin{equation}
  \Lambda_{{\rm ph}}^{{\bf t}+\bm{\lambda},W}({\bf v}_{{\rm ph}})=\Lambda_{{\rm ph}}^{{\bf t},W}({\bf v}_{{\rm ph}})\qquad({\rm for}\;\bm{\lambda}\in\Lambda).
\end{equation}
So two parameters ${\bf t}'$ and ${\bf t}$ that differ by an element $\bm{\lambda}\in\Lambda$ are equivalent, and hence the inequivalent values of ${\bf t}$ form the $d$-dimensional torus $T^{d}=V/\Lambda$.  

\item {\bf Translational equivalence.}  Next note that if we shift ${\bf t}\to{\bf t}'={\bf t}+\Delta{\bf t}$, with $\Delta{\bf t}\in V_{{\rm ph}}$, the two quasicrystals are {\it equivalent up to translation}:
\begin{equation}
  \Lambda_{{\rm ph}}^{{\bf t}+\Delta{\bf t},W}({\bf v}_{{\rm ph}})
  =\Lambda_{{\rm ph}}^{{\bf t},W}({\bf v}_{{\rm ph}}+\Delta{\bf t})\quad({\rm for}\;\Delta{\bf t}\in V_{{\rm ph}}).
\end{equation}

\item {\bf Local indistinguishability.}  Two quasicrystals $\Lambda_{{\rm ph}}^{{\bf t},W}$ with same $\Lambda$ and ${\cal W}$ (or $W$), but different ${\bf t}$ are {\it locally indistinguishable} \cite{baake1999guidemathematicalquasicrystals, baake2013aperiodic}\footnote{The term ``locally indistinguishable" used here is synonymous with the earlier term ``locally isomorphic" introduced in \cite{levine1986quasicrystals}.}  Informally, this means that any finite patch found somewhere in one quasicrystal can also be found somewhere in the other.

More precisely, in the window scheme, it means that, for any finite region $R\subset V_{{\rm ph}}$, $\exists$ a region $R'\subset V_{{\rm ph}}$ (a translation of $R$) such that $\Lambda_{{\rm ph}}^{{\bf t'},{\cal W}}$ in region $R'$ is identical to $\Lambda_{{\rm ph}}^{{\bf t},{\cal W}}$ in region $R$; or, in other words, $\exists$ a vector $\Delta{\bf v}_{{\rm ph}}\in V_{{\rm ph}}$ such that 
$V_{{\rm ph}}^{{\bf t},W}({\bf v}_{{\rm ph}})=V_{{\rm ph}}^{{\bf t}',W}({\bf v}_{{\rm ph}}+\Delta{\bf v}_{{\rm ph}})$, for ${\bf v}_{{\rm ph}}\in R$.  

And in the weighting scheme, it means that, for any finite region $R\subset V_{{\rm ph}}$, $\exists$ a vector $\Delta{\bf v}_{{\rm ph}}\in V_{{\rm ph}}$ s.t.~$V_{{\rm ph}}^{{\bf t},W}({\bf v}_{{\rm ph}})$ and $V_{{\rm ph}}^{{\bf t}',W}({\bf v}_{{\rm ph}}+\Delta{\bf v}_{{\rm ph}})$ are arbitrarily close (as distributions) for ${\bf v}_{{\rm ph}}\in R$.  

Quasicrystals $\Lambda_{{\rm ph}}^{{\bf t},W}$ with the same $\Lambda$ and ${\cal W}$ (or $W$), but different ${\bf t}$, are in the same {\bf local isomorphism class}, and the quasicrystals in this equivalence class are in one-to-one correspondence with the points on the torus ${\bf t}\in T^{d}=V/\Lambda$.

Note that, for the same $\Lambda$ and ${\cal W}$ (or $W$), but different translation parameters ${\bf t}$ and ${\bf t}'$, where ${\bf t}'$ {\it cannot} be written as ${\bf t}+\Delta{\bf t}+\bm{\lambda}$ (with $\Delta{\bf t}\in V_{{\rm ph}}$ and $\bm{\lambda}\in\Lambda$), the two quasicrystals $\Lambda_{{\rm ph}}^{{\bf t},W}$ and $\Lambda_{{\rm ph}}^{{\bf t}',W}$ are {\it not} equal up to translation: so they are \emph{locally} isomorphic but {\bf globally distinct}.  The local isomorphism class contains an uncountable infinity of locally isomorphic yet globally distinct quasicrystals.

\end{itemize}

\subsubsection{(Quasi-)Translational Invariance}
\label{subsec:TranslationalInvariance}

Since $V_{{\rm ph}}$ is totally irrational relative to $\Lambda$ ({\it i.e.}\ $V_{{\rm ph}}$ does not intersect any points in $\Lambda$, besides the origin), the resulting C\&P pattern is {\it aperiodic}: {\it i.e.}\ it has no non-zero vector $\Delta{\bf v}_{{\rm ph}}\in V_{{\rm ph}}$ s.t.~$\Lambda_{{\rm ph}}^{{\bf t},W}({\bf v}_{{\rm ph}})=\Lambda_{{\rm ph}}^{{\bf t},W}({\bf v}_{{\rm ph}}+\Delta{\bf v}_{{\rm ph}})$.  

However, although $\Lambda_{{\rm ph}}^{{\bf t},W}({\bf v}_{{\rm ph}})$ has no {\it exact} translational invariance, it still has a kind of {\bf quasi-translational invariance}. 

This means that, for any region $R\subset V_{{\rm ph}}$, $\exists$ a non-zero vector $\Delta{\bf v}_{{\rm ph}}\in V_{{\rm ph}}$ s.t.~$\Lambda_{{\rm ph}}^{{\bf t},W}({\bf v}_{{\rm ph}})$ and $\Lambda_{{\rm ph}}^{{\bf t},W}({\bf v}_{{\rm ph}}+\Delta{\bf v}_{{\rm ph}})$ are equal (in the window scheme) or are arbitrarily close as distributions (in the  weighting scheme) for ${\bf v}_{{\rm ph}}\in R$.  

\subsubsection{(Quasi-)Orientational Invariance}
\label{subsec:OrientationalInvariance}

In the sC\&P scheme (where the lattice $\Lambda$ has automorphism group ${\rm Aut}(\Lambda)$, while the projector $\Pi_{{\rm ph}}$ and the weighting function $W$ are both invariant under a subgroup $G\subset {\rm Aut}(\Lambda)$): for ${\bf t}={\bf 0}$, the resulting quasicrystal $\Lambda_{{\rm ph}}^{{\bf 0},W}$ also has {\it exact} orientational invariance under $G$:
\begin{equation}
  \Lambda_{{\rm ph}}^{{\bf 0},W}({\bf v}_{{\rm ph}})
  =\Lambda_{{\rm ph}}^{{\bf 0},W}(g{\bf v}_{{\rm ph}}),\quad
  \forall g\in G.
\end{equation}

For ${\bf t}\neq0$, $\Lambda_{{\rm ph}}^{{\bf t},W}$ may no longer have {\it exact} $G$-invariance; but (since it is locally indistinguishable from $\Lambda_{{\rm ph}}^{{\bf 0},W}$) it  still has a type of {\bf quasi-orientational invariance} (or {\bf quasi $G$ invariance}).  

That is, for any region $R\subset V_{{\rm ph}}$ and any $g\in G$, $\exists$ a vector $\Delta{\bf v}_{{\rm ph}}\in V_{{\rm ph}}$ s.t.~$\Lambda_{{\rm ph}}^{{\bf t},W}({\bf v}_{{\rm ph}})$ and $\Lambda_{{\rm ph}}^{{\bf t},W}(g{\bf v}_{{\rm ph}}+\Delta{\bf v}_{{\rm ph}})$ are equal (in the window scheme) or are arbitrarily close as distributions (in the weighting scheme) for ${\bf v}_{{\rm ph}}\in R$.  

\subsubsection{(Quasi-)Scale Invariance}
\label{subsec:ScaleInvariance}

Quasicrystals can have a type of discrete scale invariance that periodic crystals lack.  

In particular, such symmetry arises in the following context.  In the sC\&P scheme, we consider the non-crystallographic subgroup $G\subset{\rm Aut}(\Lambda)$ that preserves both the splitting $V=V_{{\rm ph}}\oplus V_{{\rm in}}$ ({\it i.e.}\ $[g,\Pi_{{\rm ph}}]=0$) and the window ${\cal W}=g{\cal W}$ or weighting function $W({\bf v}_{{\rm in}})=W(g{\bf v}_{{\rm in}}), \forall g\in G$.  But there is also a larger subgroup $\widetilde{G}$ 
\begin{equation}
  G\subset\widetilde{G}\subset{\rm Aut}(\Lambda)
\end{equation}
whose elements $\tilde{g}\in \widetilde{G}$ still preserve the splitting $V=V_{{\rm ph}}\oplus V_{{\rm in}}$; but they do {\it not} (necessarily) preserve the window ($\tilde{g}{\cal W}\neq{\cal W}$) or weighting function ($W(\tilde{g}\,{\bf v}_{{\rm in}})\neq W({\bf v}_{{\rm in}})$); and instead the group $\widetilde{G}$ also includes elements $\tilde{g}$ that act to rescale the physical and internal subspaces 
\begin{subequations}
  \label{g_tilde}
  \begin{eqnarray}
  \tilde{g}\,{\bf v}_{{\rm ph}}&=&\lambda_{{\rm ph}}{\bf v}_{{\rm ph}}\qquad({\bf v}_{{\rm ph}}\in V_{{\rm ph}}) \\
  \tilde{g}\,{\bf v}_{\;\!{\rm in}\,}&=&\lambda_{\;\!{\rm in}\,}{\bf v}_{\;\!{\rm in}\,}\qquad({\bf v}_{\;\!{\rm in}}\in V_{\;\!{\rm in}\,}).
  \end{eqnarray}
\end{subequations}
Since $\tilde{g}\in{\rm Aut}(\Lambda)\subset O(V)$ has determinant $\pm1$, the numbers $\lambda_{\rm ph}$ and $\lambda_{\rm in}$ satisfy $\lambda_{{\rm ph}}^{d_{{\rm ph}}}\lambda_{{\rm in}}^{d_{{\rm in}}}=\pm 1$ with $d_{{\rm ph}}+d_{{\rm in}}=d$.  

The special numbers $\lambda_{\rm ph}$ are called the {\bf scale factors} associated to the sC\&P scheme. These scale factors may, in turn, be written as $\lambda_{\rm ph}=\pm \varepsilon_1^{m_1} \dots \varepsilon_{r}^{m_r}$, in terms of a finite list of $r$ {\bf fundamental scale factors} $\{\varepsilon_1,\ldots,\varepsilon_r\}$ ($\varepsilon_i>1$) raised to integer powers ($m_i\in\mathbb{Z}$); and these fundamental scale factors are, in turn, the {\bf fundamental units} of an {\bf algebraic number field} $K'$ associated with the non-crystallographic group $G$ (see Appendix \ref{app:inflation} for more details).  For example, for the 2D Ammann-Beenker tiling, there is a single fundamental scale factor -- the {\bf silver ratio} $1+\sqrt{2}$; and for the 2D Penrose tiling, or the 4D Elser-Sloane quasicrystal, there is again a single fundamental scale factor -- the {\bf golden ratio} $(1+\sqrt{5})/2$.

Such transformations may lead to two types of scale invariance -- local scale invariance (associated with the window scheme) and global scale invariance (associated with the weighting scheme).  We introduce these in turn. 

\subsubsection{Local Scale Invariance}
\label{subsec:LocalScaleInvariance}

\begin{figure}
\includegraphics[width=0.95\linewidth]{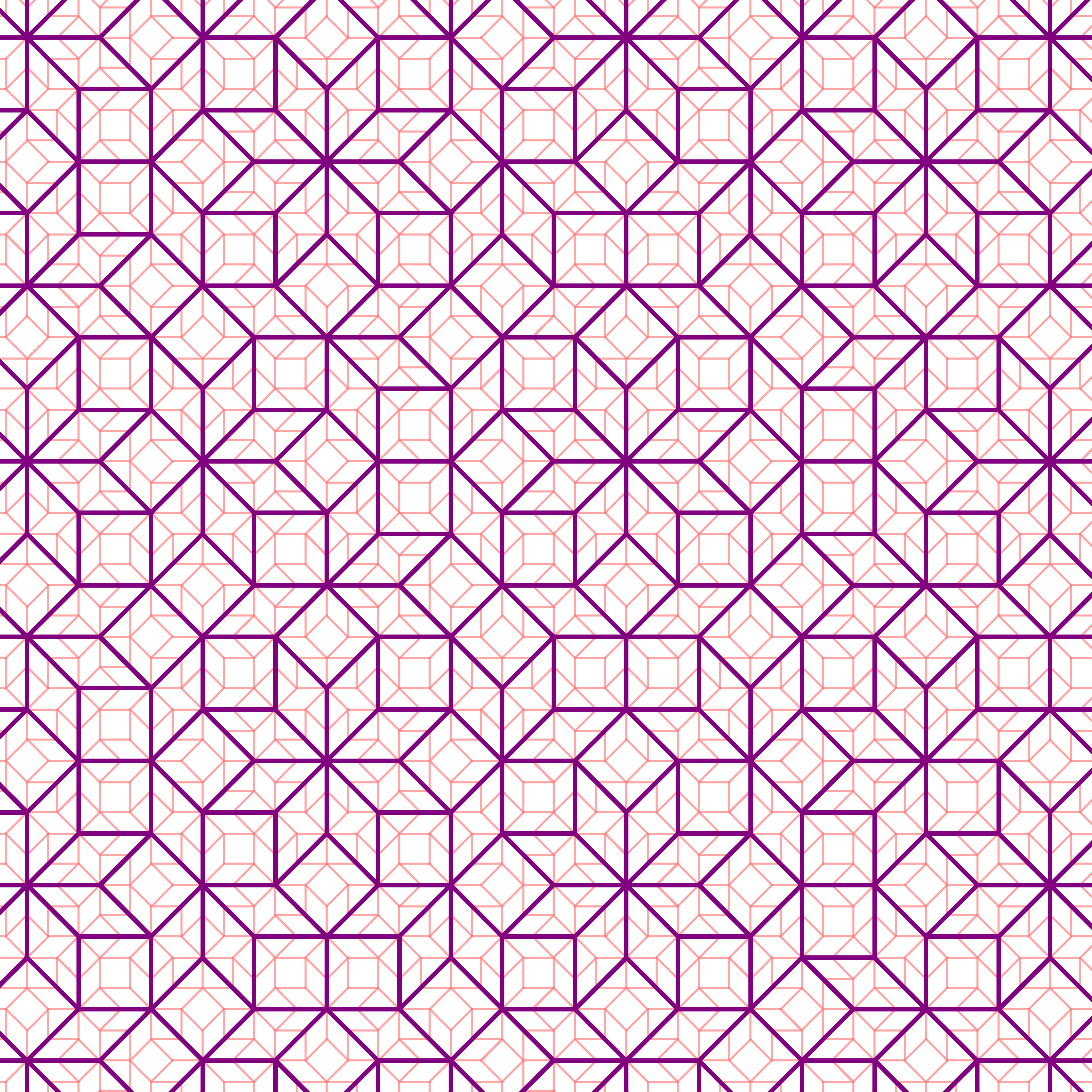}
\caption{\justifying A patch of the Ammann-Beenker tiling in purple, underlaid by its inflation in pink.} 
\label{fig:AmmannBeenkerInflation} 
\end{figure}

Informally, local scale invariance (also known as local inflation/deflation symmetry -- LIDS \cite{baake2013aperiodic}) refers to the inflation/deflation phenomenon exhibited by self-similar aperiodic tilings like the Penrose tiling or Ammann-Beenker (AB) tiling. Using the AB tiling for illustration, we define an {\bf inflation rule}, where we cut each large (purple) rhombus or square into a certain fixed pattern of smaller (pink) rhombi and squares, as shown in Fig.~\ref{fig:AmmannBeenkerInflation}.\footnote{The inflation rules for the Penrose tiling are shown in Fig.~\ref{fig:penroseTiling}.}  By cutting up all the tiles of any AB tiling (the ``parent'' AB tiling) in this way and rescaling it by the ``silver ratio''
$1+\sqrt{2}$, we obtain another legal AB tiling (the ``offspring'' AB tiling). This inflation process may be iterated arbitrarily many times.

We can also consider the inverse process, {\bf deflation}, in which we glue the tiles of an offspring AB tiling into ``supertiles,'' again according to the pattern in Fig.~\ref{fig:AmmannBeenkerInflation}, to recover the parent AB tiling from which it descended.  In a legal AB tiling, there is a \emph{unique} way to group all the tiles into supertiles in this way, so one can unambiguously determine the parent AB tiling from its offspring, and this deflation process may {\it also} be iterated indefinitely. 
 
To introduce local scale invariance more abstractly, we first review another idea: local equivalence. 

\begin{itemize}
    
    \item {\bf Local equivalence.}~Two patterns $P$ and $P'$ (which could be tilings, quasicrystals\ldots) are {\it locally equivalent} -- also called {\it mutually locally derivable} (MLD) \cite{baake2013aperiodic} -- if $P$ can be obtained from $P'$ in a unique way by local rules, and vice versa, such that any symmetry of $P'$ is also a symmetry of $P$, and vice versa.

    For example (see Fig.~\ref{fig:AmmannBeenkerInflation}): the Ammann-Beenker tiling (in purple) is locally equivalent to its ``inflation'' (in pink) \cite{baake2013aperiodic}.
    
    \item {\bf Local scale invariance.}~A pattern $P$ has local scale invariance -- also called local inflation/deflation symmetry (LIDS) \cite{baake2013aperiodic} -- with {\bf inflation factor} $\lambda>1$ if it is both locally equivalent to another pattern $P'$, {\it and} locally indistinguishable from $\lambda P'$ (where $\lambda P'$ means the pattern $P'$, expanded by the factor $\lambda$).  The pattern $\lambda P'$ is the {\bf inflation} of $P$, and the local rule for obtaining $P'$ from $P$ is the corresponding {\bf inflation rule}.  Or, viewed in the reverse direction, $P$ is called the {\bf deflation} of $\lambda P'$, and the local rule for obtaining $P$ from $P'$ is the corresponding {\bf deflation rule}.

    For example (see Fig.~\ref{fig:AmmannBeenkerInflation}):~the Ammann-Beenker tiling (in purple) is not only locally equivalent to its inflation (in pink), it is also locally indistinguishable from its inflation (after rescaling the pink tiling by the appropriate scale factor, $1+\sqrt{2}$).  Hence it has local scale invariance.
    
    Note that a periodic pattern {\it cannot} have local scale invariance with $\lambda>1$. Equivalently, the contrapositive: a pattern with local scale symmetry with $\lambda>1$ cannot be periodic.  Proof: if the two crystalline patterns $P$ and $P'$ are locally equivalent, they have the same fundamental domain; but then $P$ and $\lambda P'$ do {\it not} have the same fundamental domain, so are not locally indistinguishable.
\end{itemize}

To apply this to sC\&P , consider two patterns, $P=\Lambda_{{\rm ph}}^{{\bf t},{\cal W}}({\bf v}_{{\rm ph}})$ and $P'=\Lambda_{{\rm ph}}^{{\bf t},\widetilde{{\cal W}}}({\bf v}_{{\rm ph}})$, with $\widetilde{{\cal W}}=\lambda_{{\rm in}}{\cal W}$.  

On the one hand, using Eqs.~(\ref{eq:window_scheme}, \ref{g_tilde}) and the fact that $\tilde{g}\in{\rm Aut}(\Lambda)$, it follows that 
\begin{equation}
  \Lambda_{{\rm ph}}^{{\bf t},\widetilde{{\cal W}}}=\lambda_{{\rm ph}}\Lambda_{{\rm ph}}^{\widetilde{{\bf t}},{\cal W}}
\end{equation}
where 
\begin{equation}
  \widetilde{{\bf t}}=\frac{{\bf t}_{{\rm ph}}}{\lambda_{{\rm ph}}}+\frac{{\bf t}_{{\rm in}}}{\lambda_{{\rm in}}}.
\end{equation}
In other words, the quasicrystal $\Lambda_{{\rm ph}}^{{\bf t},{\cal W}}$ (obtained using the old translation ${\bf t}$ and the new window $\widetilde{{\cal W}}$) is the {\it same} as the quasicrystal $\lambda_{{\rm ph}}\Lambda_{{\rm ph}}^{\widetilde{{\bf t}},{\cal W}}$ (obtained by using the old window ${\cal W}$ and the new translation ${\bf t}$, and then rescaling the resulting quasicrystal by $\lambda_{{\rm ph}}$).  And this, in turn is locally isomorphic to the quasicrystal $\lambda_{{\rm ph}}\Lambda_{{\rm ph}}^{{\bf t},W}$ (by replacement $\widetilde{\bf t}\to {\bf t}$).  Hence, $P$ is locally isomorphic to $\lambda_{{\rm ph}}P'$.

On the other hand, in the window scheme, two such quasicrystals, obtained using the same $\Lambda$, ${\bf t}$, and $\Pi_{{\rm ph}}$, but two different windows ${\cal W}$ and $\widetilde{{\cal W}}$, will be locally equivalent if and only if the set $\widetilde{{\cal W}}$ may be expressed as a finite number of unions and intersections of (translated copies of) ${\cal W}$, and vice versa \cite{baake1991quasiperiodic, baake2013aperiodic}.  So, iff this condition is satisfied,
$P$ is {\it also} locally equivalent to $P'$, and hence
the quasicrystal $\Lambda_{{\rm ph}}^{{\bf t},W}$ has local scale invariance.
    
For example, in the AB tiling, the original window ${\cal W}$ is a regular octagon, while the rescaled window $\widetilde{{\cal W}}$ is a smaller regular octagon (scaled down by the factor $1+\sqrt{2}$).  Since each of these shapes may be construct from a finite number of unions and intersections of the other shape, it follows that the AB tiling has local scale invariance -- the inflation/deflation rule illustrated in Fig.~\ref{fig:AmmannBeenkerInflation}.

\subsubsection{Global Scale Invariance}
\label{subsec:GlobalScaleInvariance}

Whereas {\it local} scale invariance is a phenomenon that can occur in the {\it window} C\&P scheme, {\it global} scale invariance is a phenomenon that can occur in the {\it weighting} C\&P scheme.

First consider the set $\Lambda_{{\rm ph}}=\Pi_{{\rm ph}}\Lambda$ obtained by projecting all points in $\Lambda$ onto the physical subspace $V_{{\rm ph}}$.  $\Lambda_{{\rm ph}}$ forms a dense, but countable set of points in $V_{{\rm ph}}$.  

Next choose a pure-power-law weighting function 
\begin{equation}
  W({\bf v}_{{\rm in}})\propto v_{{\rm in}}^{-(d_{{\rm in}}+k)}.
\end{equation}
where $v_{\rm in}=|{\bf v}_{\rm in}|$ is the magnitude of ${\bf v}_{in}$, $d_{{\rm in}}=d-d_{{\rm ph}}$ is the dimension of the internal space, and $k>0$. Since $W$ is non-zero everywhere in $V_{{\rm in}}$, the resulting quasicrystal $\Lambda_{{\rm ph}}^{{\bf 0},W}$ has a dense set of delta function peaks -- one living over each point in $\Lambda_{{\rm ph}}$.  

From Eq.~(\ref{eq:weighting_scheme}), it follows that this quasicrystal has the following symmetry
\begin{eqnarray}
  \label{eq:global_scale_invariance}
  \Lambda_{{\rm ph}}^{{\bf 0},W}({\bf v}_{{\rm ph}})
  =\lambda_{{\rm in}}^{k}\Lambda_{{\rm ph}}^{{\bf 0},W}(\lambda_{{\rm ph}}{\bf v}_{{\rm ph}}),
\end{eqnarray}
where we have used $\tilde{g}\in{\rm Aut}(\Lambda)\Rightarrow\lambda_{{\rm ph}}^{d_{{\rm ph}}}\lambda_{{\rm in}}^{d_{{\rm in}}}=1$.

In other words, Eq.~(\ref{eq:global_scale_invariance}) says that, as we rescale the physical space $V_{{\rm ph}}\to\lambda_{{\rm ph}}V_{{\rm ph}}$, the underlying dense subset $\Lambda_{{\rm ph}}\subset V_{{\rm ph}}$ is mapped to itself, one-to-one; and, moreover, as we map any point ${\bf x}_{{\rm ph}}\in\Lambda_{{\rm ph}}$ to the new point ${\bf x}_{{\rm ph}}'=\lambda_{{\rm ph}}{\bf x}_{{\rm ph}}\in\Lambda_{{\rm ph}}$, we find that the delta function peak living at the original point ${\bf x}_{{\rm ph}}$ was higher (by a factor of $\lambda_{{\rm in}}^{k}$) than the delta function peak living at the new point ${\bf x}_{{\rm ph}}'$.  We say such a quasicrystal has {\bf exact global scale invariance with weight $k$}.

So far we have focused on the case ${\bf t}=0$, since it yields exact global scale invariance.  But it is also singular since the delta function peaked at the origin (${\bf v}_{{\rm ph}}=0$) is multiplied by an infinite coefficient $W({\bf 0})=\infty$.  

If we shift from this singular (${\bf t}=0$) case to a generic case -- {\it i.e.}\ a value of ${\bf t}$ that {\it cannot} be written as $\lambda+{\bf t}_{{\rm ph}}$ (with $\lambda\in\Lambda$ and ${\bf t}_{{\rm ph}}\in V_{{\rm ph}}$) -- we obtain a new quasicrystal $\Lambda_{{\rm ph}}^{{\bf t},W}$ that is non-singular: it does not contain any delta functions multiplied by infinite coefficients, and the mass contained in any finite region is finite.  

Moreover, the non-singular quasicrystal $\Lambda_{{\rm ph}}^{{\bf t},W}$ is {\bf {\it almost} locally indistinguishable} from $\Lambda_{{\rm ph}}^{{\bf 0},W}$.  By this we mean that, for any finite region $R$ in $V_{{\rm ph}}$, we can find another finite region $R'$ in $V_{{\rm ph}}$ such that $\Lambda_{{\rm ph}}^{{\bf t},W}$ in region $R$ is arbitrarily close (as a measure) to $\Lambda_{{\rm ph}}^{{\bf 0},W}$ in $R'$; and, in the other direction, for any finite region $R$ in $V_{{\rm ph}}$ {\it that does not include the origin}, we can find another finite region $R'$ in $V_{{\rm ph}}$ such that $\Lambda_{{\rm ph}}^{{\bf 0},W}$ in $R$ is arbitrarily close (as a measure) to $\Lambda_{{\rm ph}}^{{\bf t},W}$ in $R'$.  We say that generic non-singular quasicrystal $\Lambda_{{\rm ph}}^{{\bf t},W}$ has {\bf global quasi-scale invariance}.

\section{Spacetime Quasicrystals: Formalism, Examples}
\label{sec:TowardsLorentzianQuasicrystal}

Having laid the groundwork in the previous sections, the extension to \textit{spacetime} quasicrystals is now rather straightforward, and we can focus on the key novel features, and giving the first examples of such quasicrystals.

\subsection{Mathematical Preliminaries}
\label{subsec:MathematicalPreliminaries}

\subsubsection{Lorentzian Lattices}
\label{subsubsec:Lorentzian_Lattices}

In Sec.~\ref{subsec:Lattices}, we introduced lattices living in Euclidean space $V=\mathbb{R}^{d}$.  To generalize the discussion, we now replace $\mathbb{R}^{d}$ (with $d$ spatial dimensions) to $\mathbb{R}^{s,t}$ (with $s$ spatial and $t$ temporal dimensions).  In other words, we replace the positive-definite inner product $\langle {\bf v},{\bf w}\rangle=v^{\alpha}\delta_{\alpha\beta}w^{\beta}$ on $\mathbb{R}^{d}$ by the non-degenerate but indefinite symmetric bilinear form $\langle {\bf v},{\bf w}\rangle=v^{\alpha}\eta_{\alpha\beta}w^{\beta}$ on $\mathbb{R}^{s,t}$ where, in an orthonormal (Cartesian) basis, $\eta_{\alpha\beta}$ is
\begin{equation}
\eta_{\alpha\beta} = \mathrm{diag}(\underbrace{ \, -1,\dots,-1}_{\mathclap{t \text{ times}  }} \, , \, \underbrace{+ 1\dots,+1}_{\mathclap{s \text{ times }}} \,).
\label{eq:LorentzianMetricEta}
\end{equation}
The other definitions in Sec.~\ref{subsec:Lattices} remain  unchanged.  

As before, the even and odd self-dual lattices are of particularly high symmetry and interest.  Perhaps surprisingly, the classification of such lattices is much simpler in {\it indefinite} spacetimes  ({\it i.e.} when $t, s$ are both $\geq 1$):
\begin{itemize}
  \item {\bf Odd self-dual lattices}: for any indefinite $V=\mathbb{R}^{s,t}$ (with $s,t$ both $\geq 1$), there is a unique {\it odd} self-dual lattice (denoted $I_{s,t}$) which, in an orthonormal (Cartesian) basis, consists of all vectors ${\bf x}\in \mathbb{R}^{s,t}$ with integer components ${x}^{\mu}$:
  \begin{subequations}
  \begin{equation}
    I_{s,t}=\{{\bf x}|x^{\mu}\in\mathbb{Z}\;(\mu=1,\ldots,s+t)\}.
  \end{equation}
  In the Euclidean case $(s,t)=(d,0)$, this would be the usual (hyper)cubic lattice. 
  \item {\bf Even self-dual lattices}: for any indefinite $V=\mathbb{R}^{s,t}$ (with $s,t$ both $\geq 1$) whose {\bf signature} $\sigma\equiv s-t$ is also a multiple of 8, we also have a unique {\it even} self-dual lattice (denoted ${\rm II}_{s,t}$) which, in an orthonormal (Cartesian) basis, consists of all vectors ${\bf x}\in \mathbb{R}^{s,t}$ with (i) components $x^{\mu}$ all in $\mathbb{Z}$ or all in $\mathbb{Z}+\frac{1}{2}$; and (ii) {\it even} inner product with the vector $n=(1,\ldots,1$):
  \begin{eqnarray}
    {\rm II}_{s,t}&=&\{{\bf x}|\,{\it all}\;x^{\mu}\in\mathbb{Z}\;{\rm or}\,{\it all}\,x^{\mu}\in\mathbb{Z}+\frac{1}{2};\nonumber\\
    &&\quad{\rm and}\;\langle {\bf n},{\bf x}\rangle\in2\mathbb{Z}\;{\rm for}\;{\bf n}=(1,\ldots,1)\}.
  \end{eqnarray}
  \end{subequations}
  Note that in the Euclidean case $(s,t)=(8,0)$ this also describes the usual $E_{8}$ lattice.
\end{itemize}

When $V={\rm R}^{s,1}$ (with a single time direction) we say spacetime is {\bf Lorentzian}, and $\Lambda$ is a {\bf Lorentzian lattice}. We focus on this case for the rest of this paper.  

\subsubsection{Symmetries of Lorentzian Lattices}
\label{subsubsec:Symmetries_of_Lorentzian_Lattices}

\begin{figure}
\includegraphics[width=0.99\linewidth]{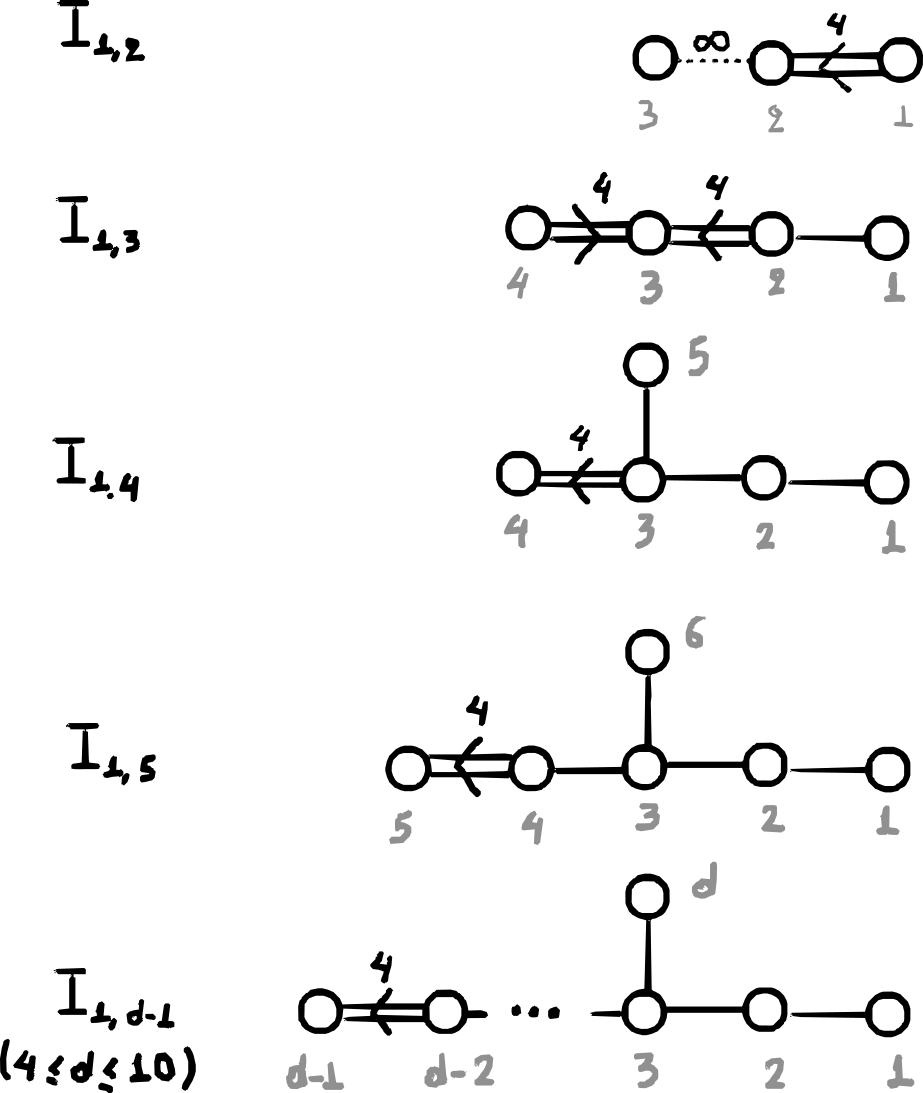}
\caption{\justifying The Coxeter-Dynkin diagrams describing the reflective symmetries of the odd and even self-dual lattices $\mathrm{I}_{s,1}$ for $3 \leq s+1 \leq 10$. The corresponding fundamental roots are given in Table \ref{tab:SimpleRootsReflectiveOddLatticesN2to17}.} 
\label{fig:VinbergRootsI1nUpto10} 
\end{figure} 

\begin{figure}
\includegraphics[width=0.95\linewidth]{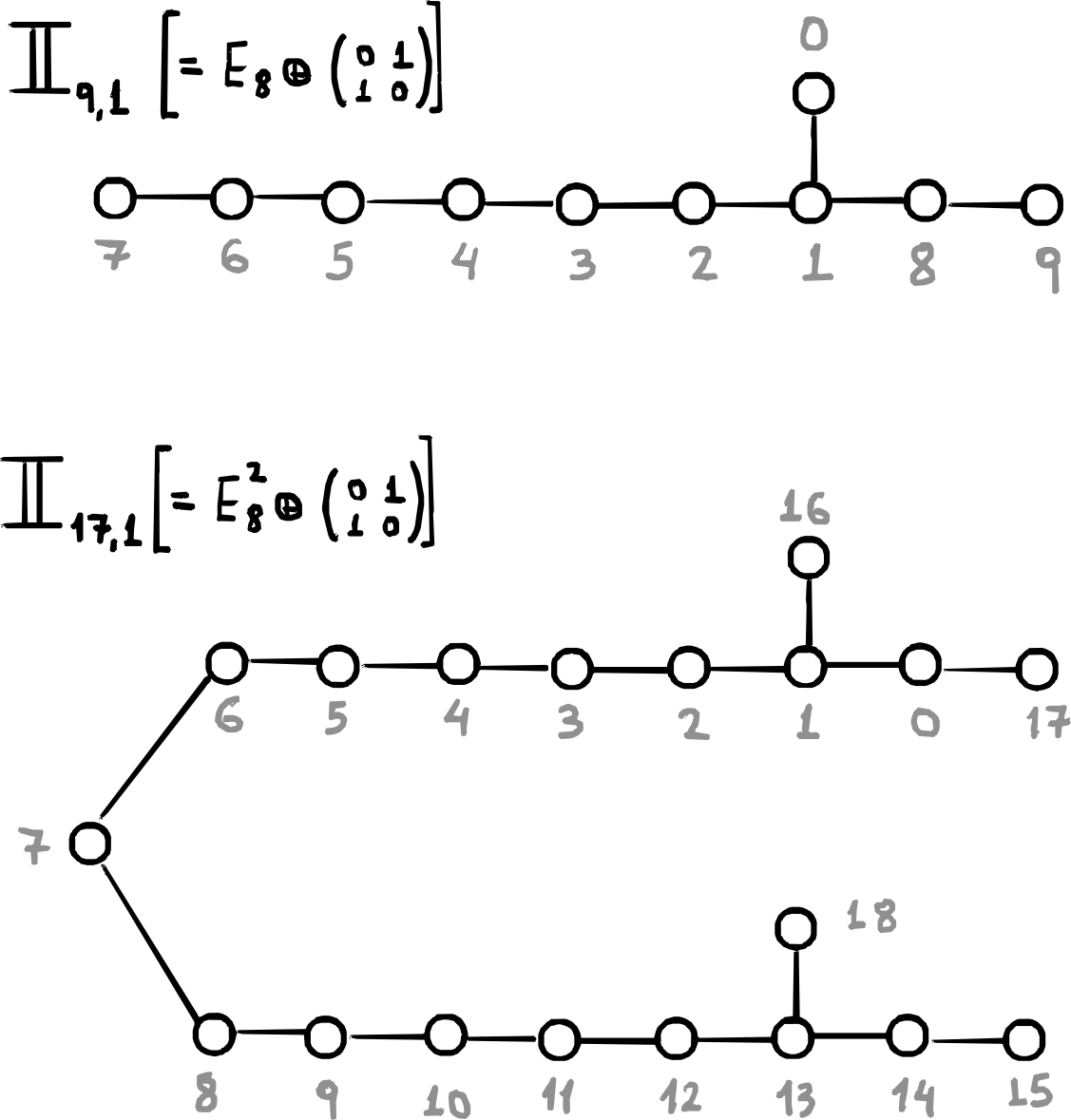}
\caption{\justifying The Coxeter-Dynkin diagrams describing the reflection groups of the even self-dual lattices ${\rm II}_{9,1}$ and ${\rm II}_{17,1}$.~In Cartesian coordinates, and numbered as in the figure, the fundamental roots for ${\rm II}_{9,1}$ are: $i:(0|0^i,+1,-1,0^{7-i})$ for $0\leq i\leq 7$; $8:\,(0|-1^{2},0^{7})$, $9:\,(\frac{1}{2}|(\frac{1}{2})^{9})$.  And the fundamental roots for ${\rm II}_{17,1}$ are: $i:(0|0^i,+1,-1,0^{15-i})$ for $0\leq i\leq 15$, $16:(0|-1^2,0^{15})$, $17:(\frac{3}{2}|-\frac{1}{2},(\frac{1}{2})^{16})$, $18:(1|0^{14},1^{3})$.} 
\label{fig:CoxeterDynkinDiagramsEvenSelfDual} 
\end{figure}

{\bf Vinberg's results.}~For a Lorentzian lattice $\Lambda$, the reflection subgroup of ${\rm Aut}(\Lambda)$ is a hyperbolic Coxeter group.  In dimension $s+1\geq3$, the odd and even self-dual Lorentzian lattices $I_{s,1}$ and (for $s-1=0$ mod 8) ${\rm II}_{s,1}$ have infinite automorphism groups, with infinite reflection subgroups (which are hyperbolic Coxeter groups).  Vinberg, and Vinberg and Kaplinskaja \cite{vinberg1967discrete, vinberg1972groups, vinberg1975some, kaplinskaja1978groups} studied the symmetries of these lattices and: (i) showed that they were reflective iff $s+1\leq20$; and (ii) in these reflective cases, determined the Coxeter diagrams of the reflection subgroups (using an algorithm now known as ``Vinberg's algorithm'').    Their results for the odd self-dual lattices ${\rm I}_{s,1}$ are shown in Fig.~\ref{fig:VinbergRootsI1nUpto10} for dimensions $3\leq s+1\leq 10$, and in Appendix \ref{app:SymmetriesLorentzianLattices} for dimensions 
$11\leq s+1\leq 20$.  Their results for the even self-dual lattices ${\rm II}_{s,1}$ ({\it i.e.} ${\rm II}_{9,1}$ and ${\rm II}_{17,1}$) are given in Fig.~\ref{fig:CoxeterDynkinDiagramsEvenSelfDual}.\footnote{Subsequently, Conway succeeded in also working out the automorphism group of the even self-dual lattice ${\rm II}_{25,1}$ and its reflection subgroup: in this case, even the number of {\it fundamental} roots is infinite and, in fact, the fundamental roots are a copy of the remarkable 24-dimensional Leech lattice \cite{conway1983automorphism}.}

\textbf{Generalized Graphical Notation}.~To describe hyperbolic Coxeter groups, we must slightly generalize the graphical notation for Coxeter diagrams introduced Sec.~\ref{subsec:Coxeter_Theory}.  As before, for two generators $R_i$ and $R_j$ with Coxeter exponent $m_{ij}$, the corresponding vertices $i$ and $j$ in the Coxeter graph are connected by an edge labelled by the integer $m_{ij}$ (and, as before, we use the short hand that, for an edge with $m_{ij}=2$, $3$, $4$ or $6$, we draw no edge, a single edge, a double edge, or a triple edge, respectively).  Geometrically, the two corresponding roots ${\bf r}_{i}$ and ${\bf r}_{j}$ have inner product $\frac{\langle {\bf r}_{i},{\bf r}_{j}\rangle}{(r_{i}^{2}r_{j}^{2})^{1/2}}=-{\rm cos}(\pi/m_{ij})$, corresponding to two $s$-dimensional mirrors in $\mathbb{R}^{s,1}$ intersecting at angle $\pi/m_{ij}$ and hence -- from restricting to the intersection of these mirrors with the upper unit hyperboloid $H_{s}$ in $\mathbb{R}^{s,1}$ -- two $(s-1)$-dimensional hyperbolic mirrors in $s$-dimensional hyperbolic space $H_{s}$ (also intersecting at angle $\pi/m_{ij}$).  For hyperbolic Coxeter groups, though, two more possibilities appear for the exponents $m_{ij}$. First, two roots can define mirrors that \textit{meet only asymptotically at infinity} in the hyperbolic plane, corresponding to the Coxeter exponent $m_{ij}=\infty$.  To reflect the fact that their mirrors meet only asymptotically, we connect such nodes with a \textbf{dotted line} (instead of Vinberg's notation which is to use a thick line).  Second, two roots can define mirrors that {\it do not intersect at all} (even at infinity) in the hyperbolic plane.  Such pairs of roots start to appear in $\mathbb{R}^{s,1}$ with $s\geq 18$, and we connect them with a \textbf{dashed line} (following Vinberg -- see {\it e.g.}\ \cite{vinberg1975some}).

{\bf Remarks.}  Let us add a few remarks about the symmetries of the self-dual lattices:
\begin{itemize} 
\item For ${\rm I}_{s,1}$, the first $s$ fundamental roots form a sub-diagram of (spherical/finite) type $B_s$, which corresponds to the fact that the odd Lorentzian lattices $\mathrm{I}_{s,1}$ are the Lorentzian analogues of the $B_s$ Euclidean root lattice, which is isomorphic to the hypercubic lattice $\mathbb{Z}^{s}$.
\item In dimension $s+1\leq 10$, the lattices ${\rm I}_{s,1}$ or ${\rm II}_{s,1}$ have $n=s+1$ fundamental roots; these fundamental roots are a (linearly-independent) basis for the lattice (or the Minkowski space $\mathbb{R}^{s,1}$ in which it is embedded); and the fundamental domain of the corresponding hyperbolic Coxeter group is a {\it simplex} in hyperbolic space $H^{s}$ (so these are among the complete list of such hyperbolic Coxeter groups given in Figs. 2\&3 of Sec.~6.9 in \cite{humphreys_1990}).
\item By contrast, in dimension $s+1>10$, the lattices ${\rm I}_{s,1}$ or ${\rm II}_{s,1}$ have $n>s+1$ fundamental roots; so these fundamental roots are {\it over-complete} as a basis for the lattice (or the Minkowski space $\mathbb{R}^{s,1}$ in which it is embedded); and the fundamental domain of the corresponding hyperbolic Coxeter group is {\it not} a simplex, but rather a shape with more codimension-one faces (bounding mirrors).  Note that, for a simplex, any pair of faces intersects in the boundary of the simplex; but for a more complex fundmamental domain, a given pair of mirror faces might only intersect {\it outside} the shape ({\it i.e.} if the faces are extended beyond the shape), or else not at all (in which case the corresponding dots are connected by a dashed line in the Coxeter diagram).
\end{itemize}

\subsubsection{Hyperbolic Coxeter Eigensystems, Salem Numbers, and Newman-Penrose Frames}
\label{subsubsec:Hyperbolic_Coxeter_Elements}

Given a hyperbolic Coxeter group with a finite number of fundamental roots ${\bf r}_{i}$, we can take the product (in some order) of all the fundamental $d\times d$ reflection matrices $R_{i}$ in Minkowski space to obtain a $d\times d$ Coxeter element $C$.  

If the group's Coxeter diagram is {\bf bicolorable} ({\it i.e.}\ if the vertices in the diagram can be colored red and blue, such that no two vertices of the same color are joined by an edge; see Fig.~\ref{fig:non-crystallographicEuclideanFoldings}), then we can define $C_{{\rm red}}=\prod_{i\in{\rm red}}R_{i}$ as the product of all red fundamental reflections, $C_{{\rm blue}}=\prod_{i\in{\rm blue}}R_{i}$ as the product of all blue fundamental reflections, and the Coxeter element as $C=C_{{\rm red}}C_{{\rm blue}}$.

For any hyperbolic Coxeter group whose Coxeter-Dynkin diagram has a finite number of nodes and is bicolorable, the eigensystem of $C$ has the following properties:

\begin{itemize}
\item \textbf{Eigenvalues:} When the dimension is even ($d=2k$), the spectrum of a hyperbolic Coxeter element \cite{a1976valeurs} consists of $k$ pairs of reciprocal eigenvalues:
\begin{equation}
  \lambda_{j}^{\pm}=1/\lambda_{j}^{\mp}\quad(j=0,\ldots,k-1).
\end{equation}
The first pair ($\lambda_{0}^{+}$ and $\lambda_{0}^{-})$ are real and positive with 
\begin{equation}
  \lambda_{0}^{+}>1,\quad\lambda_{0}^{-}<1,
\end{equation}
while the remaining $k-1$ pairs are complex-conjugates lying on the unit circle in $\mathbb{C}$
\begin{equation*}
  \lambda_{j}^{\pm}={\rm e}^{\pm i\varphi_{j}},
  \quad (j=1,\ldots,k-1).
\end{equation*}
If the dimension is odd ($d=2k+1$) there is also an additional eigenvalue $\lambda_{2k+1}=-1$.

The largest eigenvalue $\lambda_{0}^{+}$ is a {\bf Salem number} (a real algebraic integer $s>1$ whose conjugate roots all have absolute value no greater than 1, and at least one of which has absolute value exactly 1).  Salem numbers and their connections to hyperbolic Coxeter elements are of number-theoretic interest (see {\it e.g.}\ \cite{mcmullen2002coxeter, mukunda2007pisot}).  

In particular, {\bf Lehmer's conjecture} is that the {\it smallest} real Salem number is {\bf Lehmer's number} $\lambda_{{\rm out}}\approx 1.17628$, which is the largest real eigenvalue of the Coxeter element obtained from the first non-trivial even self-dual Lorentzian lattice $\mathrm{II}_{1,9}$.  For more about the Coxeter polynomial for $\mathrm{II}_{9,1}$ ({\bf Lehmer's polynomial}), see \cite{GROSS20091034}.

\item \textbf{Eigenbasis:} When the dimension is even ($d=2k$), the corresponding eigenvectors are a pair of real vectors $\{ \mathbf{v}_{0}^{+}, \mathbf{v}_{0}^{-}\}$ and $k-1$ pairs of complex-conjugate vectors $\{ \mathbf{v}_j^{+},\mathbf{v}_j^{-}\}$ with ${\bf v}_{j}^{-}=({\bf v}_{j}^{+})^{\ast}$ $(j=1,\dots,k-1)$; and when the dimension is odd ($d=2k+1$), there is an additional real (in fact, rational) eigenvector $\mathbf{v}_{2k+1}$ (corresponding to eigenvalue $\lambda_{2k+1}=-1$).  (We henceforth restrict attention to the case of even dimension $d=2k$, as all statements have trivial extensions to the odd case.)  

Any vector ${\bf w}\in \mathbb{R}^{s,1}$ in Minkowski space may be expressed in this Coxeter eigenbasis as
\begin{equation}
  {\bf w}=\alpha_{0}^{+}{\bf v}_{0}^{+}+
  \alpha_{0}^{-}{\bf v}_{0}^{-}+\sum_{j=1}^{k-1}
  (\alpha_{j}{\bf v}_{j}^{+}+\alpha_{j}^{\ast}{\bf v}_{j}^{-})
\end{equation}
where the first two coefficients are real ($\alpha_{0}^{\pm}\in\mathbb{R}$) while the other coefficients are complex 
($\alpha_{j}\in\mathbb{C}$).

It is straightforward to show that the Coxeter eigenbasis is a special type of basis called a {\bf Newman-Penrose frame} \cite{Newman:1961qr}; that is, in terms of Minkowksi bilinear form $\langle {\bf v}|{\bf w}\rangle=\eta_{\mu\nu}v^{\mu}w^{\nu}=v^{\mu}w_{\mu}$, they have the following properties:
\begin{itemize}
\item \textit{Nullness} of each eigenvector
\begin{equation*}
  \langle {\bf v}_{j}^{\pm},{\bf v}_{j}^{\pm}\rangle=0\quad
  (j=0,\ldots,k-1),
\end{equation*}
\item \textit{Normalization} of each eigenvector pair: 
\begin{eqnarray*}
\langle {\bf v}_{0}^{+},{\bf v}_{0}^{-}\rangle&=&-1 \\
\langle {\bf v}_{j}^{+},{\bf v}_{j}^{-}\rangle&=&+1\quad(j=1,\ldots,k-1),
\end{eqnarray*}
\item \textit{Orthogonality} of different eigenvector pairs:
\begin{equation*}
\langle {\bf v}_{j}^{\pm},{\bf v}_{k}^{\pm}\rangle=
\langle {\bf v}_{j}^{\pm},{\bf v}_{k}^{\mp}\rangle=0
\quad (j\neq k).
\end{equation*}
\end{itemize}

Let $G$ denote the group generated by $C$ (which, in the bicolorable case, is a subgroup of the group $G'$ generated separately by $C_{\rm red}$ and $C_{\rm blue}$).  Minkowski space $\mathbb{R}^{s,1}$ transforms as a reducible representation of $G$ (or $G'$): we can split it into $k$ different two-dimensional real irreducible representations (or ${\bf 2}$'s in particle physics notation) called {\bf Coxeter planes}.  The $j$th Coxeter plane $V_{j}$ is the plane spanned by the two real basis vectors
${\bf u}_{j}^{\pm}=({\bf v}_{j}^{+}\pm{\bf v}_{j}^{-})/\sqrt{2}$.  The Coxeter plane $V_{0}$ has an induced metric of signature $(1,1)$, while the other Coxeter planes $V_{j}$ ($j=1,\ldots,k-1$) have induced metrics of signature $(2,0)$, so we have 
\begin{equation}
\begin{array}{ccccccccc}
V & = & V_0 & \oplus & V_1 & \oplus & \dots & \oplus & V_{k-1} \\
d & = & \mathbf{2} & \oplus & \mathbf{2} & \oplus & \dots & \oplus & \mathbf{2} \\
(s,1) & = & (1,1) & \oplus & (2,0) & \oplus & \dots & \oplus & (2,0). 
\end{array}
\label{eq:VDecompositionHyperbolicCoxeter}
\end{equation}
The Coxeter element $C$ acts as an \textit{irrational rotation} on each Coxeter plane: on $V_0$, it is a rotation by a hyperbolic angle $\vartheta_0 = \mathrm{cosh}^{-1}(\lambda_{0}^{+} + \lambda_{0}^{-})/2$ (i.e. \textit{a discrete Lorentz boost}); on each of the $V_j$'s, it is an ordinary (Euclidean) discrete rotation by $\vartheta_j = \cos^{-1}{(\lambda_j^{+} + \lambda_j^{-})/2}$. In other words, in the basis $\{{\bf u}_{0}^{+},{\bf u}_{0}^{-},\ldots,{\bf u}_{k-1}^{+},{\bf u}_{k-1}^{-}\}$, the Coxeter element takes the 2-by-2 block-diagonal form 
\begin{equation}
  \setlength{\arraycolsep}{0pt}
  \setlength{\delimitershortfall}{0pt}
  \newcommand*{\myfbox}[1]{%
    \fcolorbox{red}{red!20!white}{$#1$}%
  }
  C = 
   \begin{bmatrix}
    \,\myfbox{\mathrm{B}\big(\vartheta_0\big)}\, & 0  & \,\cdots \,  \\
     0 & \,\myfbox{\mathrm{R}\big(\vartheta_1\big)}\, &  & \\
    \, \vdots \,& & \, \myfbox{\ddots} \, & \\
    & & & \,\myfbox{\mathrm{R}\big(\vartheta_{k-1}\big)}\,
  \end{bmatrix}
  \label{eq:CoxeterActionLorentzianCase}
\end{equation}
where
\begin{equation*}
    \mathrm{B}\big(\vartheta_0\big) = \begin{pmatrix}
        \cosh \left( \vartheta_0 \right) & \sinh \left( \vartheta_0 \right) \\[2ex]
        \sinh \left( \vartheta_0 \right) & \cosh \left( \vartheta_0 \right)\\
    \end{pmatrix}
\end{equation*}
are the 2-by-2 discrete Lorentz boosts
\begin{equation*}
    \mathrm{R}\big(\vartheta_j\big) = \begin{pmatrix}
        \cos \left( \vartheta_j\right) & -\sin \left( \vartheta_j\right) \\[2ex]
        \sin \left( \vartheta_j\right) & \cos \left( \vartheta_j\right)\\
    \end{pmatrix}
\end{equation*}
are the 2-by-2 discrete Euclidean rotations. 
\end{itemize}

\subsection{Spacetime C\&P Scheme}
\label{subsec:SpacetimeCNP}

In the previous section, we first introduced the ordinary cut-and-project (``C\&P'') scheme, followed by the more refined Symmetric C\&P scheme, in which quasi-periodicity, orientational symmetry and discrete scale symmetry (inflation/deflation symmetry) are tightly interconnected.  The most interesting quasicrystals are produced by the sC\&P scheme.  As discussed in Sec.~\ref{subsec:sCNP}, an sC\&P quasicrystal is determined by three choices, so in this section we review these three choices (in blue), and discuss any novelties associated with generalizing each one to the Lorentzian spacetime context.

\subsubsection{The Lattice $\Lambda$}

\textcolor{blue}{{\bf Choice 1:} a $d$-dimensional lattice $\Lambda^{{\bf t}}$ in $V=\mathbb{R}^{d}$ with automorophism group ${\rm Aut}(\Lambda)$}.  

Now $\Lambda$ is a {\it Lorentzian} lattice in $V=\mathbb{R}^{s,1}$.  Whereas ${\rm Aut}(\Lambda)$ for a Euclidean lattice was a {\it finite} group, ${\rm Aut}(\Lambda)$ for a Lorentzian lattice in dimension $s+1\geq3$ can be an {\it infinite} group: in particular, 
\begin{enumerate} 
  \item for an odd or even self-dual lattice $I_{s,1}$ and $II_{s,1}$ (see Figs.~\ref{fig:VinbergRootsI1nUpto10} and \ref{fig:CoxeterDynkinDiagramsEvenSelfDual});
  \item for a {\bf Lorentzian root lattices}, {\it i.e.}\ the integer span of the fundamental roots of any of the crystallographic hyperbolic Coxeter groups enumerated in Secs. 6.7-6.9 and Figs. 2 \& 3 of Ref.~\cite{humphreys_1990}.
\end{enumerate}

\subsubsection{The Non-Crystallographic Symmetry}

\textcolor{blue}{{\bf Choice 2:} non-crystallographic subgroup $G\in{\rm Aut}(\Lambda)$, and a decomposition $V=V_{{\rm ph}}\oplus V_{{\rm in}}$ that is invariant under $G$: $\Pi_{{\rm ph}}=g\Pi_{{\rm ph}}g^{-1}$ (or, equivalently $[\Pi_{{\rm ph}},g]=0$), $\forall g\in G$.} 

Whereas for Euclidean quasicrystals, the relevant (finite/spherical) non-crystallographic subgroups have all been enumerated (see Fig.~\ref{fig:Irreducible_Root_Systems}), for Lorentzian quasicrystals, the relevant (hyperbolic) non-crystallographic subgroups have not been (although the non-crystallographic groups with simplicial fundamental domains {\it have} been enumerated in Figs. 2\&3 of Sec.~6.9 in \cite{humphreys_1990}).  It would be interesting, for each Lorentzian lattice $\Lambda$ mentioned above, to determine the non-crystallographic subgroups of ${\rm Aut}(\Lambda)$ -- we leave this for future work.

In the present paper, we focus for concreteness on a particular non-crystallographic subgroup $G\in{\rm Aut}(\Lambda)$ which can be constructed for any Lorentzian lattice $\Lambda$ listed above: namely, the group $G$ generated by a Coxeter element $C$ of $\Lambda$ ({\it i.e.}\ the product of the fundamental reflections corresponding to the fundamental roots of $\Lambda$).  

As explained in Sec.~\ref{subsubsec:Hyperbolic_Coxeter_Elements}, the action of $G$ splits $V$ into 2D subspaces $V_{j}$ ($j=0,\ldots,k-1$) that transform as 2D irreducible representations of $G$.  To obtain a {\it spacetime} quasicrystal, we must take the physical subspace $V_{\rm ph}$ to consist of the subspace $V_{0}$ which has Lorentzian signature $(1,1)$, and some (but not all) of the remaining subspaces $V_{j}$ with Euclidean signature $(2,0)$.  

This amounts to choosing the projector $\Pi_{\rm ph}$ to be
\begin{eqnarray}
  (\Pi_{\rm ph})^{\mu}_{\;\;\nu}&=&
  \sum_{j=0}^{k_{\rm ph}<k}\left(\frac{(v_{j}^{+})^{\mu}(v_{j}^{-})_{\nu}}{\langle {\bf v}_{j}^{+},{\bf v}_{j}^{-}\rangle}+\frac{(v_{j}^{-})^{\mu}(v_{j}^{+})_{\nu}}{\langle {\bf v}_{j}^{-},{\bf v}_{j}^{+}\rangle}\right)  
\end{eqnarray}
This induces a decomposition $V=V_{\rm ph}\oplus V_{\rm in}$ that is invariant under $G$ and (in the case where the original Coxeter group is bicolorable) also under the larger subgroup 
$G'$ generated by the two separate generators $C_{{\rm red}}$ and $C_{{\rm blue}}$ (which includes $G$ as a subgroup).

Note that this subgroup $G$ generated by the Coxeter element (or, in the bicolorable case, the larger group $G'$), is the Lorentzian generalization of the group that is used in the Euclidean case to generate well-known quasicrystals like the Ammann-Beenker tiling (starting from $\Lambda=\mathbb{Z}^{4}$) or the Penrose tiling (starting from $\Lambda=\mathbb{Z}^{5}$ or $\Lambda=A_{4}$).  

But, whereas $G$ (or $G'$) is a {\it finite} group in the Euclidean case, it is an {\it infinite} group in the Lorentzian case (since the Coxeter number $h$ is finite for finite/spherical Coxeter groups, but infinite for hyperbolic Coxeter groups).  So Lorentzian quasicrystals have infinitely more orientational symmetry than their Euclidean cousins!

\subsubsection{The Window}

\textcolor{blue}{{\bf Choice 3}: a window ${\cal W}\subset V_{{\rm in}}$ or weighting function $W:V_{{\rm in}}\to\mathbb{C}$ invariant under $G$: {\it i.e.}\ ${\cal W}=g{\cal W}$ or $W({\bf v}_{{\rm in}})=W(g\,{\bf v}_{{\rm in}})$, $\forall g\in G$.} 

In the Euclidean case, since the non-crystallographic group $G$ was a {\it finite} group, the window ${\cal W}\in V_{\rm in}$ or weighting function $W:V_{\rm in}\to\mathbb{C}$ only needed to have finite/discrete rotational symmetry.  For example, the 8-fold symmetric Ammann-Beenker tiling was produced by an 8-fold symmetric window ${\cal W}$ (see Fig.~\ref{fig:sCNPWindowABTiling}).

By contrast, in the Lorentzian case, since the non-crystallographic group $G$ has infinite order, a symmetric window ${\cal W}$ or weighting function must also have a infinite rotational symmetry.  This implies that, in each ``internal'' Coxeter plane $V_{j}$ ($k_{\rm ph}<j\leq k$), the window must be a {\it circularly-symmetric disk} of radius $\rho_{j}$, or the weighting function must be circularly-symmetric -- {\it i.e.}\ a function of the form
\begin{equation}
  W\left(\frac{r_{(k_{\rm ph}^{}+1)}}{\rho_{(k_{\rm ph}^{}+1)}},\ldots,\frac{r_{k}^{}}{\rho_{k}^{}}\right)
\end{equation}
where $r_{j}$ is the radial coordinate in the $j$th Coxeter plane $V_{j}$ ({\it i.e.}\ the coordinate that measures the distance from the origin in that plane, according to the induced metric).

\subsection{Other implications}
\label{subsec:implications}

Let us mention some other novel features of Lorentzian quasicrystals (compared to their Euclidean cousins).

\subsubsection{Point Set, But No Tiling} 
 
So far have described the sC\&P algorithm with symmetric window ${\cal W}$ as a way to generate a $G$-symmetric quasi-crystalline point set (for some non-crystallographic group $G$).  But in the Euclidean case (where $G$ is a finite group), one can often (as in the Penrose tiling, or Ammann-Beenker tiling) regard these points as the endpoints of edges; these edges as the boundaries of faces, etc, so that the $G$-symmetric point-set is promoted to a $G$-symmetric {\it tiling} or {\it tessellation}.  

By contrast, in the Lorentzian case, one can't promote a $G$-symmetric Lorentzian sC\&P point set to a $G$-symmetric tiling (with a finite number of polygonal tiles). To see this, note that if the $G$-symmetric tiling contains any tile $T$, it must also contain the ``rotated'' tile $gT$ ($\forall g\in G$); but if $G$ is an infinite group, this will either imply that the tile $T$ is accompanied by an infinite number of other, differently-oriented tiles; or else, $T$ must be invariant under a finite-index subgroup of $G$ (in which case the tile itself is ``spherically symmetric'').  

\subsubsection{No Preferred Window Size}

In the Euclidean case, if the window ${\cal W}\in V_{\rm in}$ is chosen to have a certain natural size, it uniquely lifts to a cell ${\cal C}\in V$ whose vertices have rational coordinates in the lattice basis, and whose $n$-faces project onto $V_{\rm ph}$ to give the $n$-tiles in the quasi-crystalline tiling.  

Is there a natural size for the spherically-symmetric window ${\cal W}$ in the Lorentzian case?  The answer appears to be negative: the larger the radius, the denser the projected set $\Lambda_{\rm phys}(\mathsf{W})$, but there seems to be no natural or intrinsic way to choose a particular radius as the best one.  Relatedly, a spherically-symmetric window does not have a unique lift from $V_{\rm ph}$ to $V$.

\subsubsection{No Local Scale Invariance}

In Section \ref{subsec:ScaleInvariance}, we explained two notions of scale invariance: local and global.  And, in particular, in \ref{subsec:LocalScaleInvariance} we explained that a C\&P quasicrystal has local scale invariance -- also known as ``local inflation/deflation symmetry'' (LIDS) -- if and only if the inflated window $\widetilde{{\cal W}}$ can be obtained via a finite number of unions and intersections of translated copies of the original window ${\cal W}$; and vice versa \cite{baake1991quasiperiodic, baake2013aperiodic}.  For a spherically-symmetric window this is impossible (a rescaled ball cannot be expressed as a finite number of unions and intersections of translated copies of the original ball).  So a $G$-symmetric spacetime quasicrystal (with infinite non-crystallographic symmetry group $G$) cannot have local inflation invariance.  

\subsubsection{Self-Duality and Global Scale-invariance} 

So far in Sec.~\ref{subsec:implications} we have focused on spacetime sC\&P quasicrystals produced via the {\it window} scheme; and we have seen that (in the spacetime context) there is no $G$-invariant tiling associated with the resulting quasicrystalline point set, no preferred size of the window ${\cal W}$ (and hence no preferred density of the point set), and no local inflation-deflation symmetry. 

Let us turn to consider spacetime sC\&P quasicrystals produced via the {\it weighting} scheme:
\begin{itemize}
\item If we take the weighting function $W$ to be a Gaussian, there {\it is} a preferred width of the Gaussian -- namely the width that makes the quasicrystal self-dual, just as explained in Sec.~\ref{subsec:SelfDualQuasicrystals}.
\item If we take the weighting function $W$ to be a power law, we obtain a quasicrystal with global scale invariance, just as explained in Sec.~\ref{subsec:GlobalScaleInvariance}.  
\item In particular, if we take the particular case of a trivial power law, $W({\bf v}_{\rm in})=1$, so we project each point in the lattice $\Lambda$ with equal weight, we effectively obtain the dense point set $\Lambda_{\rm ph}=\Pi_{\rm ph}\Lambda$.  This does indeed have global scale invariance: when this dense set is rescaled by certain special numbers, it is precisely mapped to itself, one-to-one.  These special numbers are the real units of the algebraic number field associated with the eigenvalues of the Coxeter element.  
\end{itemize}
Given all of these observations, we see that in the spacetime context, the weighting scheme is more natural than the window scheme; and the self-dual or globally-scale-invariant quasicrystals produced by the weighting scheme are most natural of all.

\subsection{Examples of Spacetime Quasicrystals}
\label{subsec:ExamplesSpacetimeQuasicrystals}

In this section, we finally apply the machinery developed so far to construct the first examples of quasicrystals in Lorentzian spacetime. 

\subsubsection{Example 1: $(1+1)$-dimensional Quasicrystals from $\mathrm{I}_{3,1}$}
\label{subsec:2DLorentzianQuasiCrystal}

\begin{figure*}
    \begin{tabular}{cc}
        \subfloat[\label{fig:SpacetimeCNP-I13SelfDualPattern}]{
        \includegraphics[width=0.41\linewidth]{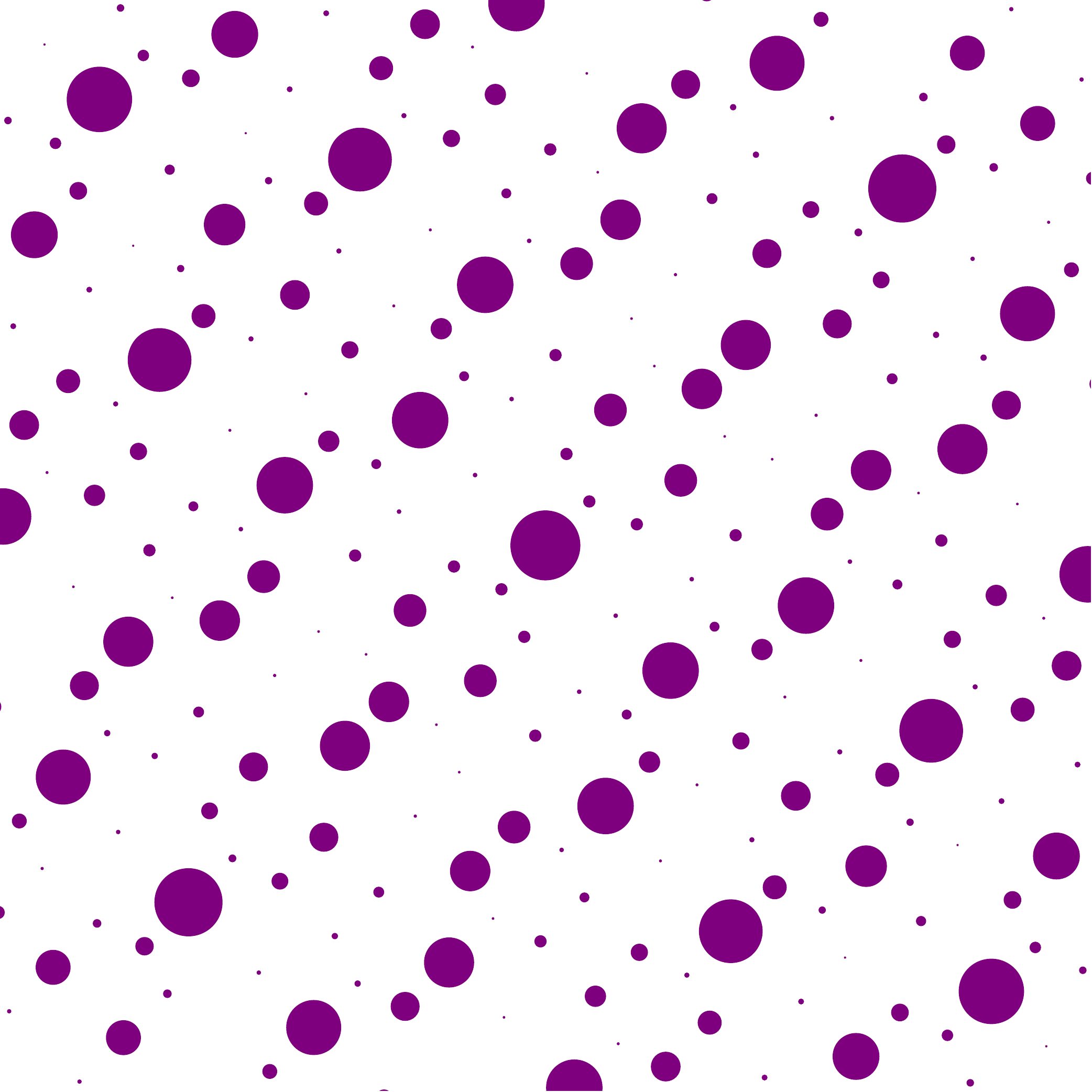}
            \hspace{1em}}
        \subfloat[\label{fig:SpacetimeCNP-I13SphericalWindow}]{    
            \includegraphics[width=0.41\linewidth]{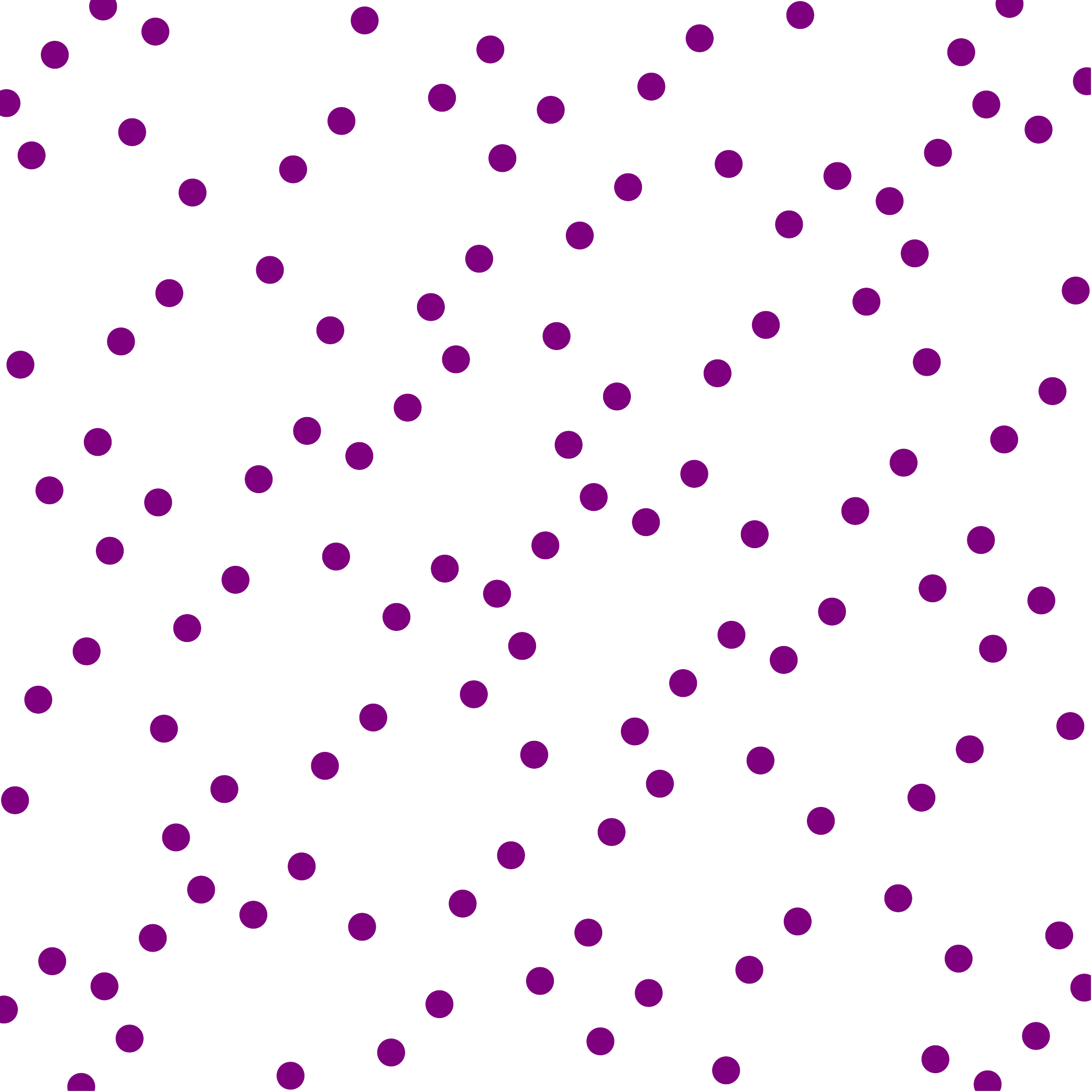}
        } \\
         \rule{0pt}{3ex}\\   
        \subfloat[\label{fig:SpacetimeCNP-I13ScaleInvPattern}]{\includegraphics[width=0.41\linewidth]{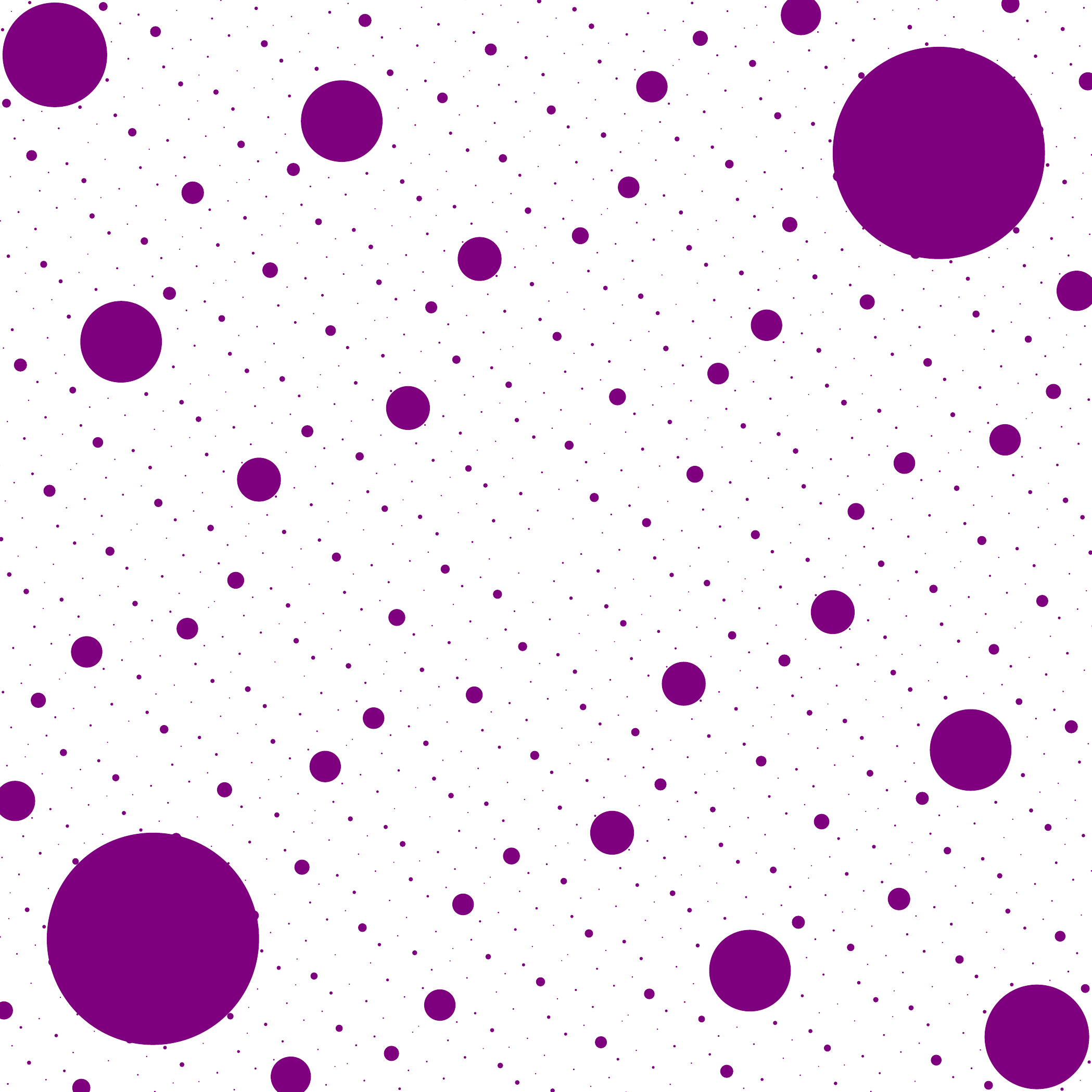}}
        \hspace{1em}
        \subfloat[\label{fig:SpacetimeCNP-I13ScaleInvPattern2}]{\includegraphics[width=0.41\linewidth]{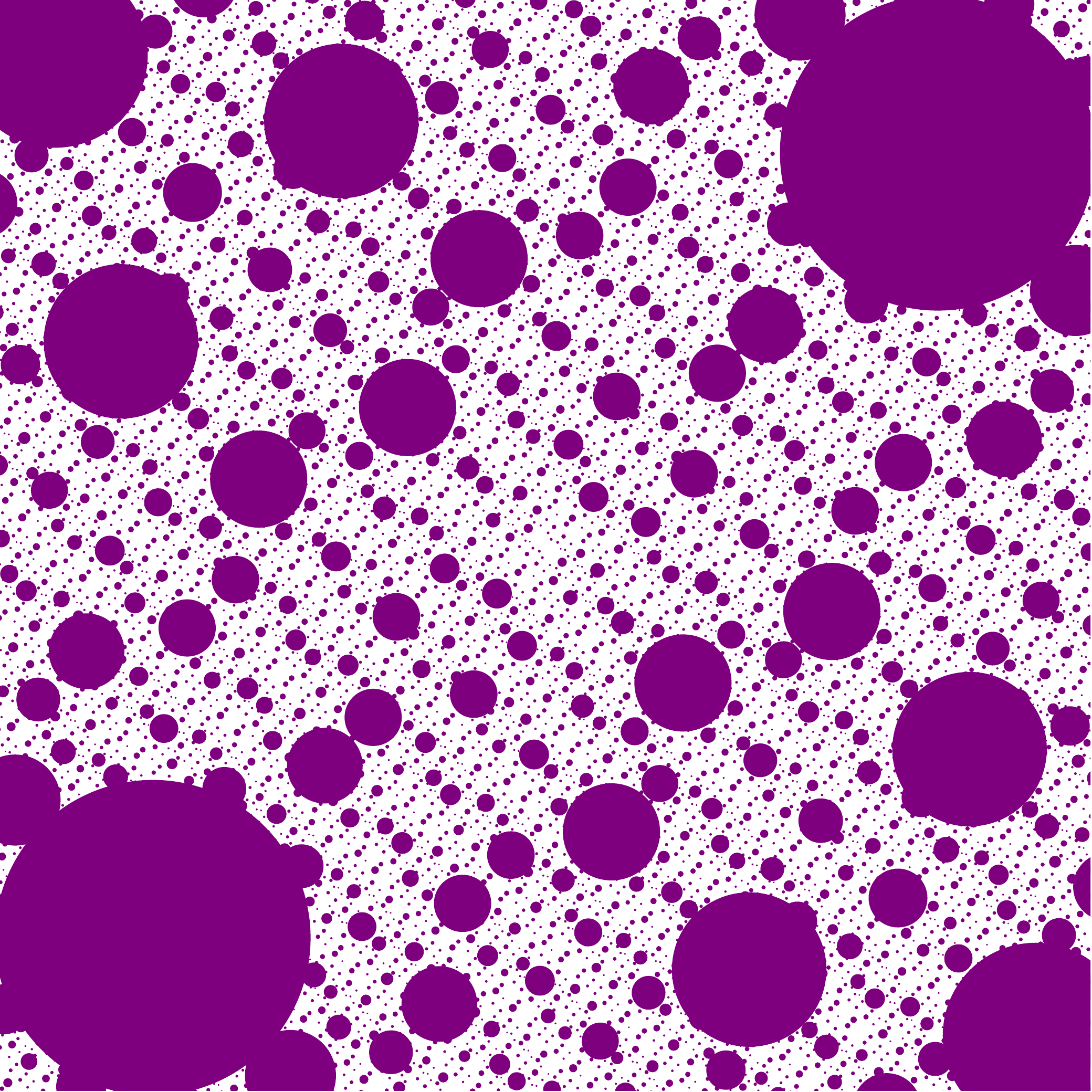}}
    \end{tabular}

\caption{\justifying Four variants of the $(1+1)$D spacetime quasicrystal constructed from the odd self-dual lattice $\textrm{I}_{3,1}$ in Sec.~\ref{subsec:2DLorentzianQuasiCrystal}. Upper left: A patch of the self-dual variant of the quasicrystal ({\it i.e.}\ based on the Gaussian weighting function $W({\bf v}_{\rm in})\propto {\rm exp}(-\pi v_{in}^{2})$ with $v_{\rm in}^2\equiv\langle {\bf v}_{\rm in},{\bf v}_{\rm in}\rangle$).  Upper right: The same patch in the variant based on a circular window of radius $3\sigma$, where $\sigma=(2\pi)^{-1/2}$ is the standard deviation of the self-dual Gaussian weighting function.  Lower left: The same patch in a variant of the quasicrystal which is scale-invariant with weight $k=1.01$ (so constructed using weighting function $W({\bf v}_{\rm in})\propto v_{in}^{-(2+1.01)}$).  Lower right: The same patch in a variant of the quasicrystal which is scale-invariant with weight $k=0.25$ (so constructed using weighting function $W({\bf v}_{\rm in})\propto v_{in}^{-(2+0.25)}$). 
In both lower panels, a singular central peak has been removed from the origin (this singular peak would be absent if we use any shifted lattice $\Lambda^{\bf t}$ with ${\bf t}\notin\Lambda$.  Both of the power laws (characterized by $k=1.01$ and $k=0.25$) fall steeply enough to give a finite mass in a finite region; but only the first ($k=1.01$) satisfies the criterion for yielding a translation-bounded complex measure (see Lemma 9.5 in \cite{baake2013aperiodic}).  Note the upper left panel only includes peaks with amplitude $>10^{-3}$, while the lower left and right panels only include peaks with amplitude $>10^{-6}$ (since, in these three panels, if all peaks are included, they densely fill the plane, though most have negligibly tiny coefficients).}
\label{fig:I13quasicrystal}
\end{figure*}

For simplicity, we will start with a low dimensional example: a $(1+1)$ dimensional quasicrystal, obtained from the lattice $\mathrm{I}_{3,1}$. (Actually, we construct three variants of this quasicrystal: windowed, self-dual, and scale-invariant of weight one.)  This low dimensional example has the advantage that we can display the resulting quasicrystals explicitly (see Fig.~\ref{fig:I13quasicrystal}).

The Coxeter diagram for $\mathrm{I}_{3,1}$ is shown in Fig. \ref{fig:VinbergRootsI1nUpto10}, and the corresponding Schl\"{a}fli matrix reads
\begin{equation*}
S = \begin{pmatrix}
2 & -1 & 0 & 0\\
-1 & 2 & -\sqrt{2} & 0 \\
0 & -\sqrt{2} & 2 &-\sqrt{2} \\
0 & 0 & -\sqrt{2}&2 
\end{pmatrix}.
\end{equation*}
Expressed in an orthonormal basis (Cartesian coordinates), the fundamental roots (from Table \ref{tab:SimpleRootsReflectiveOddLatticesN2to17}) are:
\begin{equation*}
\begin{aligned}
& \mathbf{r}_1 = (0,-1,1,0), \quad \mathbf{r}_2 = (0,0,-1,1), \\
& \mathbf{r}_3 = (0,0,0,-1), \quad \mathbf{r}_4 = (1,1,1,1).
\end{aligned}.
\end{equation*}
and the Coxeter element $C=R_1 R_2 R_3 R_4$ is 
\begin{equation*}
C = 
\left(\begin{array}{cccc}
2 & -1 & -1 & -1 \\
-1 & 1 & 1 & 0 \\
1 & 0 & -1 & -1 \\
1 & -1 & 0 & -1 
\end{array}\right).
\end{equation*}
Its eigenvalues (rounded to 5 digits) are
\begin{eqnarray*}
\bigg\{
\lambda_{0}^{\pm}&=&\frac{1\!+\!\sqrt{17}}{4} \pm \frac{1}{2}\sqrt{\frac{\sqrt{17}\!+\!1}{2}} \approx (2.08102)^{\pm1}, \\ 
\lambda_1^{\pm}&=& \frac{1\!-\!\sqrt{17}}{4} \pm \frac{i}{2}\sqrt{\frac{\sqrt{17}\!-\!1}{2}} \approx -0.78078\!\pm\!0.62481\,i \bigg\} \nonumber
\end{eqnarray*}
and the corresponding projection operators are
\begin{eqnarray*}
  (\Pi_{\rm ph})^{\mu}_{\;\;\nu}&=&\frac{(v_{0}^{+})^{\mu}(v_{0}^{-})_{\nu}}{\langle{\bf v}_{0}^{+},{\bf v}_{0}^{-}\rangle}
  +\frac{(v_{0}^{-})^{\mu}(v_{0}^{+})_{\nu}}{\langle{\bf v}_{0}^{-},{\bf v}_{0}^{+}\rangle} \\
  &=& \tfrac{+1}{\sqrt{17}}
  \left(
  \begin{array}{cccc} 
  \tfrac{7+\sqrt{17}}{2} & 0 & -2 & -2 \\
  0 & \tfrac{3+\sqrt{17}}{2} & 1 & -1 \\
  2 & 1 & \tfrac{-5+\sqrt{17}}{2} & -1 \\
  2 & -1 & -1 & \tfrac{-5+\sqrt{17}}{2}
  \end{array}\right), \\
  (\Pi_{\rm in})^{\mu}_{\;\;\nu}&=&\frac{(v_{1}^{+})^{\mu}(v_{1}^{-})_{\nu}}{\langle{\bf v}_{1}^{+},{\bf v}_{1}^{-}\rangle}
  +\frac{(v_{1}^{-})^{\mu}(v_{1}^{+})_{\nu}}{\langle{\bf v}_{1}^{-},{\bf v}_{1}^{+}\rangle}=\delta^{\mu}_{\;\;\nu}-(\Pi_{\rm ph})^{\mu}_{\;\;\nu} \\
  &=&\tfrac{-1}{\sqrt{17}}
  \left(
  \begin{array}{cccc} 
  \tfrac{7-\sqrt{17}}{2} & 0 & -2 & -2 \\
  0 & \tfrac{3-\sqrt{17}}{2} & 1 & -1 \\
  2 & 1 & \tfrac{-5-\sqrt{17}}{2} & -1 \\
  2 & -1 & -1 & \tfrac{-5-\sqrt{17}}{2}
  \end{array}\right).
\end{eqnarray*} 
Now, in Fig.~\ref{fig:I13quasicrystal} we illustrate the resulting quasicrystal that results from several natural choices of the weighting function (or window): first, the self-dual quasicrystal obtained by choosing a Gaussian weighting function $W({\bf v}_{\rm in})$ with the self-dual width, {\it i.e.}\ $W({\bf v}_{\rm in})\propto {\rm exp}(-\pi v_{\rm in}^{2})$ with $v_{\rm in}^{2}\equiv\langle{\bf v}_{\rm in},{\bf v}_{\rm in}\rangle$ (top left panel), the quasicrystal obtained by taking a circular-disk-shaped window ${\cal W}$ with radius $3\sigma$ (top right panel) where $\sigma=(2\pi)^{-1/2}$ is the standard deviation of the self-dual Gaussian, and two globally-scale-invariant quasicrystals (bottom panel) of weight $k=1.01$ (left) and $k=0.25$ (right), obtained by taking a weighting function $W({\bf v}_{\rm in})\propto v_{\rm in}^{-(2+k)}$ for these two values of $k$.  

\subsubsection{Example 2: $(3+1)$-dimensional Quasicrystals from $\mathrm{II}_{9,1}$}
\label{subsec:4DLorentzianQuasicrystal}

Now we turn to a higher-dimensional example that will discuss in the following section when we speculate about its possible physical interest.

As our lattice $\Lambda$, we now choose the remarkable 10-dimensional lattice $\mathrm{II}_{9,1}$ (the first non-trivial even unimodular lattice) and the $E_{10}$ root lattice. Its Coxeter diagram is shown in Fig.~\ref{fig:CoxeterDynkinDiagramsEvenSelfDual}, and the corresponding Schl\"{a}fli matrix reads
\begin{equation*}
S = \begin{pmatrix}
2 & -1 & 0 & 0 & 0 & 0 & 0 & 0 & 0 & 0 \\
-1 & 2 & -1 & 0 & 0 & 0 & 0 & 0 & -1 & 0 \\
0 & -1 & 2 & -1 & 0 & 0 & 0 & 0 & 0 & 0 \\
0 & 0 & -1 & 2 & -1 & 0 & 0 & 0 & 0 & 0 \\
0 & 0 & 0 & -1 & 2 & -1 & 0 & 0 & 0 & 0 \\
0 & 0 & 0 & 0 & -1 & 2 & -1 & 0 & 0 & 0 \\
0 & 0 & 0 & 0 & 0 & -1 & 2 & -1 & 0 & 0 \\
0 & 0 & 0 & 0 & 0 & 0 & -1 & 2 & 0 & 0 \\
0 & -1 & 0 & 0 & 0 & 0 & 0 & 0 & 2 & -1 \\
0 & 0 & 0 & 0 & 0 & 0 & 0 & 0 & -1 & 2 \\
\end{pmatrix}.
\end{equation*}
Expressed in an orthonormal basis (Cartesian coordinates), the fundamental roots are given explicitly in the caption of Fig.~\ref{fig:CoxeterDynkinDiagramsEvenSelfDual}, and the Coxeter element $C=R_{0}R_{1}\ldots R_{9}$ is
\begin{equation*}
C = \frac{1}{4}\left(\begin{array}{rrrrrrrrrr}
5 & -1 & -1 & -1 & -1 & -1 & -1 & -1 & -1 & -1 \\
1 & -1 & -1 & -1 & -1 & -1 & -1 & -1 & -1 & 3 \\
-1 & 1 & -3 & 1 & 1 & 1 & 1 & 1 & 1 & 1 \\
-1 & -3 & 1 & 1 & 1 & 1 & 1 & 1 & 1 & 1 \\
1 & -1 & -1 & 3 & -1 & -1 & -1 & -1 & -1 & -1 \\
1 & -1 & -1 & -1 & 3 & -1 & -1 & -1 & -1 & -1 \\
1 & -1 & -1 & -1 & -1 & 3 & -1 & -1 & -1 & -1 \\
1 & -1 & -1 & -1 & -1 & -1 & 3 & -1 & -1 & -1 \\
1 & -1 & -1 & -1 & -1 & -1 & -1 & 3 & -1 & -1 \\
1 & -1 & -1 & -1 & -1 & -1 & -1 & -1 & 3 & -1 \\
\end{array}\right).
\end{equation*}
Rounded to 5 digits, its eigenvalues are
\begin{eqnarray*}
&\big\{&\lambda_{0}^{\pm}=(1.17628)^{\pm 1}, \\ 
&&\lambda_{1}^{\pm}=-0.29233 \pm 0.95632\,i, \\
&&\lambda_{2}^{\pm}=-0.94331 \pm 0.33193\,i, \\
&&\lambda_{3}^{\pm}=+0.45687 \pm 0.88954\,i, \\
&&\lambda_{4}^{\pm}=-0.73444 \pm 0.67868\,i\big\}.
\end{eqnarray*}
To construct a $(3+1)$-dimensional quasicrystal that is invariant under the group $G$ generated by $C$, we must choose the physical space to be spanned by the real pair of eigenvectors ${\bf v}_{0}^{\pm}$ and one of the four pairs of complex conjugate eigenvectors. For example, if we choose the first pair of complex conjugate eigenvectors ${\bf v}_{1}^{\pm}$ we obtain the projection operators
\begin{eqnarray*}
  (\Pi_{\rm ph})^{\mu}_{\;\;\nu}&=&\frac{(v_{0}^{+})^{\mu}(v_{0}^{-})_{\nu}}{\langle{\bf v}_{0}^{+},{\bf v}_{0}^{-}\rangle}
  +\frac{(v_{0}^{-})^{\mu}(v_{0}^{+})_{\nu}}{\langle{\bf v}_{0}^{-},{\bf v}_{0}^{+}\rangle} \nonumber\\
  &+&\frac{(v_{1}^{+})^{\mu}(v_{1}^{-})_{\nu}}{\langle{\bf v}_{1}^{+},{\bf v}_{1}^{-}\rangle}
  +\frac{(v_{1}^{-})^{\mu}(v_{1}^{+})_{\nu}}{\langle{\bf v}_{1}^{-},{\bf v}_{1}^{+}\rangle}, \\
  (\Pi_{\rm in})^{\mu}_{\;\;\nu}&=&\delta^{\mu}_{\;\;\nu}-
  (\Pi_{\rm ph})^{\mu}_{\;\;\nu}.
\end{eqnarray*} 
But, instead of the pair ${\bf v}_{1}^{\pm}$, we could alternatively have chosen the pair ${\bf v}_{2}^{\pm}$, ${\bf v}_{3}^{\pm}$ or ${\bf v}_{4}^{\pm}$.  So there are four inequivalent $G$-invariant ways to decompose $V=\mathbb{R}^{9,1}$ into $V_{\rm ph}=\mathbb{R}^{3,1}$ and $V_{\rm in}=\mathbb{R}^{6,0}$, with each splitting giving rise to a different type of quasicrystal.

Finally, for any of these four inequivalent decompositions, our final choice is the weighting function: once again, the two most natural choices are the Gaussian-self-dual weighting function, or the power-law weighting function which is globally scale invariant with weight $w$.

Although these $(3+1)$ dimensional quasicrystals are too high dimensional to display explicitly, we can construct them numerically in exactly the same way as we did in the previous $(1+1)$ dimensional examples.  

\section{Spacetime Quasicrystals: Physical Speculations}
\label{sec:PhysicalSpeculations}

So far, this paper has focused on developing the mathematical formalism necessary to construct spacetime quasicrystals, and constructing the first examples.  But, in this final section, we can't resist offering some admittedly-half-baked speculations about how such spacetime quasicrystals might relate to the real world.  

\subsection{The Universe in a Nutshell}
\label{subsec:FittingUniverseNutshell}

Superstring theory \cite{Green:1987sp, Green:1987mn, Polchinski:1998rq, Polchinski:1998rr} has been much-studied as a candidate unified theory; but mathematical consistency requires the superstring to live in a $(9+1)$ dimensional target space (to cancel the worldsheet Weyl anomaly).  

To reconcile this with the $(3+1)$ dimensional universe we observe, the standard approach is to imagine (a modern variant of) the Kaluza-Klein picture in which four of the ten dimensions are large and non-compact, while the remaining six ``extra'' spatial dimensions are ``compactified'' ({\i.e.}\ curled up into an unobservably tiny 6D compact manifold). Alas, this picture so far does not have any observational evidence in its favour, and arguably creates more problems than its solves. In particular, stabilizing the six extra dimensions in a way that agrees with observations has been a notoriously difficult, and has driven theorists to consider awkwardly complex mechanisms that are often compared to Rube-Goldberg machines, and lead to a vast ``string landscape'' of possible vacua for 4D spacetime.

Here we propose a new geometric picture for how our 4D universe might be embedded in a higher-dimensional spacetime (completely different from the usual Kaluza-Klein picture.)~Although our proposal is logically independent of string theory, superstrings provides an intriguing motivation, as we now explain.

\subsubsection{Classical Picture: Irrational Spacetime}
\label{subsubsec:IrrationalityOfSpacetime}
\vspace{-4mm}
\begin{figure}[ht]
    \centering
    \includegraphics[width=0.99\linewidth]{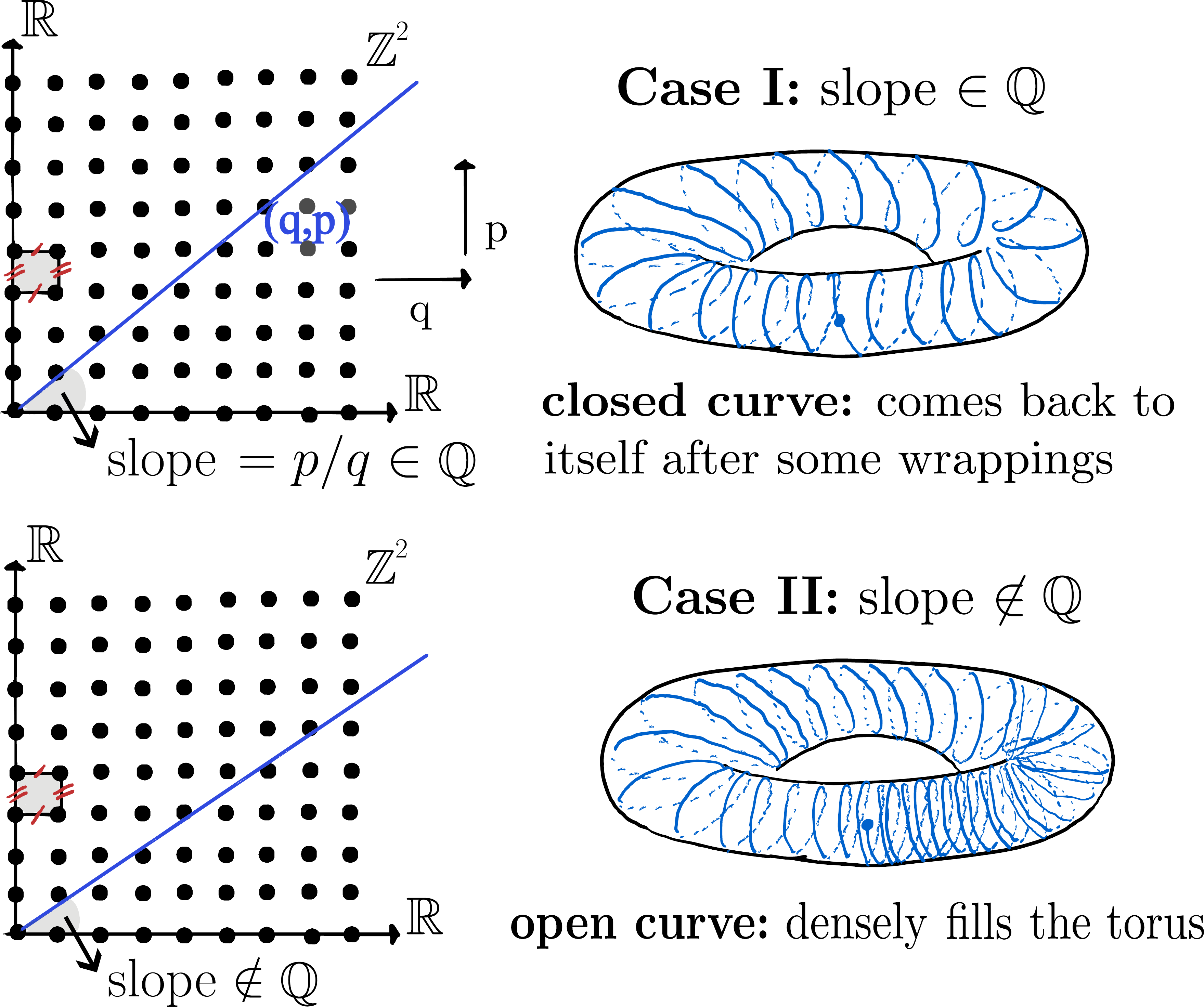}
    \caption{\justifying Case I: the line's slope is a rational number $p/q$, with $p$ and $q$ relatively prime integers. Then, if one moves $q$ steps in the $x$-direction and $p$ steps in the $y$-direction, one returns to the same point on the torus $\mathbb{R}^{2}/\mathbb{Z}^{2}$. Upon compactification (identification of the red edges on the square fundamental cell for the lattice), the origin becomes identified with any other lattice point. Thus, after compactification, the rational line will eventually come back to itself.  However, if the line is irrationally sloped (case II), then no point of the lattice will belong to the line. Such a line densely fills the torus, extends indefinitely on it and avoids ever crossing itself as it wraps around the torus.}
\label{fig:IrrLineCompactification}
\end{figure}

The most well studied compactifications of string theory are the {\it toroidal} compactifications: here one takes $m$ of the target-space dimensions to be curled up in an $m$-dimensional torus $T^{m}$, which is constructed by taking $m$-dimensional flat space $\mathbb{R}^{m}$ modulo an $m$-dimensional lattice $\Lambda_{m}$: $T^{m}=\mathbb{R}^{m}/\Lambda_{m}$. Due to stringy effects (in particular $T$-duality), such compactifications can have much higher symmetry than one would naively expect based on the isometry group of the compactified target spacetime.  In Ref.~\cite{moore1993finitedirections}, Moore studied such toroidal compactifications in generality, including the possibility of compactifying the time direction! He found that the {\it most symmetric} toroidal compactification of the superstring was obtained when {\it all 10 dimensions} of the target space were compactified on the 10D Lorentzian torus $T^{9,1}$ constructed by taking 10D Minkowski  space $\mathbb{R}^{9,1}$ modulo the 10D even self-dual Lorentzian lattice $\Lambda_{9,1}$: $T^{9,1}=\mathbb{R}^{9,1}/\mathrm{II}_{9,1}$.  

This most-symmetric superstring compactification was assumed to be unphysical, and of purely academic interest, for two reasons: (i) first, because if all ten dimensions are compactified at an ultra-microscopic scale, then one seems to be left without any remaining non-compact directions to serve as the $(3+1)$ macroscopic directions we observe; and (ii) second, because all directions (including time) are compactified, then the target spacetime has closed time-like curves which are usually assumed to be pathological, since they seem to be paradoxical and at odds with the causality properties of our universe.  

However, the previous section's spacetime quasicrystal construction suggests a way around this conclusion.

To see why, let us first warm up with a simplified 2D example: Consider the 2D square lattice $\Lambda_{2}$ in the 2D Euclidean plane $\mathbb{R}^{2}$; and a 1D line of slope $\sigma$ cutting through the lattice.  Now if we mod out $\mathbb{R}^{2}$ by the 2D lattice $\Lambda_{2}$, we obtain the torus $T^{2}=\mathbb{R}^{2}/\Lambda_{2}$; and the 1D line in $\mathbb{R}^{2}$ is mapped to a 1D curve on the torus $T^{2}$ (see the blue line in Fig.~\ref{fig:IrrLineCompactification}). If the original 1D line in $\mathbb{R}^{2}$ had a \textit{rational} slope $\sigma=p/q$ (where $p$ and $q$ are relatively prime integers), it is mapped to a {\it closed} curve of {\it finite} length on $T^{2}$, since the curve comes back joins up with itself after wrapping $p$ times {\it around} the $T^2$'s hole, and $q$ times {\it through} the $T^2$'s hole (top panel in Fig.~\ref{fig:IrrLineCompactification}).  By contrast, if the original line in $\mathbb{R}^{2}$ had an {\it irrational} slope $\sigma$, it is mapped to an {\it open} curve of {\it infinite} length on $T^{2}$, since the curve never joins up with itself, and instead wraps around and around the $T^2$, densely covering it (bottom panel in Fig.~\ref{fig:IrrLineCompactification}).  In this case, an ant confined to the line would inhabit a 1D world of infinite extent, even though it is embedded in a compact 2D torus of finite area!

\begin{figure}[]
    \centering
    \includegraphics[width=0.85\linewidth]{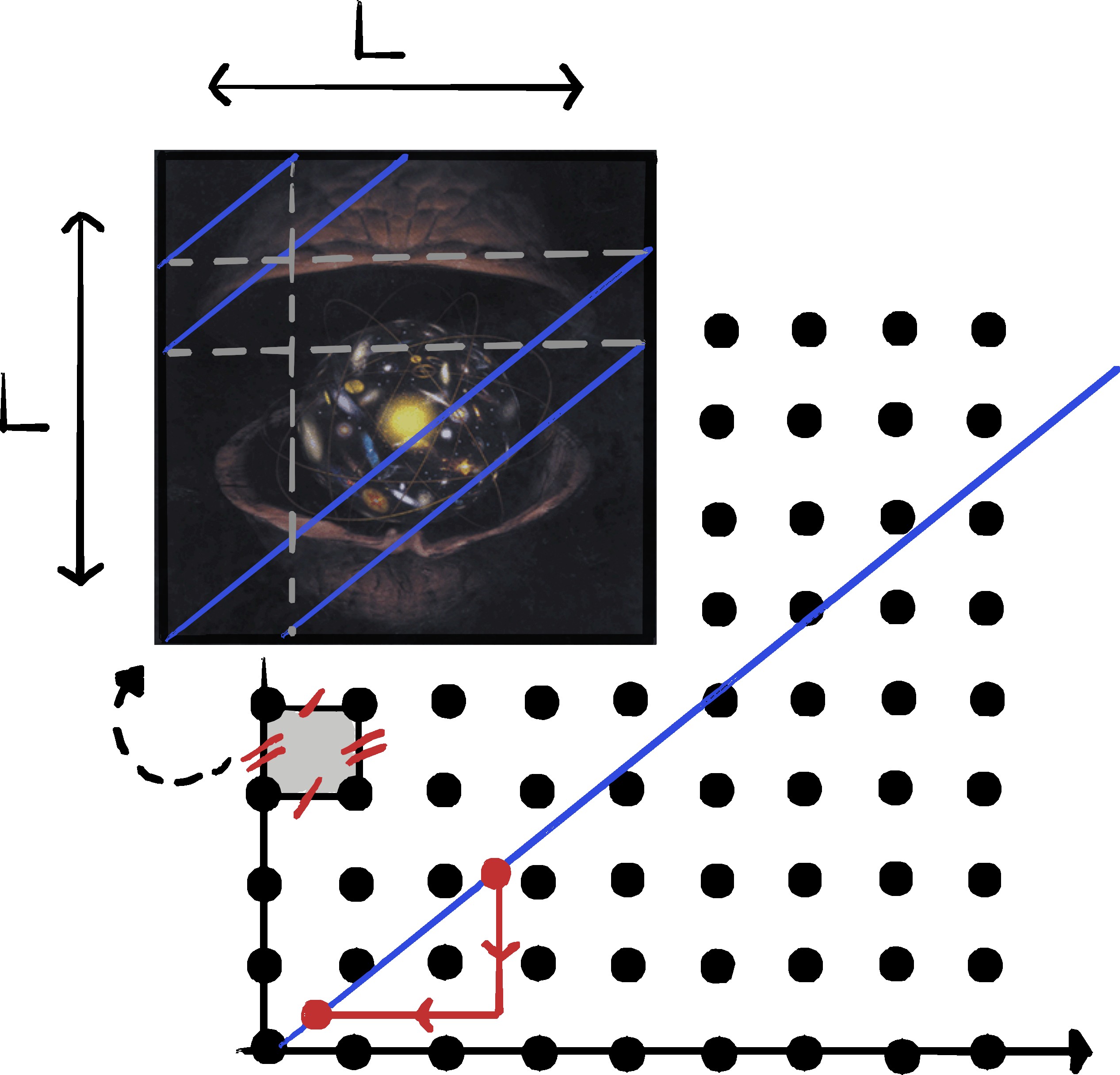}
    \caption{\justifying~Upon compactification, all points along the blue line can be identified with a point within the first square, which has been turned into a 2D torus.~By extension, the full infinite blue line is mapped within the square.~The game resembles ``Pac-Man'': whenever the line hits the boundary of the square, it restarts in the opposite edge from its equivalent point (up to edge-identification).~This is depicted at the background within the zoomed-in square. Thus, the whole infinite 1D blue line can fit inside a tiny 2D torus.~The no self-crossing condition follows from irrationality.~Extrapolating to the idea that the blue line represents our 4D spacetime living within 10 dimensions, the whole infinite universe can fit in a tiny nutshell! Figure adapted from Hawking's \cite{Hawking2001-HAWTUI}.}
    \label{fig:UniverseInNutshell}
\end{figure}

Now return to the 10D example of interest: consider the 10D Lorentzian lattice $\mathrm{II}_{9,1}$ in 10D Minkowski space $\mathbb{R}^{9,1}$.  We have seen in Sec.~\ref{subsec:4DLorentzianQuasicrystal} that the Coxeter element $C$ of $\mathrm{II}_{9,1}$ generates a non-crystallographic group $G$ that group induces a decomposition $V=V_{\rm ph}\oplus V_{\rm in}$, where $V_{\rm ph}=\mathbb{R}^{3,1}$ (a copy of 4D Minkowski space) and $V_{\rm in}=\mathbb{R}^{6,0}$ (a copy of 6D Euclidean space) transform under two different representations ($\rho_{3,1}$ and $\rho_{6,0}$) of $G$.  Crucially, since $G$ is non-crystallographic, this 4D Minkowski subspace $V_{\rm ph}=\mathbb{R}^{3,1}$ is {\it completely irrational} with respect to the lattice $\mathrm{II}_{9,1}$ ({\it i.e.}\ it does not intersect any points in the lattice $\mathrm{II}_{9,1}$, except for the origin).  So now, if we mod out $\mathbb{R}^{9,1}$ by the lattice $\mathrm{II}_{9,1}$, we obtain the 10D Lorentzian torus $T^{9,1}=\mathbb{R}^{9,1}/\mathrm{II}_{9,1}$; and the irrational 4D Minkowski subspace $V_{\rm ph}=\mathbb{R}^{3,1}$ is mapped to a $(3+1)$D surface on the $T^{9,1}$, which densely fills the torus, wrapping around and around it without ever intersecting itself (like the blue line in the lower panel of Fig.~\ref{fig:IrrLineCompactification}).

Wigner realized that different types of elementary particles are characterized by the way in which they transform under the fundamental symmetries of the theory.  In particular, elementary particles that transform as the representation $\rho_{3,1}$ of $G$ would be confined to the irrational $(3+1)$-dimensional surface that transforms in the appropriate way.  A being made of such particles would regard themselves as living in an {\it infinite} (non-compact) 4D Minkowski space, with {\it no closed timelike curves}, despite the fact that this 4D spacetime is embedded in a 10D torus $T^{9,1}$ of finite volume, with all 10 dimensions compactified (including the time direction)!  

In other words, this picture fits an infinite $(3+1)$D universe in a $(9+1)$D nutshell (see Fig. \ref{fig:UniverseInNutshell}).

\subsubsection{Hierarchy Puzzles: Cartoon Picture}
\label{subsubsec:CartoonPicture}

Now let us see how this picture might be helpful in explaining some of the large hierarchies between three fundamental dimensional scales in Nature: the Planck scale $M_{\rm Pl}\sim 10^{18}~{\rm GeV}$, the electroweak scale $M_{\rm EW}\sim 10^{3}~{\rm GeV}$ and the vacuum energy scale $M_{\rm vac}\sim 10^{-12}~{\rm GeV}$.

To set the stage, consider the action for the standard model of particle physics coupled to gravity.  Written in its symmetric form, almost all of the terms have dimension{\it less} coefficients, and only three have dimension{\it ful} coefficients. Consider these three  dimensionful terms:
\begin{equation}  
    S =\int d^4x \sqrt{-g} \left[ \underbrace{\mspace{+5mu} \textcolor{black}{M_{\rm EW}^2} (h^{\dagger} h) \mspace{+5mu}}_{\mathclap{\text{Higgs}}}+ \underbrace{\mspace{5mu} \textcolor{black}{M_{\rm Pl}^2} R \mspace{5mu}}_{\mathclap{\text{Bending}}} \mspace{+5mu} - \underbrace{\mspace{5mu} \textcolor{black}{M_{\rm vac}^4} \mspace{5mu}}_{\mathclap{\text{Stretching}}} \mspace{5mu} \right]
    \label{eq:ActionHierarchyProblem}
\end{equation}
namely, the quadratic self-coupling $M_{\rm EW}^{2}(h^{\dagger}h)$ of the Higgs doublet $h$ (responsible for the Higgs mechanism), the Einstein-Hilbert term $M_{\rm Pl}^{2}R$ (responsible for gravity) and the vacuum energy term $M_{\rm vac}^{4}$ (responsible for the universe's slowly accelerating expansion today).

Historically, gravity was discovered before the standard model of particle physics; but for motivational purposes, imagine a counterfactual history where it was the other way around.  To a physicist who only knew about the standard model, it would seem that there was only {\it one} dimensionful term in the action (the Higgs term), and hence only {\it one} fundamental scale in nature: $M_{\rm EW}$.  

If we then informed this physicist of the existence of gravity, and two additional dimensionful terms in the action, they would probably guess the order of magnitude of these new coefficients was set by the only scale they were aware of ($M_{\rm EW}$) -- {\it i.e.} they might guess the new terms would be of order $\int d^{4}x\sqrt{-g}(M_{\rm EW}^{2}R+M_{\rm EW}^{4})$.  They would be shocked to discover that, in reality, the coefficient of 
the Einstein-Hilbert term (proportional to the Ricci curvature $R$) was so much {\it larger} than expected (by a factor of $10^{30}$), while the coefficient of the vacuum energy term (proportional to the overall volume of spacetime) was so much {\it smaller} than expected (by a factor of $10^{60}$).  In other words, they would be shocked that there was such a {\it high cost} (in terms of action) for {\it bending} spacetime, and such a {\it low cost} for {\it stretching} spacetime.

From a ``cartoon'' standpoint, the picture presented in Sec.~\ref{subsubsec:IrrationalityOfSpacetime} seems like it might be a promising starting point for explaining these surprising qualitative features.  On the one hand, the fact that the irrationally-sloped 4D Minkowski slice $\mathbb{R}^{3,1}$ {\it densely} fills the 10D torus $T^{9,1}$ seems like it might explain why $\mathbb{R}^{3,1}$ is so {\it hard} to bend (because any given portion of the $\mathbb{R}^{3,1}$ is infinitely tightly ``hemmed in'' in the extra dimensions between other infinitely-close neighbouring portions of the $\mathbb{R}^{3,1}$, like the blue curve in the lower panel of Fig.~\ref{fig:IrrLineCompactification}). On the other hand, the fact that the $\mathbb{R}^{3,1}$ slice is infinite in extent (like the blue curve in the bottom panel of Fig.~\ref{fig:IrrLineCompactification}) rather than rejoining itself after a finite distance (like the blue curve in the top panel of Fig.~\ref{fig:IrrLineCompactification}) might explain why it is so {\it easy} to stretch (since it has no intrinsic length scale, and nothing to obstruct or ``push back against'' rescaling it).  Moreover, we have seen that the associated C\&P quasicrystals -- and in particular the relevant $(3+1)$D quasicrystals constructed in Sec.~\ref{subsec:4DLorentzianQuasicrystal} -- have a global discrete scale invariance (which, again, corresponds to low cost in action for rescaling spacetime).

\subsubsection{Hierarchy Puzzles: Seesaw Mechanism}
\label{subsubsec:SeesawMechanism}

We now attempt to go beyond the previous section's cartoon picture, to suggest a (semi-)quantitative argument for how this picture might help explain the observed seesaw relation between $M_{\rm vac}$ and $M_{\rm Pl}$:
\begin{equation}
  \label{M_seesaw}
  M_{\rm EW}^{2}\approx M_{\rm vac}M_{\rm Pl}.
\end{equation}

\begin{figure}
    \centering
    \includegraphics[width=0.85\linewidth]{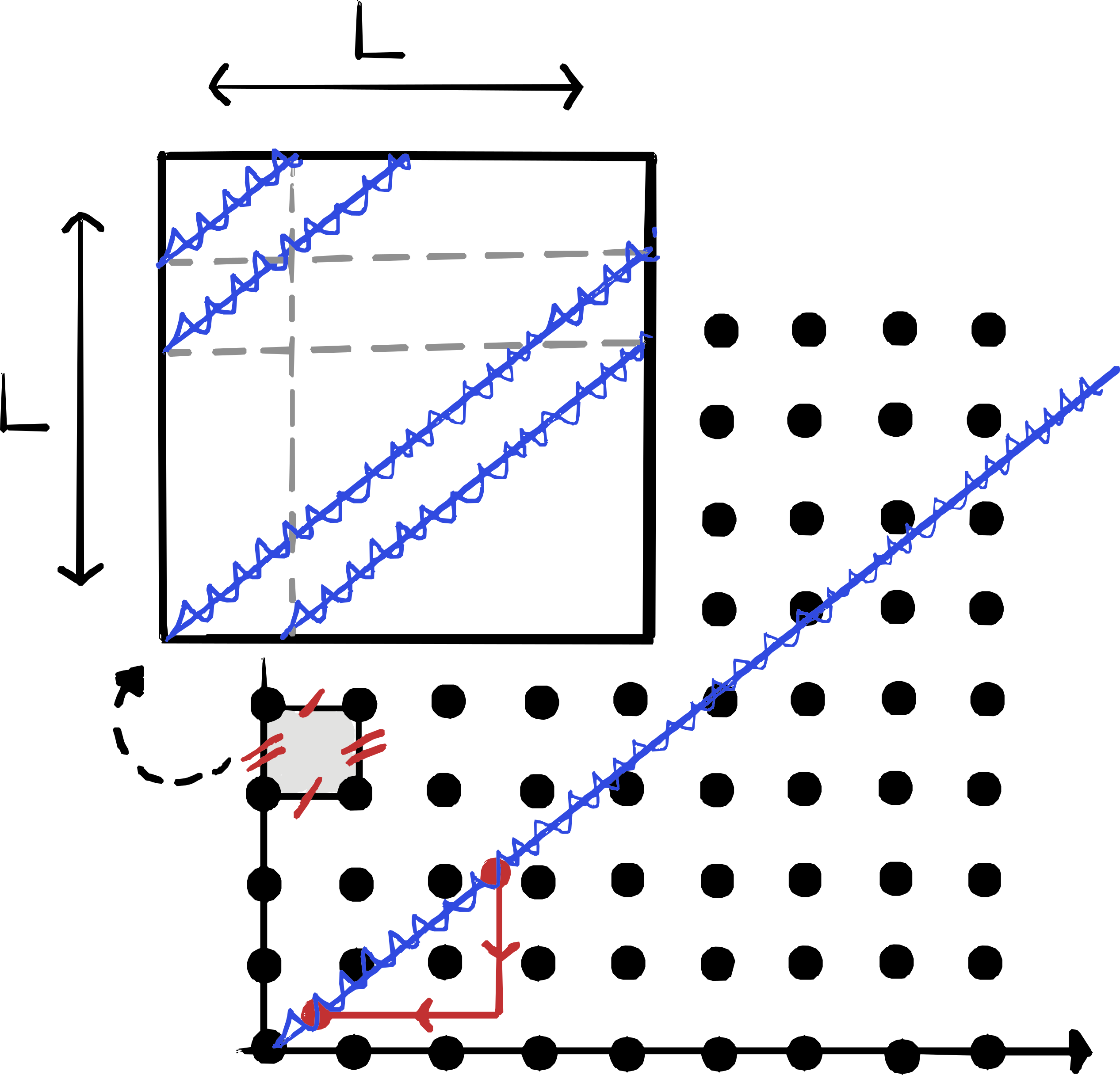}
    \caption{\justifying Enter Quantum Fluctuations.~Now the line is not infinitely thin so it will eventually touch a point of the higher-dimensional lattice, and hence the corresponding curve on the torus will eventually intersect/overlap itself.}
    \label{fig:FiniteThicknessLine}
\end{figure}

The discussion in Secs.~\ref{subsubsec:IrrationalityOfSpacetime}, \ref{subsubsec:CartoonPicture} considered an idealized, classical picture, in which the 4D Minkowski slice (densely filling the 10D spacetime torus) was taken to be infinitely ``thin'' in the six extra dimensions.  But a real object won't be infinitely thin; it will have some finite thickness due to quantum fluctuations (see Fig.~\ref{fig:FiniteThicknessLine}).

To analyze the impact of such fluctuations, consider the 10D torus at some moment in time: then we have a 3D irrationally-sloped spatial slice (with 6 extra/transverse directions) densely filling a 9D spatial torus. We will refer to this 3D slice as a ``3-brane'' for short.  And instead of assuming 6 extra dimensions from the outset, it will be instructive to initially analyze a more general picture with $n$ extra dimensions, so the 3-brane densely fills a $(3+n)$D spatial torus.

Quantum fluctuations effectively give the 3-brane a finite non-zero thickness in the extra dimensions.  Let us imagine this transverse quantum thickness to be the Planck length $l_{pl}=M_{\rm Pl}^{-1}$; and hence, since the slice has $n$ transverse dimensions, its $n$D transverse volume ${\cal V}_{n}$ is 
\begin{equation}
  \label{V_n}
  {\cal V}_{n}\approx M_{\rm Pl}^{-n}.
\end{equation}
This 3-brane lives within a $(3+n)$D spatial torus $T^{3+n}$ of linear size $L$, and hence $(3+n)$D volume 
\begin{equation}
  \label{V_3+n}
  {\cal V}_{3+n}\approx L^{3+n}.
\end{equation}
Since the 3-brane is {\it densely} packed within the torus $T^{3+n}$, its 3-volume ${\cal V}_{3}$ (along the 3-brane) times its $n$D volume ${\cal V}_{n}$ (transverse to the 3-brane) must equal the torus's volume ${\cal V}_{3+n}$:
\begin{equation}
  \label{V_seesaw}
  {\cal V}_{3+n}={\cal V}_{3}{\cal V}_{n}.
\end{equation} 
Thus, we have a seesaw relation between ${\cal V}_{3}$ and ${\cal V}_{n}$ (for fixed ${\cal V}_{3+n}$): when the 3-brane is infinitely thin (${\cal V}_{n}\to0$) its 3-volume can be infinite (${\cal V}_{3}\to\infty$) -- indeed, this was our previous classical picture; but if quantum fluctuations give the 3-brane a finite non-zero thickness ($V_{n}>0$), its 3-volume must also become finite ($V_{3}<\infty$) in order to still fit inside the torus's compact volume $V_{3+n}$; and the larger the transverse thickness $V_{6}$, the smaller the spatial 3-volume ${\cal V}_{3}$ of ``our 3-brane.''  

Let us now see how this ``volume seesaw'' (\ref{V_seesaw}) might be converted into the observed ``mass seesaw'' (\ref{M_seesaw}).  The conversion requires two key physical inputs:
\begin{enumerate}
    \item For starters, how should we interpret the finite 3-volume of our 3-brane? The standard $\Lambda$CDM model of cosmology \cite{mukhanov2005physical, weinberg2008cosmology} suggests a natural and intriguing answer.  Since our universe is observed to have positive vacuum energy density $M_{\rm vac}^{4}$, Einstein's equations imply that we are surrounded by a ``de Sitter horizon'' of radius $H_{dS}^{-1}$, and hence are limited to only ever observe or communicate with a finite portion of our universe, with finite spatial 3-volume $H_{dS}^{-3}$, where $H_{dS}$ is the de Sitter expansion rate, determined by the Friedmann equation
\begin{equation}
  H_{dS}^{2}=\frac{8\pi G}{3}\rho_{\rm vac}\approx M_{\rm Pl}^{-2}M_{\rm vac}^{4}.
\end{equation} 
In other words, cosmological observations have led us to the surprising conclusion that our universe effectively has finite spatial 3-volume.  Could this be the {\it same} as the finite spatial 3-volume ${\cal V}_{3}$ of our 3-brane?  With this identification,
\begin{equation}
  \label{V_3}
  {\cal V}_{3}\approx H_{dS}^{-3}\approx M_{\rm Pl}^{3}/M_{\rm vac}^{6},
\end{equation}
and we can substitute Eqs.~(\ref{V_n}, \ref{V_3+n}, \ref{V_3}) into the volume seesaw (\ref{V_seesaw}) to obtain
\begin{equation}
     \label{L1}
     L^{3+n}\approx M_{\rm Pl}^{3-n}M_{\rm vac}^{-6}.
\end{equation}
\item The next crucial step is to imagine (following {\it e.g.}\ \cite{arkani1998hierarchy}), that -- in contrast to the rest of the standard model fields -- gravity is not confined to the 3-brane, so that the higher-dimensional ($4+n$ dimensional) Planck mass $M_{4+n}$, with associated higher-dimensional Einstein-Hilbert term $\int d^{4+n}x\sqrt{-g_{4+n}}M_{4+n}^{2+n}R_{4+n}$ is related to the lower-dimensional ($4$ dimensional) Planck mass $M_{4}$, with associated lower-dimensional Einstein-Hilbert term $\int d^{4}x\sqrt{-g_{4}}M_{4}^{2}R_{4}$, by integrating over the full extra-dimensional volume of the torus $L^{n}$ to obtain
\begin{equation}
  \label{L2}
  M_{4}^{2}\approx M_{4+n}^{2+n}L^{n}.
\end{equation}
Here $M_{4}$ is the observed (4D) Planck mass $M_{\rm Pl}$, but what should we take for the higher-dimensional Planck mass $M_{4+n}$?~Again following {\it e.g.}\ \cite{arkani1998hierarchy}, we imagine that the enormous hierarchy between the $M_{\rm Pl}$ and $M_{\rm EW}$ is a manifestation of the fact that we are confined to a lower-dimensional slice of spacetime, rather than perceiving the higher-dimensional spacetime. In other words, whereas the lower-dimensional Planck mass has an enormous hierarchy relative to the electroweak scale ($M_{4}=M_{\rm Pl}\gg M_{\rm EW}$), the higher-dimensional Planck mass has no such hierarchy ($M_{4+n}\approx M_{\rm EW}$). With this identification, Eq.~(\ref{L3}) becomes
\begin{equation}
  \label{L3}
  M_{\rm Pl}^{2}\approx M_{\rm EW}^{2+n}L^{n}.
\end{equation}
\end{enumerate}
Finally, we can combine Eqs.~(\ref{L1}) and (\ref{L3}), to eliminate $L$ and obtain
\begin{equation}
    M_{\rm vac}M_{\rm Pl}^{\frac{(n^{2}-n+6)}{6n}}\approx M_{\rm EW}^{\frac{(2+n)(3+n)}{6n}}.
\end{equation}
Strikingly, this reduces to the observed seesaw relation (\ref{M_seesaw}) precisely when there are $n=6$ extra dimensions, just as in the scenario proposed here!

\subsection{Other Speculations and Suggestions}
\label{subsec:AdditionalRemarks}

Finally, let us end by briefly mentioning some other connections of spacetime quasicrystals to physics, as interesting topics for future investigation:
\begin{itemize}
    \item Given their discrete scale invariance, quasicrystals provide natural mathematical structures on which to discretize and study systems in which (exact or approximate) scale or conformal invariance plays an important role ({\it e.g.}\ QFTs near their fixed points, condensed matter systems near their critical points, or our universe near the Big Bang). 
    \item Scale-symmetric quasicrystals, in particular the Ammann-Beenker tilings, have recently shown to be efficient in solving the general-case \textsf{NP}-complete problem of finding Hamiltonian cycles (\textit{i.e.}~closed loops on a graph visiting each vertex precisely once)  \cite{HamCycles_Felix2024}.~Since many interesting problems in physics fall in the \textsf{NP} complexity class, this result hints towards the fact that quasicrystals are essentially as powerful (if not more) as periodic lattices in simplifying certain physical problems. It would be interesting to understand the graph-theoretic properties of spacetime quasicrystals and see what kind of problems their special properties can help us solve.
    \item Spacetime quasicrystals may be of interest in studying quantum gravity, as explicit, mathematically natural models for discretized spacetime, with many interesting, non-trivial, and counterintuitive properties, and a high degree of (orientational and scale) symmetry (and quasi-symmetry).  
    \item In particular, they provide a novel class of causal sets \cite{Bombelli:1987aa, Surya:2019ndm}, very different from either Lorentzian lattices or random sets.
    \item Due to their discrete scale invariance, they may be especially relevant to quantum gravity approaches where spacetime has a fractal, scale-invariant character ({\it e.g.}~asymptotic safety \cite{Weinberg:1976xy, Weinberg:1980gg, Niedermaier:2006wt, Dupuis:2020fhh}).
    \item There may be interesting connections between the compactification scenario we suggest in Secs.~\ref{subsubsec:IrrationalityOfSpacetime}, \ref{subsubsec:CartoonPicture}, \ref{subsubsec:SeesawMechanism} and the ``quasicrystalline string compactifications'' idea discussed in \cite{Harvey:1987da, Baykara2024Quasicrystalline}.
    \item There may be interesting connections to the much-studied IKKT matrix model of Type IIB string theory (which may also be thought of as arising from D=10 super Yang-Mills theory, by compactifying all 10 dimensions of spacetime) \cite{Ishibashi:1996xs, Aoki:1998bq}.
    \item There may be interesting connections between our compactification scenario (based on compactifying 10D spacetime using $\mathrm{II}_{9,1} = E_{10}$ root lattice) and other investigations of the role of $E_{10}$ in string theory, cosmology and the standard model \cite{Damour:2000hv, Damour:2002cu, Damour:2002et, Kleinschmidt:2015sfa, Meissner:2025qrx}.
    \item Finally, spacetime quasicrystals may yield new insights into the recently noticed, but still incompletely understood, triangle of connections between (discrete) holography and quantum error correction \cite{Pastawski:2015qua}, between quantum error correction and quasicrystals \cite{li2024penrosetilingquantumerrorcorrecting}, and between quasicrystals and (discrete) holography \cite{Boyle:2018uiv, Boyle:2024qzn}.
\end{itemize}

\begin{acknowledgments}
We thank Nikolay Bogachev, 
Mikhail Belolipetsky,
Richard Borcherds,
Gerald H\"ohn, Tobias Hartnick, 
Henna Koivusalo,  
Felix Flicker, 
Jury Radkovski, 
Ben Webster,
and Kendrick Smith for helpful discussions. SM would like to thank the School of Physics and Astronomy and the Higgs Center for Theoretical Physics at the University of Edinburgh for the hospitality in the following occasions (Fall 2023, July 2024, Fall 2024, July 2025, Fall 2025), without which finishing this work would have been impossible. He would also like to thank the Savvas Chamberlain Family Foundation and the Hellenic Heritage Foundation for their continuous support through the Anaximandros Fellowship (and personally Savvas and Christine Chamberlain, Maria Harper, Tula Alexopoulos, Michael Lagopoulos and Tony Lee) and the AHEPA Educational Foundation for the 2024/25 Mavroyiannis Scholarship. LB is supported by the STFC Consolidated Grant ``Particle Physics at the Higgs Centre.''  Research at Perimeter Institute is supported by the Government of Canada through Industry Canada and by the Province of Ontario through the Ministry of Economic Development and Innovation.
\end{acknowledgments}

\appendix

\section{Reflection Symmetries of Odd Self-Dual Lorentzian Lattices $I_{s,1}$}
\label{app:SymmetriesLorentzianLattices}

In this appendix, we summarize the reflection symmetries of the odd self-dual lattices $I_{s,1}$ for dimension $s+1\leq 20$ ({\it i.e.}\ the dimensions for which the Coxeter group has a finite number of fundamental roots).

For $3\leq s+1\leq 18$, the fundamental roots and Coxeter-Dynkin diagrams for the lattices $\mathrm{I}_{s,1}$ and $\mathrm{II}_{s,1}$ were determined by Vinberg in \cite{vinberg1967discrete, vinberg1972groups, vinberg1975some} (see Table 4A and 4B in \cite{vinberg1975some} in particular).  The cases $s+1=19,20$ were then treated by Vinberg and Kaplinskaja in \cite{kaplinskaja1978groups} (see also \cite{meyer1977prasentation}).  And for $s+1>20$ there are infinitely many fundamental roots \cite{ESSELMANN1996103}. For convenience, since some of these references are only available in Russian or German, or contain errors in the printed diagrams (e.g.~in \cite{vinberg1975some}), we have carefully re-checked their results, and present all the graphs here, along with a list of the fundamental roots, for all the case with a finite number of fundamental roots ({\it i.e.}\ $s+1\leq 20$).

We follow Vinberg's approach in Refs.~\cite{vinberg1975some,kaplinskaja1978groups} with the difference that our fundamental roots are denoted by $\mathbf{r}_i$ (instead of $\mathbf{e}_j$), while our standard orthonormal (Cartesian) basis vectors are denoted by ${\bf e}_{i}$ (instead of ${\bf u}_{i}$); and, to align with the physicists' notation, we label the Cartesian basis in $\mathbb{R}^{s,1}$ as $\{\mathbf{e}_j\}$ ($j=0,\dots,s$), with $\mathbf{e}_0$ corresponding to the temporal basis vector. 

For $3\leq s+1\leq 18$, the fundamental roots for the lattices $\mathrm{I}_{s,1}$ are given in Table \ref{tab:SimpleRootsReflectiveOddLatticesN2to17}.  For $s+1=19$ or $20$, the first $s+8$ fundamental roots are given in Table \ref{tab:SimpleRootsReflectiveOddLatticesN1819}; while for $s+1=19$, the remaining fundamental roots are given in Table \ref{tab:AllSimpleRootsReflectiveOddLattices18}; and for $s+1=20$, the remaining fundamental roots are given in Table \ref{tab:AllSimpleRootsReflectiveOddLattices19}.

\begin{table*}[t]
    \centering
    \begin{tabular}{c | c | c | c }
         Root Index $i$ & Root Vector $\mathbf{r}_i$ & $\langle \mathbf{r}_i,\mathbf{r}_i \rangle$ & for which $s$ \\
         \vspace{-2ex}& & & \\
         \hline \hline 
         \vspace{-1.75ex}& & & \\
         $1 \leq i \leq s-1$ & $-\mathbf{e}_i + \mathbf{e}_{i+1}$ & $2$ & $s\geq 2$ \\[+1ex]
         \hline 
         \vspace{-1.75ex}& & & \\
         $s$ & $-\mathbf{e}_s$ & $1$ & $s\geq 2$ \\[1ex] \hline
         \vspace{-1.75ex}& & & \\
         $s+1$ & $\mathbf{e}_0 + \mathbf{e}_1 + \mathbf{e}_2$ & 1 & $s=2$ \\[+1ex]
         & $\mathbf{e}_0 + \mathbf{e}_1 + \mathbf{e}_2 + \mathbf{e}_3$ & 2 & $s\geq 3$ \\[+1ex] \hline
        \vspace{-1.75ex}& & & \\
         $s+2$ & $3\mathbf{e}_0 + \mathbf{e}_1 +  \dots + \mathbf{e}_{10}$ & 1 & $s=10$ \\[1ex]
         &$3\mathbf{e}_0 + \mathbf{e}_1 +  \dots + \mathbf{e}_{11}$ & 2 & $s\geq 11$ \\[+1ex] \hline
        \vspace{-1.75ex}& & & \\
         $s+3$ & $4\mathbf{e}_0 + 2\mathbf{e}_1 + \mathbf{e}_2 + \dots + \mathbf{e}_{14}$ & 1 & $s=14$ \\[1ex]
               & $4\mathbf{e}_0 + 2\mathbf{e}_1 + \mathbf{e}_2 + \dots + \mathbf{e}_{15}$ & 2 & $s\geq 15$ \\[+1ex] \hline
         \vspace{-1.75ex}& & & \\      
         $s+4$ & $6\mathbf{e}_0 + 2(\mathbf{e}_1 + \dots + \mathbf{e}_7) + (\mathbf{e}_8 + \dots + \mathbf{e}_{16})$ & 2 & $s=16$ \\[1ex] 
        & $4\mathbf{e}_0 + \mathbf{e}_1 +  \dots + \mathbf{e}_{17}$ & 1 & $s=17$ \\[1ex]\hline
         \vspace{-1.75ex}& & & \\
         $s+5$ & $6\mathbf{e}_0 + 2(\mathbf{e}_1 + \dots + \mathbf{e}_7) + (\mathbf{e}_8 + \dots + \mathbf{e}_{17})$ & 2  & $ s=17$ 
    \end{tabular}
    \caption{\justifying A list of the fundamental roots for all odd self-dual lattices $\mathrm{I}_{s,1}$ with $3 \leq s+1\leq 18$. The indices in the first column match the labelling of the roots in the Coxeter-Dynkin graphs of Figs.~\ref{fig:VinbergRootsI1nUpto10}, \ref{fig:OddGraphsN10to13}, and \ref{fig:OddGraphsN14to17}.  The second column expresses each root in terms of the standard Cartesian basis $\{{\bf e}_{0},\ldots,{\bf e}_{s}\}$ for $\mathbb{R}^{s,1}$, where ${\bf e}_0$ is the unit vector along the timelike direction. The third column shows the root's norm ({\it i.e.}\ whether it is a short or a long root).  And the fourth column explains for which $s$ ({\it i.e.}\ in which number of spatial dimensions) a given fundamental root exists.  } 
\label{tab:SimpleRootsReflectiveOddLatticesN2to17}
\end{table*}

\begin{table*}[t]
    \centering
    \begin{tabular}{c | c | c }
         Root Index $i$ & Root Vector $\mathbf{r}_i$ & $\langle \mathbf{r}_i,\mathbf{r}_i \rangle$ \\
         \vspace{-2ex}& & \\
         \hline \hline 
         \vspace{-1.75ex}& & \\
         $1 \leq i \leq s-1$ & $-\mathbf{e}_i + \mathbf{e}_{i+1}$ & $2$ \\[+1ex]
         \hline 
         \vspace{-1.75ex}& & \\
         $s$ & $-\mathbf{e}_s$ & $1$ \\[1ex] \hline
         \vspace{-1.75ex}& & \\
         $s+1$ & $\mathbf{e}_0 + \mathbf{e}_1 + \mathbf{e}_2 + \mathbf{e}_3$ & 2 \\[+1ex] \hline
        \vspace{-1.75ex}& & \\
         $s+2$ &$3\mathbf{e}_0 + \mathbf{e}_1 +  \dots + \mathbf{e}_{11}$ & 2 \\[+1ex] \hline
        \vspace{-1.75ex}& & \\
         $s+3$ & $4\mathbf{e}_0 + 2\mathbf{e}_1 + \mathbf{e}_2 + \dots + \mathbf{e}_{15}$ & 2 \\[+1ex] \hline
         \vspace{-1.75ex}& & \\      
         $s+4$ & $4\mathbf{e}_0 + \mathbf{e}_1 +  \dots + \mathbf{e}_{18}$ & 2 \\[1ex]\hline
         \vspace{-1.75ex}& & \\
         $s+5$ & $5\mathbf{e}_0 + 3\mathbf{e}_1 + \mathbf{e}_2 + \dots + \mathbf{e}_{s}$ & 1,2 
         \\[1ex] \hline
         \vspace{-1.75ex}& & \\
         $s+6$ & $6\mathbf{e}_0 + 2(\mathbf{e}_1 + \dots + \mathbf{e}_7) + \mathbf{e}_8 + \dots + \mathbf{e}_{17}$ & 2  
         \\[1ex] \hline
         \vspace{-1.75ex}& & \\
         $s+7$ & $7\mathbf{e}_0 + 3\mathbf{e}_1 + 2(\mathbf{e}_2 + \dots + \mathbf{e}_9) + \mathbf{e}_{10} + \dots + \mathbf{e}_{s}$ & 1,2 
         \\[1ex]\hline
         \vspace{-1.75ex}& & \\
         $s+8$ & \, $9\mathbf{e}_0 + 3(\mathbf{e}_1 + \dots + \mathbf{e}_5) + 2(\mathbf{e}_6 + \dots + \mathbf{e}_{13}) + \mathbf{e}_{14} + \dots + \mathbf{e}_{s}$ \, & 1,2 \\[1ex]
    \end{tabular}
    \caption{\justifying A list of the first $s+8$ fundamental roots for the odd self-dual lattices $\mathrm{I}_{s,1}$ when $s+1=19$ or $20$. The indices in the first column match the labelling of the roots in the Coxeter-Dynkin graphs of Fig.~\ref{fig:OddGraphsN18to19}.  The second column expresses each root in terms of the standard Cartesian basis $\{{\bf e}_{0},\ldots,{\bf e}_{s}\}$ for $\mathbb{R}^{s,1}$, where ${\bf e}_0$ is the unit vector along the timelike direction. The third column shows the root's norm ({\it i.e.}\ whether it is a short or a long root).} 
\label{tab:SimpleRootsReflectiveOddLatticesN1819}
\end{table*}

Using the fundamental roots in these tables, one can construct the corresponding Coxeter-Dynkin graphs for the lattice $\mathrm{I}_{s,1}$ in Minkowski space of dimension $d=s+1$, with $3\leq d \leq 20$.  In order to describe the structure of the remaining roots, we present these graphs in four batches of increasing complexity, commenting on certain properties and interesting aspects along the way.  (In any dimension $d=s+1$, the first $s$ roots form a copy of the spherical/finite Coxeter graph of type $B_{s}$.)
\begin{itemize}
\item \textbf{Dimensions $3 \leq d \leq 10$}: see Fig. \ref{fig:VinbergRootsI1nUpto10}.  For these graphs, the number of fundamental roots 
(the rank) equals the dimension, $n=d$, so the fundamental roots form a (linearly-independent) basis for $\mathbb{R}^{s,1}$, and define a fundamental domain in hyperbolic space $H^{s}$ bounded by $s+1$ mirrors (a hyperbolic {\it simplex}).

\item \textbf{Dimensions $11 \leq d \leq 14$}: see Fig. \ref{fig:OddGraphsN10to13}. Now the rank exceeds the dimension by one, $n=d+1$, so the fundamental roots are over-complete, and the fundamental domain in $H^{s}$ is {\it not} a simplex \cite{vinberg1975some}. 

\begin{figure}[h]
\includegraphics[width=0.99\linewidth]{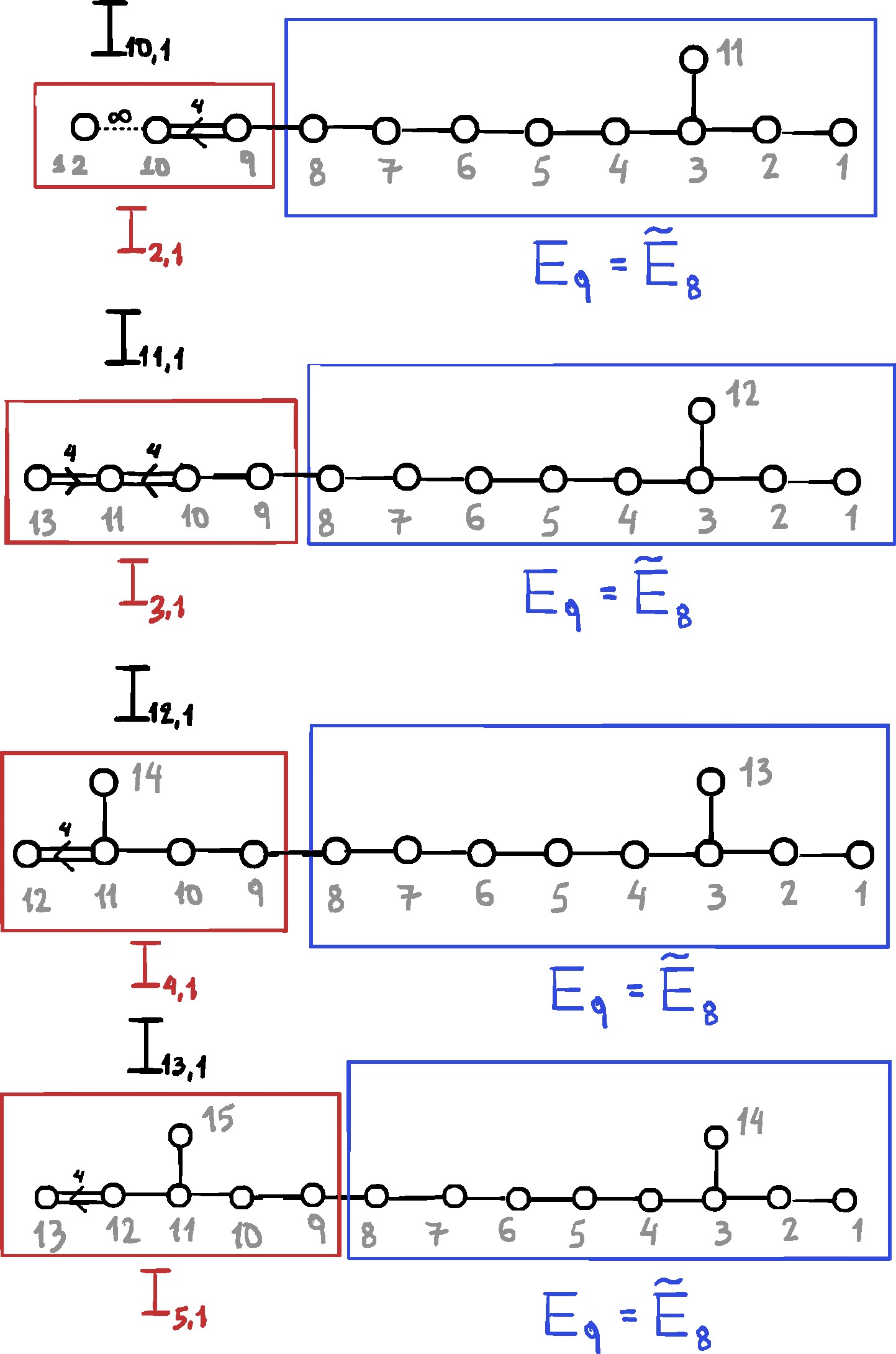}
\caption{\justifying The Coxeter-Dynkin diagrams for $I_{s,1}$ for $11\leq s+1\leq 14$ (with fundamental roots given in Table \ref{tab:SimpleRootsReflectiveOddLatticesN2to17}).}
\label{fig:OddGraphsN10to13}
\end{figure}

Note a kind of 8-fold periodicity here: the graphs for $11 \leq d \leq 14$ are obtained from those in dimensions $3 \leq d \leq 6$ by `gluing' them (see \S 4 in \cite{conway2013sphere}) to a copy of $\tilde{E}_8 = E_9$ ({\it i.e.}\ the affine extension of $E_8$).

Note a misprint in \cite{vinberg1975some}: the $\mathrm{I}_{11,1}$ graph incorrectly has a triple (rather than double) line in its $\mathrm{I}_{3,1}$ component, obscuring this 8-fold periodicity.

\item \textbf{Dimensions $15 \leq d \leq 18$}: 
see Fig.~\ref{fig:OddGraphsN14to17}. The rank continues to grow relative to the dimension : $n=d+2=s+3$ in dimension $d=15,16$; $n=d+3=s+4$ in dimension $d=17$; and $n=d+4=s+5$ in dimension $d=18$.  

Note that in dimensions $d\geq15$, the Coxeter graphs have non-trivial symmetries ({\it e.g.} the horizontal reflection symmetry of the graph for $I_{14,1}$), are no longer trees (as they contain closed loops), and are no longer bicolorable (except when $d=16$) since they contain at least one closed loop of odd length.  For $d\geq18$ they are also no longer {\it planar} graphs.

\begin{figure*}
    \centering
    \subfloat[]{%
        \includegraphics[width=0.95\linewidth]{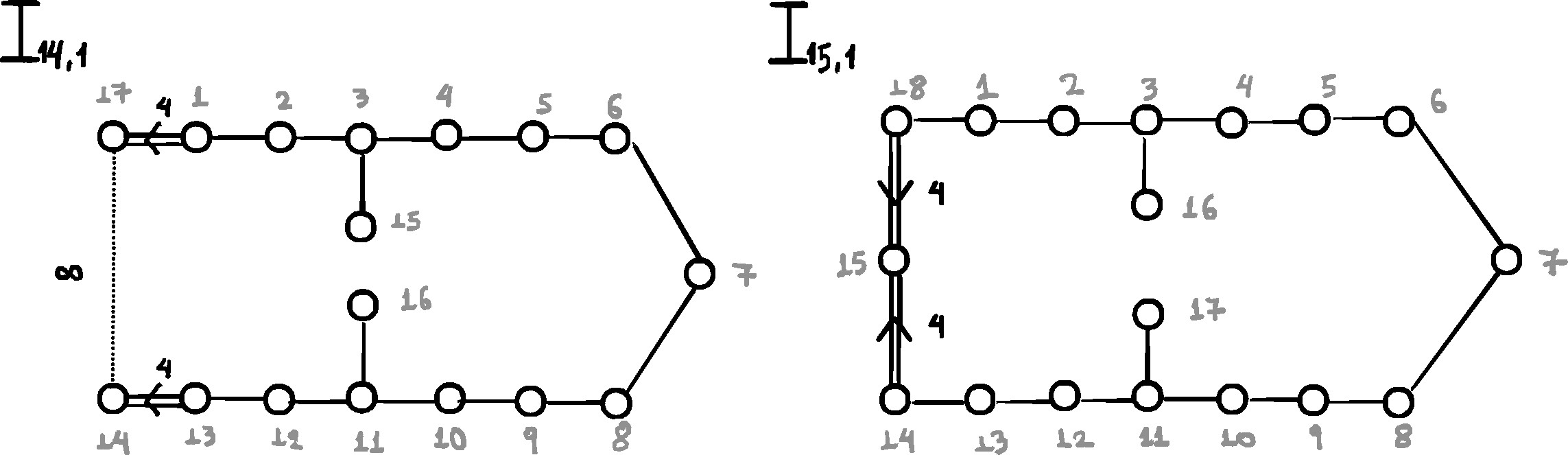}%
        \label{fig:OddGraphsN14and15}%
    }\par
    \subfloat[]{%
        \includegraphics[width=0.80\linewidth]{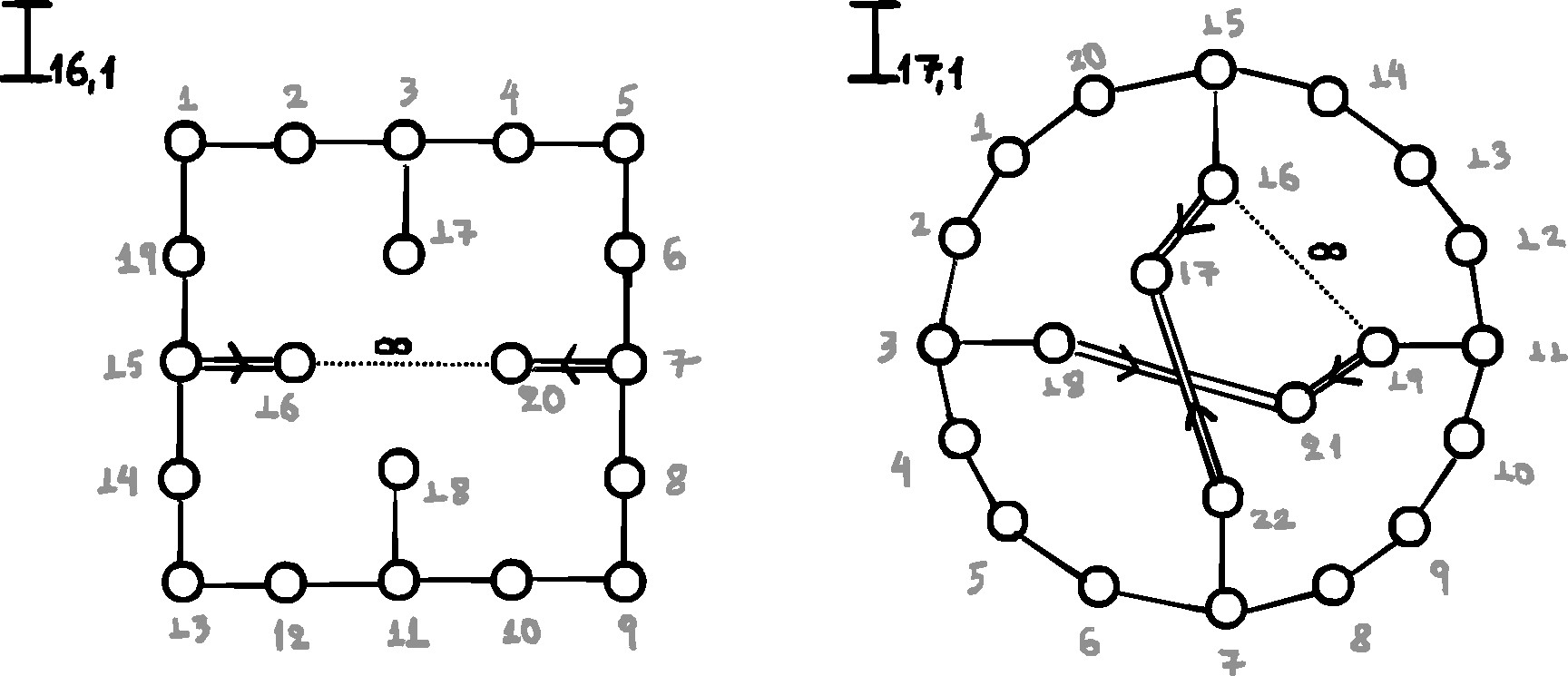}%
        \label{fig:OddGraphsN16and17}%
    }
    \caption{\justifying The Coxeter-Dynkin diagrams for $I_{s,1}$ for $15\leq s+1\leq 18$ (with fundamental roots given in Table \ref{tab:SimpleRootsReflectiveOddLatticesN2to17}).}
    \label{fig:OddGraphsN14to17}
\end{figure*} 

\item \textbf{Dimensions $19 \leq d \leq 20$}:
These last two graphs are the most complex, and were first worked out by Vinberg and Kaplinskaja in\cite{kaplinskaja1978groups}.

\begin{figure*}
    \centering
    \subfloat[\justifying The Coxeter-Dynkin diagram for $I_{18,1}$.  The left graph contains the 22 roots in the skeleton subset $\Sigma_{18}^*$ and is symmetric under the full permutation group $S_4$. The middle graph illustrates one (out of three) vertices of the 1st kind, while the right graph illustrates one (out of 12) vertices of the 2nd kind. In total, $\Sigma_{18}$ has the $22+3+12=37$ fundamental roots given in Tables \ref{tab:SimpleRootsReflectiveOddLatticesN1819},  \ref{tab:AllSimpleRootsReflectiveOddLattices18}.]{%
        \includegraphics[width=0.90\linewidth]{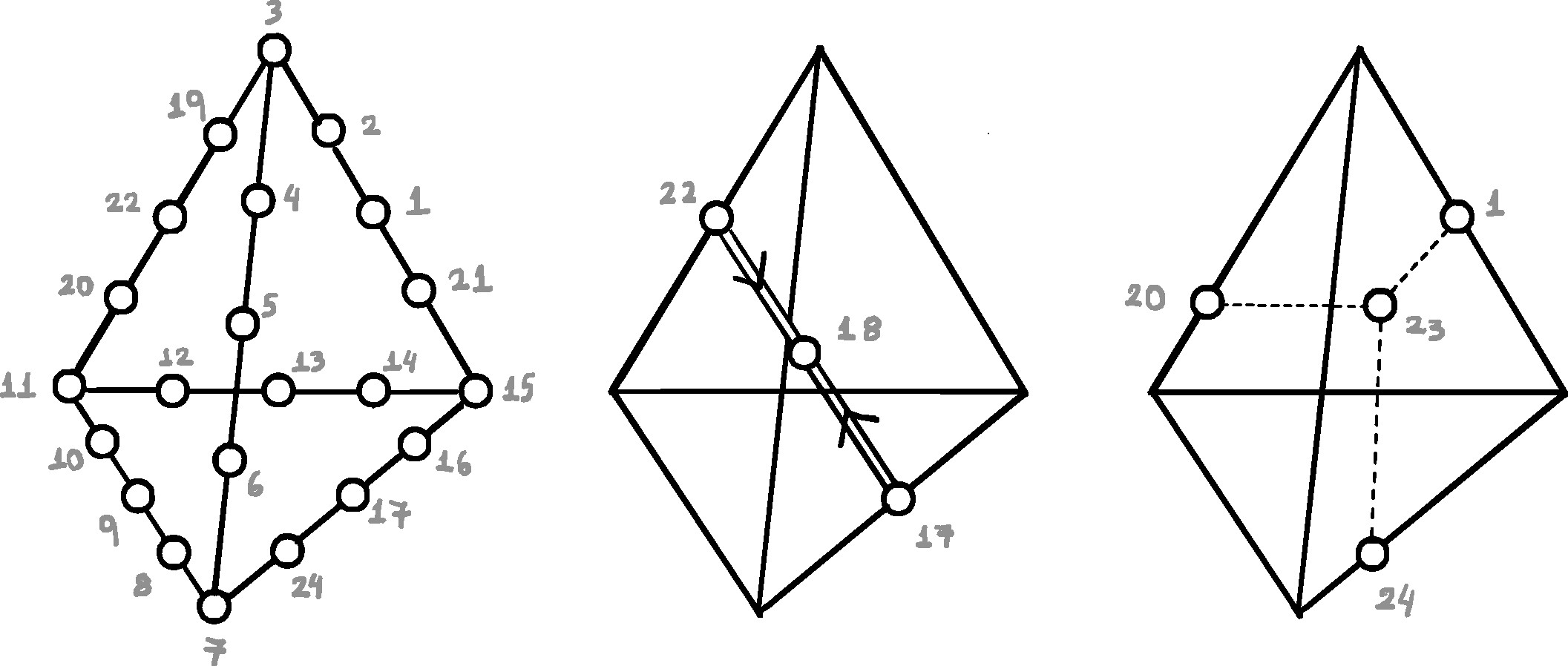}%
        \label{fig:OddGraphsN18}%
    }\par
    \subfloat[\justifying The Coxeter-Dynkin diagram for $I_{19,1}$.  The left graph contains the 25 roots in the skeleton subset $\Sigma_{19}^*$ and is symmetric under the full permutation group $S_5$. The middle graph illustrates one (out of five) vertices of the 1st kind, while the right graph illustrates one (out of 20) vertices of the 2nd kind. In total, $\Sigma_{19}$ has the $25+5+20=50$ fundamental roots given in Tables \ref{tab:SimpleRootsReflectiveOddLatticesN1819}, \ref{tab:AllSimpleRootsReflectiveOddLattices19}. ]{%
        \includegraphics[width=0.95\linewidth]{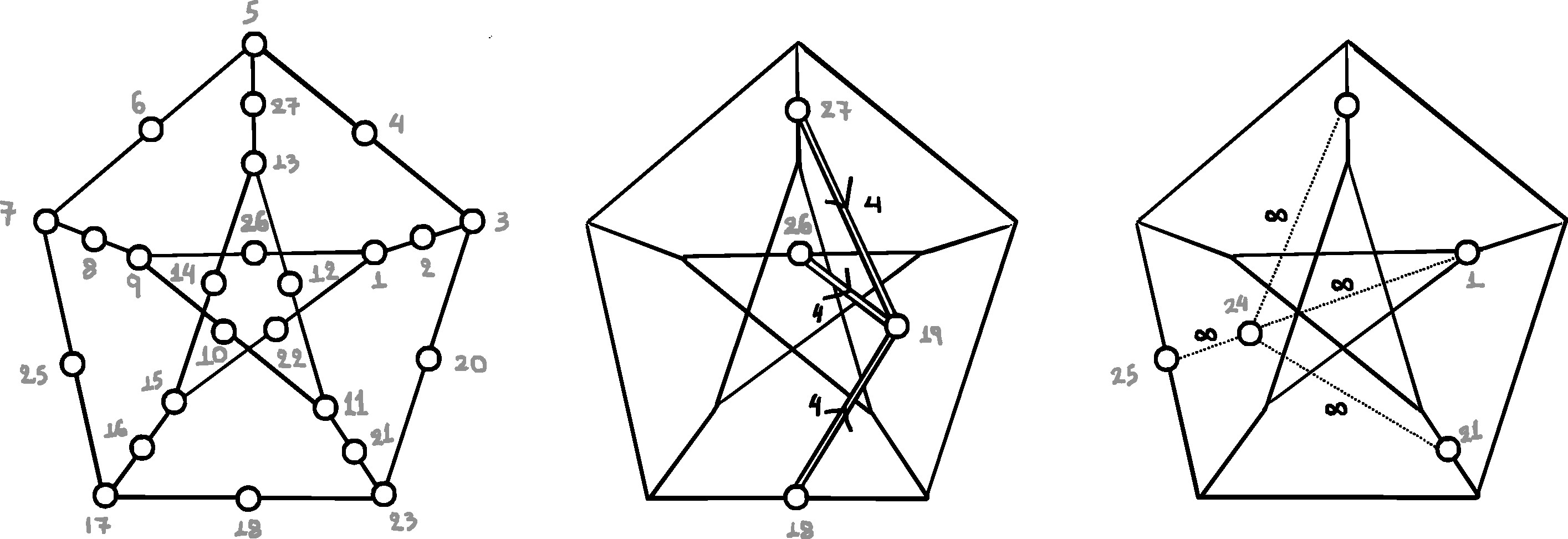}%
        \label{fig:OddGraphsN19}%
    }
    \caption{\justifying Coxeter-Dynkin diagrams for (the skeletons of) $\textrm{I}_{18,1}$ and $\textrm{I}_{19,1}$ (see text for more information).} 
\label{fig:OddGraphsN18to19}
\end{figure*}

To explain the set of fundamental roots in these two cases, first consider the $s+8$ fundamental roots given in Table \ref{tab:SimpleRootsReflectiveOddLatticesN1819} for both $\mathrm{I}_{18,1}$ and $\mathrm{I}_{19,1}$.  These are sets of 26 and 27 roots, which we call $\Sigma_{18}^{(0)}$ and $\Sigma_{19}^{(0)}$, respectively.  Then, define the subsets
\begin{equation*}
\begin{aligned}
& \Sigma_{18}^* = \{1,\dots,24\} - \{ 18,23\} \quad \text{and} \\[1ex]
& \Sigma_{19}^* = \{ 1,\dots,27\} - \{ 19,24\},
\end{aligned}
\end{equation*}
consisting of only the long roots (of norm 2).  These subsets $\Sigma_{18}^*$ and $\Sigma_{19}^*$ form, respectively, a {\it skeleton} Coxeter-Dynkin diagram in the shape of a tetrahedron (left panel of Fig.~\ref{fig:OddGraphsN18}), or a pentagon with an inscribed star (known as the \textit{Petersen graph}), left panel of Fig.~\ref{fig:OddGraphsN19}).  Note that these skeleton diagrams each have non-trivial symmetries described, respectively, by the \textit{full} permutation groups $S_4$ (of order $4!=24$) and $S_5$ (of order $5!=120$).  

Now, we obtain the {\it full} set of fundamental roots ($\Sigma_{s}$) by adding `extra' roots to the skeleton set ($\Sigma_{s}^*$) in such a way that the corresponding Coxeter diagram's ($S_{4}$ or $S_{5}$) symmetry is maintained.  In particular, the extra roots are the roots we previously excluded from $\Sigma_{s}^{(0)}$ to obtain $\Sigma_{s}^{\ast}$, as well as any additional roots needed to maintain the symmetry ($S_{4}$ or $S_{5}$, respectively) of the Coxeter-Dynkin diagram.  In particular, all the extra vertices are either \textbf{vertices of the 1st kind}, related by ($S_{4}$ or $S_{5}$) symmetry to the root $\mathbf{r}_s$, or \textbf{vertices of the 2nd kind}, related by symmetry to the root $\mathbf{r}_{s+5}$.  

For $\Sigma_{18}$, there are three vertices of the first kind (one of which is illustrated in the middle panel of Fig.~\ref{fig:OddGraphsN18}) and twelve vertices of the second kind (one of which is illustrated in the right panel of (Fig.~\ref{fig:OddGraphsN18}), yielding $22+3+12=37$ fundamental roots in total.  Finally, the full set of fundamental roots  $\Sigma_{18}$ consists of the roots ${\bf r}_{1},\ldots,{\bf r}_{26}$ given in Table \ref{tab:SimpleRootsReflectiveOddLatticesN1819}, along with the remaining fundamental roots ${\bf r}_{27},\ldots,{\bf r}_{37}$ given in Table \ref{tab:AllSimpleRootsReflectiveOddLattices18}.

For $\Sigma_{19}$, there are five vertices of the first kind (one of which is illustrated in the middle panel of Fig.~\ref{fig:OddGraphsN19}) and 25 vertices of the second kind (one of which is illustrated in the right panel of Fig.~\ref{fig:OddGraphsN19}), yielding $25+5+20=50$ fundamental roots in total.  Fundamental roots ${\bf r}_{1},\ldots,{\bf r}_{27}$ are given in Table \ref{tab:SimpleRootsReflectiveOddLatticesN1819}, while the remaining fundamental roots ${\bf r}_{28},\ldots,{\bf r}_{50}$ are given in Table \ref{tab:AllSimpleRootsReflectiveOddLattices19}.

\end{itemize}

\begin{table*}[t]
    \centering
        \begin{tabular}{c | c | c | c }
         Root Index $i$ & Root Vector $\mathbf{r}_i$ & $\langle \mathbf{r}_i,\mathbf{r}_i \rangle$ \\
         \vspace{-2ex}& & & \\
         \hline \hline 
         \vspace{-1.75ex}& & & \\
         27 & $9\mathbf{e}_0 + 3(\mathbf{e}_1 + \dots + \mathbf{e}_8) + \mathbf{e}_9 + \dots + \mathbf{e}_{18}$  & 1 \\[1ex]
         \vspace{-1.75ex}& & & \\
         28 & $11\mathbf{e}_0 + 3(\mathbf{e}_1 + \dots + \mathbf{e}_{13}) + \mathbf{e}_{14} + \dots + \mathbf{e}_{18}$  & 1 & \\[1ex]
         \vspace{-1.75ex}& & & \\
         29 & $15\mathbf{e}_0 + 5(\mathbf{e}_1 + \dots + \mathbf{e}_{5}) + 3(\mathbf{e}_{6} + \dots + \mathbf{e}_{16}) + \mathbf{e}_{17} + \mathbf{e}_{18} $  & 1 \\[1ex]
         \vspace{-1.75ex}& & & \\
        30 & $15\mathbf{e}_0 + 5(\mathbf{e}_1 + \dots + \mathbf{e}_{6}) + 3(\mathbf{e}_{7} + \dots + \mathbf{e}_{14}) + \mathbf{e}_{15} + \dots + \mathbf{e}_{18} $  & 1 \\[1ex]
         \vspace{-1.75ex}& & & \\
        31 & $17\mathbf{e}_0 + 5(\mathbf{e}_1 + \dots + \mathbf{e}_{9}) + 3(\mathbf{e}_{10} + \dots + \mathbf{e}_{16}) +\mathbf{e}_{17} + \mathbf{e}_{18} $ & 1 \\[1ex]
         \vspace{-1.75ex}& & & \\
        32 & $19\mathbf{e}_0 + 7(\mathbf{e}_1 + \mathbf{e}_2) + 5(\mathbf{e}_3 + \dots + \mathbf{e}_{10}) + 3(\mathbf{e}_{11} + \dots + \mathbf{e}_{17}) + \mathbf{e}_{18} $ & 1 \\[+1ex] 
        \vspace{-1.75ex}& & & \\
        33 & $21\mathbf{e}_0 + 7(\mathbf{e}_1 + \dots + \mathbf{e}_4) + 5(\mathbf{e}_5 + \dots + \mathbf{e}_{12}) + 3(\mathbf{e}_{13} + \dots + \mathbf{e}_{17}) + \mathbf{e}_{18}$ & 1 \\[+1ex]
         \vspace{-1.75ex}& & & \\
        34 & \,\, $21\mathbf{e}_0 + 7(\mathbf{e}_1 + \dots + \mathbf{e}_5) + 5(\mathbf{e}_6 + \dots + \mathbf{e}_{10}) + 3(\mathbf{e}_{13} + \dots + \mathbf{e}_{18}) \,\, $ & 1 \\[+1ex]
         \vspace{-1.75ex}& & & \\
       35 & \,\, $23\mathbf{e}_0 + 9 \mathbf{e}_1 +  7(\mathbf{e}_2 + \dots + \mathbf{e}_6) + 5(\mathbf{e}_7 + \dots + \mathbf{e}_{12}) + 3(\mathbf{e}_{13} + \dots + \mathbf{e}_{18}) \,\, $ & 1 \\[+1ex]
        \vspace{-1.75ex}& & & \\
        36 & \,\, $25\mathbf{e}_0 + 9 (\mathbf{e}_1 + \mathbf{e}_2) +   7(\mathbf{e}_3 + \dots + \mathbf{e}_8) + 5(\mathbf{e}_9 + \dots + \mathbf{e}_{13}) + 3(\mathbf{e}_{14} + \dots + \mathbf{e}_{18}) \,\, $ & 1 \\[+1ex]
         \vspace{-1.75ex}& & & \\
         37 & \, $27\mathbf{e}_0 + 9 (\mathbf{e}_1 + \dots + \mathbf{e}_4) +   7(\mathbf{e}_5 + \dots + \mathbf{e}_9) + 5(\mathbf{e}_{10} + \dots + \mathbf{e}_{14}) + 3(\mathbf{e}_{15} + \dots + \mathbf{e}_{18})$ \, & 1 
\end{tabular}
    \caption{\justifying The remaining fundamental roots for $\mathrm{I}_{18,1}$.} 
\label{tab:AllSimpleRootsReflectiveOddLattices18}
\end{table*}

\begin{table*}[t]
    \centering
        \begin{tabular}{c | c | c | c }
         $i$ & Root Vector $\mathbf{r}_i$ & $\langle \mathbf{r}_i,\mathbf{r}_i \rangle$ \\
        \vspace{-2ex}& & & \\
        \hline \hline 
        \vspace{-1.75ex}& & & \\
        28 & $6\mathbf{e}_0 + 2(\mathbf{e}_1 + \dots + \mathbf{e}_6) + (\mathbf{e}_7 + \dots + \mathbf{e}_{19})$  & 1 \\[1ex]
         \vspace{-1.75ex}& & & \\
         29 & $9\mathbf{e}_0 + 3(\mathbf{e}_1 + \dots + \mathbf{e}_8) + (\mathbf{e}_9 + \dots + \mathbf{e}_{19})$  & 2 \\[1ex]
         \vspace{-1.75ex}& & & \\
        30 & $11\mathbf{e}_0 + 3(\mathbf{e}_1 + \dots + \mathbf{e}_{13}) + (\mathbf{e}_{14} + \dots + \mathbf{e}_{19})$  & 2 \\[1ex]
        \vspace{-1.75ex}& & & \\
        31 & $8\mathbf{e}_0 + 3(\mathbf{e}_1 + \mathbf{e}_{2}) + 2(\mathbf{e}_{3} + \dots + \mathbf{e}_{12}) + (\mathbf{e}_{13} + \dots + \mathbf{e}_{19} $  & 1 \\[1ex]
         \vspace{-1.75ex}& & & \\
        32 & $10\mathbf{e}_0 + 3(\mathbf{e}_1 + \dots +\mathbf{e}_{8}) + 2(\mathbf{e}_{9} + \dots + \mathbf{e}_{14}) + (\mathbf{e}_{15} + \dots + \mathbf{e}_{19} $  & 1 \\[1ex]
         \vspace{-1.75ex}& & & \\
        33 & $15\mathbf{e}_0 + 5(\mathbf{e}_1 + \dots + \mathbf{e}_{5}) + 3(\mathbf{e}_{6} + \dots + \mathbf{e}_{16}) + (\mathbf{e}_{17} + \mathbf{e_{18}} + \mathbf{e}_{19}) $  & 2 \\[1ex]
         \vspace{-1.75ex}& & & \\
        34 & $15\mathbf{e}_0 + 5(\mathbf{e}_1 + \dots + \mathbf{e}_{6}) + 3(\mathbf{e}_{7} + \dots + \mathbf{e}_{14}) + (\mathbf{e}_{15} + \dots + \mathbf{e}_{19}) $  & 2 \\[1ex]
         \vspace{-1.75ex}& & & \\
        35 & $12\mathbf{e}_0 + 4(\mathbf{e}_1 + \dots + \mathbf{e}_{4}) + 3(\mathbf{e}_{5} + \dots + \mathbf{e}_{10}) + 2(\mathbf{e}_{11} + \dots + \mathbf{e}_{16}) + (\mathbf{e}_{17}+\mathbf{e}_{18}+\mathbf{e}_{19}) $  & 1 \\[1ex]
         \vspace{-1.75ex}& & & \\ 
        36 & $17\mathbf{e}_0 + 5(\mathbf{e}_1 + \dots + \mathbf{e}_{9}) + 3(\mathbf{e}_{10} + \dots + \mathbf{e}_{16}) + (\mathbf{e}_{17} + \mathbf{e}_{18} +\mathbf{e}_{19}) $  & 2 \\[1ex]
         \vspace{-1.75ex}& & & \\
        37 & $19\mathbf{e}_0 + 7(\mathbf{e}_1 + \mathbf{e}_2) + 5(\mathbf{e}_3 + \dots + \mathbf{e}_{10}) + 3(\mathbf{e}_{11} + \dots + \mathbf{e}_{17}) + (\mathbf{e}_{18} + \mathbf{e}_{19}) $ & 2 \\[+1ex]
        \vspace{-1.75ex}& & & \\
        38 & $21\mathbf{e}_0 + 7(\mathbf{e}_1 + \dots + \mathbf{e}_4) + 5(\mathbf{e}_5 + \dots + \mathbf{e}_{11}) + 3(\mathbf{e}_{12} + \dots + \mathbf{e}_{19})$ & 2 \\[+1ex]
        \vspace{-1.75ex}& & & \\
        39 & $21\mathbf{e}_0 + 7(\mathbf{e}_1 + \dots + \mathbf{e}_4) + 5(\mathbf{e}_5 + \dots + \mathbf{e}_{12}) + 3(\mathbf{e}_{13} + \dots + \mathbf{e}_{17}) + (\mathbf{e}_{18} + \mathbf{e}_{19})$ & 2 \\[+1ex]
        \vspace{-1.75ex}& & & \\
        40 & $21\mathbf{e}_0 + 7(\mathbf{e}_1 + \dots + \mathbf{e}_5) + 5(\mathbf{e}_6 + \dots + \mathbf{e}_{10}) + 3(\mathbf{e}_{11} + \dots + \mathbf{e}_{18}) + \mathbf{e}_{19}$ & 2 \\[+1ex]
        \vspace{-1.75ex}& & & \\
        41 & \,\, $23\mathbf{e}_0 + 7(\mathbf{e}_1 + \dots + \mathbf{e}_7) + 5(\mathbf{e}_8 + \dots + \mathbf{e}_{12}) + 3(\mathbf{e}_{13} + \dots + \mathbf{e}_{19}) \,\, $ & 2 \\[+1ex]
        \vspace{-1.75ex}& & & \\
        42 & \,\, $23\mathbf{e}_0 + 9 \mathbf{e}_1 +  7(\mathbf{e}_2 + \dots + \mathbf{e}_6) + 5(\mathbf{e}_7 + \dots + \mathbf{e}_{12}) + 3(\mathbf{e}_{13} + \dots + \mathbf{e}_{18}) + \mathbf{e}_{19} \,\, $ & 2 \\[+1ex]
        \vspace{-1.75ex}& & & \\
        43 & \,\, $25\mathbf{e}_0 + 9 (\mathbf{e}_1 + \mathbf{e}_2) +   7(\mathbf{e}_3 + \dots + \mathbf{e}_7) + 5(\mathbf{e}_8 + \dots + \mathbf{e}_{14}) + 3(\mathbf{e}_{15} + \dots + \mathbf{e}_{19}) \,\, $ & 2 \\[+1ex]
        \vspace{-1.75ex}& & & \\
        44 & \,\, $25\mathbf{e}_0 + 9 (\mathbf{e}_1 + \mathbf{e}_2) +   7(\mathbf{e}_3 + \dots + \mathbf{e}_8) + 5(\mathbf{e}_9 + \dots + \mathbf{e}_{13}) + 3(\mathbf{e}_{14} + \dots + \mathbf{e}_{18}) + \mathbf{e}_{19} \,\, $ & 2 \\[+1ex]
        \vspace{-1.75ex}& & & \\
        45 & \,\, $27\mathbf{e}_0 + 9 (\mathbf{e}_1 + \dots + \mathbf{e}_3) +  7(\mathbf{e}_4 + \dots + \mathbf{e}_{10}) + 5(\mathbf{e}_{11} + \dots + \mathbf{e}_{14}) + 3(\mathbf{e}_{15} + \dots + \mathbf{e}_{19}) \,\, $ & 2 
        \\[+1ex]
        \vspace{-1.75ex}& & & \\
        46 & \,\, $27\mathbf{e}_0 + 9 (\mathbf{e}_1 + \dots + \mathbf{e}_4) +  7(\mathbf{e}_5 + \dots + \mathbf{e}_{8}) + 5(\mathbf{e}_{9} + \dots + \mathbf{e}_{15}) + 3(\mathbf{e}_{16} + \dots + \mathbf{e}_{19}) \,\, $ & 2
        \\[+1ex]
        \vspace{-1.75ex}& & & \\
        47 & \,\, $27\mathbf{e}_0 + 9 (\mathbf{e}_1 + \dots + \mathbf{e}_4) +  7(\mathbf{e}_4 + \dots + \mathbf{e}_{9}) + 5(\mathbf{e}_{10} + \dots + \mathbf{e}_{14}) + 3(\mathbf{e}_{15} + \dots + \mathbf{e}_{18}) + \mathbf{e}_{19} \,\, $ & 2 \\[+1ex]
        \vspace{-1.75ex}& & & \\
        48 & \,\, $29\mathbf{e}_0 + 9 (\mathbf{e}_1 + \dots + \mathbf{e}_6) +  7(\mathbf{e}_{7} + \dots + \mathbf{e}_{10}) + 5(\mathbf{e}_{10} + \dots + \mathbf{e}_{15}) + 3(\mathbf{e}_{16} + \dots + \mathbf{e}_{19}) \,\, $ & 2 \\[+1ex]
        \vspace{-1.75ex}& & & \\
        49 & \,\, $31\mathbf{e}_0 +11 (\mathbf{e}_1 + \mathbf{e}_2) + 9 (\mathbf{e}_3 + \dots + \mathbf{e}_6) +  7(\mathbf{e}_7 + \dots + \mathbf{e}_{10}) + 5(\mathbf{e}_{11} + \dots + \mathbf{e}_{15}) + 3(\mathbf{e}_{16} + \dots + \mathbf{e}_{19}) \,\, $ & 2
        \\[+1ex]
        \vspace{-1.75ex}& & & \\
        50 & \,\, $33\mathbf{e}_0 + 11 (\mathbf{e}_1 + \mathbf{e}_2 + \mathbf{e}_3) + 9 (\mathbf{e}_4 + \dots + \mathbf{e}_8) +  7(\mathbf{e}_9 + \dots + \mathbf{e}_{13}) + 5(\mathbf{e}_{14} + \dots + \mathbf{e}_{17}) + 3(\mathbf{e}_{16} + \dots + \mathbf{e}_{19}) \,\, $ & 2 
\end{tabular}
    \caption{\justifying The remaining fundamental roots for $\mathrm{I}_{19,1}$} 
\label{tab:AllSimpleRootsReflectiveOddLattices19}
\end{table*}

\section{Scale Factors as Algebraic Numbers}
\label{app:inflation}

In this appendix, we explain some points about how to determine a quasicrystal's scale factors algebraically from an associated number field; and we determine this number field for the self-dual Lorentzian lattices up to dimension $d=10$.

Consider an sC\&P quasicrystal obtained by starting from a lattice $\Lambda$, constructing its Coxeter element $C$, and using the non-crystallographic subgroup $G\in {\rm Aut}(\Lambda)$ generated by $C$ to decompose $V$ as $V=V_{\rm ph}\oplus V_{\rm in}$.  

The characteristic polynomial of $C$ -- called the {\bf Coxeter polynomial} $\chi_{C}(\lambda)$ -- is a polynomial with integer coefficients, which we take to be irreducible ({\it i.e.}\ not factorizable over $\mathbb{Q}$), which corresponds to the fact that $G$ is non-crystallographic.  Let $K$ denote the {\bf splitting field} of $\chi_{C}(\lambda)$ -- {\it i.e.}\ the smallest field extension of $\mathbb{Q}$ over which $\chi_{C}(\lambda)$ ``splits'' ({\it i.e.} decomposes into linear factors).  And let $K'$ denote the maximal real subfield of $K$ ({\it i.e.}\ the largest subfield of $K$ contained in $\mathbb{R}$).  

The (fundamental) units of $K'$ will be the quasicrystal's (fundamental) inflation factors.

For example, consider the Ammann-Beenker tiling (see Sec. \ref{subsec:AmmannBeenkerTilingsExample}): the Coxeter polynomial $\chi_C(\lambda)\in \mathbb{Z}[\lambda]$ reads
\begin{equation*}
\begin{aligned}
\chi_C(x) &= \lambda^4+1 = (\lambda - \zeta_8)(\lambda-\zeta_8^*)(\lambda-\zeta_8^3)(\lambda-\zeta_8^{3*})\\
& = (\lambda^2 - 2\mathrm{Re}(\zeta_8) \lambda + 1)(\lambda^2 - 2\mathrm{Re}(\zeta_8^3) \lambda + 1)
\end{aligned}.
\end{equation*}
where $\zeta_{8}={\rm exp}(2\pi i/8)$.  We see that $\chi_{C}(\lambda)$ cannot be factored over $\mathbb{Q}$, can be fully factored over the splitting field $K=\mathbb{Q}(\zeta_{8})$ (a degree four extension of $\mathbb{Q}$), and can be partly factored over its maximal real subfield $K'=\mathbb{Q}(\zeta_{8}+\zeta_{8}^{\ast})=\mathbb{Q}(\sqrt{2})$.  The inflation factors of the corresponding quasicrystal will be the units of $K'$; for the Ammann-Beenker tilings, the fundamental inflation factor -- {\it i.e.}\ the fundamental unit of $K'=\mathbb{Z}(\sqrt{2})$ -- is the silver ratio $\epsilon=(1+\sqrt{2})$, and the general unit may be expressed as $\pm(1+\sqrt{2})^{m}$ ($m\in\mathbb{Z}$).

In fact, for all the most well-known and widely-studied Euclidean quasicrystals, the associated real algebraic number fields $K'$ are also quadratic fields $\mathbb{Q}(\sqrt{D})$ -- and for such fields, the units always have this simple structure (that, up to a sign, they are simply all integer powers of a single fundamental unit $\lambda$).

Once we turn to spacetime quasicrystals, we will (usually) no longer be dealing with quadratic field extensions, but with more complicated field extensions (of higher degree, and with more than one fundamental unit); but otherwise, the above considerations still apply.

For illustration, consider the spacetime quasicrystal obtained via sC\&P from the odd self-dual lattice  $\mathrm{I}_{3,1}$ (see Section \ref{subsec:2DLorentzianQuasiCrystal}).  The Coxeter polynomial $\chi_{C}(\lambda)$ is
\begin{eqnarray*}
\chi_C(\lambda)&=&\lambda^4 - \lambda^3 - 2\lambda^2 - \lambda + 1 \\
&=&(\lambda-\lambda_{0}^+)(\lambda-\lambda_{0}^-)(\lambda-\lambda_1^+)(\lambda-\lambda_1^-)
\end{eqnarray*}
where two of these roots ($\lambda_{0}^{\pm}$) are real, two of them ($\lambda_{1}^{\pm}$) are complex, and the larger real root $\lambda_{0}^{+}$ is a Salem number.  All four roots lie in the splitting field $K=\mathbb{Q}[\lambda_{1}^{+}]$ ({\it i.e.}\ the smallest field containing both $\mathbb{Q}$ and one of the complex roots of $\chi_{C}$ -- {\it e.g.} $\lambda_{1}^{+}$).  However, finding the exact form of $\lambda_{1}^{+}$ would require solving a relatively high-order polynomial equation (in this simple example it is just a quartic equation, which is still possible to solve, but higher-dimensional examples would involve higher-order equations that are impossible to solve in closed form).  

However, if we are just interested in our quasicrystal's scale factors, then solving this higher-order equation is unnecessary since, as explained above, we don't need the full splitting field $K$, but rather its maximal real subfield $K'$.  The subfield $K'$ is not generated by a single root (like $\lambda_{1}^{+}$), but rather by the sum of a pair of such roots: $K'=\mathbb{Q}[\mu_{j}]$ where $\mu_{j}\equiv\lambda_{j}^{+}+\lambda_{j}^{-}$; and solving for $\mu_{j}=\lambda_{j}^{+}+\lambda_{j}^{-}$ is easier (requires solving a lower-order polynomial) than solving for either $\lambda_{j}^{+}$ or $\lambda_{j}^{-}$ separately.  

To see this, note that $\chi_{C}(\lambda)$ is (i) monic, (ii) irreducible over $\mathbb{Q}$, and (iii) palindromic ({\it i.e.}\ its coefficients $a_0,\ldots,a_4$ satisfy $a_i=a_{n-i}$), so we can write it as
\begin{equation}
  \chi_{C}(\lambda)=\lambda^{2}Q(\mu)\qquad\mu\equiv\lambda+\frac{1}{\lambda}
\end{equation}
where
\begin{equation}
    Q(\mu)=\mu^{2}-\mu-4
\end{equation}
is a lower order polynomial with roots
\begin{eqnarray*}
  \mu_{0}&=&\lambda_{0}^{+}+\frac{1}{\lambda_{0}^{+}}
  =\lambda_{0}^{+}+\lambda_{0}^{-}=\frac{1+\sqrt{17}}{2}>2, \\
  \mu_{1}&=&\lambda_{1}^{+}+\frac{1}{\lambda_{1}^{+}}
  =\lambda_{1}^{+}+\lambda_{1}^{-}=\frac{1-\sqrt{17}}{2}<2.
\end{eqnarray*}
So this quasicrystal's fundamental scale factor is the fundamental unit of $\mathbb{Q}[\sqrt{17}]$ (namely, $4+\sqrt{17}$).   

We can analogously identify the roots $\mu_{j}$ and field $K'=\mathbb{Q}[\mu_{0}]$ for $I_{s,1}$ up to dimensions $s+1=10$, as summarized in Table \ref{tab:FieldExtensionsHyperbolicCases}.  In all cases, $\mu_{0}=\lambda_{0}^{+}+\lambda_{0}^{-}>2$, while the $\mu_{j}=\lambda_{j}^{+}+\lambda_{j}^{-}<2$ for $j\geq 1$.  But as we climb up the ladder of dimensions, the relevant field extensions $K$ and $K'$ become of higher and higher degree, with more and more fundamental units.  We have only shown the units for the first three cases (the simplest, quadratic cases), but the units in the remaining cases, may be computed from the given information via standard methods of algebraic number theory.

\begin{table*}[]
    \centering
    \begin{tabular}{c || c | c | c | c}
         $\Lambda$ & $\chi_C(\lambda)$ & $\mu_0 \equiv \lambda_{0}^{+} + \lambda_{0}^{-} > 2$ & $ \mathbb{Q}(\mu_0)$ & Fund. Units \\ \hline \hline 
         $\textrm{I}_{2,1}$ & $- (\lambda+1)(\lambda^2 - 4\lambda + 1)$ & $2+\sqrt{3} \approx 3.73205$ & $\mathbb{Q}(\sqrt{3})$ & $2\pm\sqrt
         3$  \\[+1ex]
         $\textrm{I}_{3,1}$ & $\lambda^2 (\mu^2 - \mu - 4)$ & $\frac{1}{2}(1+\sqrt{17}) \approx 2.56155$ & $\mathbb{Q}(\sqrt{17})$ & $4\pm\sqrt{17}$ \\[+1ex]
         $\textrm{I}_{4,1}$ & $-\lambda^2(\lambda+1)(\mu^2 - \mu - 3)$ & $\frac{1}{2}(1+\sqrt{13}) \approx 2.30277$ & $\mathbb{Q}(\sqrt{13})$ & $\frac{1}{2}(3 \pm \sqrt{13})$\\[+1ex]
         $\textrm{I}_{5,1}$ & $\lambda^3 (\mu^3 - 4\mu - 2)$ & $ \frac{1}{3^{1/3}}\big( \beta_6^{1/3} + \beta_6^{* 1/3} \big) \approx 2.21432 \qquad \big(\beta_6 = \frac{1}{3}(9+i \sqrt{111})\big) $ & $\cdot$ & $\cdot$ \\[1ex]
         $\textrm{I}_{6,1}$ & $- \lambda^3(\lambda+1)(\mu^3-\mu^2-3\mu+1) $ & $\frac{1}{3}\big( 1 + \beta_7^{1/3} + \beta_7^{* 1/3}\big) \approx 2.17009 \qquad \big( \beta_7 = 1 + 3i\sqrt{111}\big) $ & $\cdot$ & $\cdot$ \\[1ex]
         $\textrm{I}_{7,1}$ & $\lambda^4(\mu^4 - 5\mu^2 - \mu + 4)$ & [see below\footnote{$$\mu_0 = \frac{1}{4}\bigg( 2\sqrt{\tilde{\beta}_8^+} + \sqrt{2} \sqrt{\tilde{\beta}_8^+ + 3 \tilde{\beta}_8^- + \frac{4}{\sqrt{\tilde{\beta}_8^+}}}  \bigg)\approx 2.14386, \qquad \tilde{\beta}_8^\pm = \frac{1}{3}\big(10 \pm (\beta_8^{1/3} + \beta_8^{* 1/3} )\big)$$}] $\qquad \big( \beta_8 = \frac{1}{2}(1217+3i\sqrt{8331})\big)$ & $\cdot$ & $\cdot$ \\[1ex]
         $\textrm{I}_{8,1}$ & $ -\lambda^4(\lambda+1)(\mu^4 - \mu^3 - 4\mu^2 + 2\mu + 3)$ & [see below\footnote{$$ \mu_0 = \frac{1}{4}\bigg(1 + \sqrt{\tilde{\beta}_9^+} + \frac{1}{\sqrt{2}} \sqrt{\tilde{\beta}_9^+ + 3\tilde{\beta}_9^- + \frac{4}{\sqrt{\tilde{\beta}_9^+}}}\bigg) \approx 2.12676, \qquad \tilde{\beta}_9^\pm = \frac{1}{3}\big( 35 \pm 4 (\beta_9^{1/3} + \beta_9^{* 1/3})\big)$$}] $\qquad \big(\beta_9 = \frac{1}{2}(853+3i\sqrt{5871})\big)$ & $\cdot$ & $\cdot$ \\[1ex]
         $\mathrm{I}_{9,1}$ & $\lambda^5(\mu^2-2)(\mu^3-4\mu-1)$ & $ \frac{1}{3}\big( \beta_{10}^{1/3} + \beta_{10}^{* 1/3}\big) \approx 2.11491, \qquad \big( \beta_{10} = \frac{3}{2}(9+i\sqrt{687})\big)$ & $\cdot$ & $\cdot$ \\[1ex]
         $\mathrm{II}_{9,1}$ & $\lambda^5(\mu^5 + \mu^4 - 5\mu^3 - 5\mu^2 + 4\mu + 3) $ & [no closed form for the quintic] = 2.02642 & $\cdot$ & $\cdot$
    \end{tabular}
    \caption{\justifying The real field extension $K'=\mathbb{Q}(\mu_{0})$ for $I_{s,1}$, up to dimension $s+1=10$.  We have only given the fundamental units for the simplest (first three) cases; the fundamental units for the remaining cases may be obtained by standard computational methods in algebraic number theory.}
    \label{tab:FieldExtensionsHyperbolicCases}
\end{table*}


\bibliography{apssamp}

@PREAMBLE{
 "\providecommand{\noopsort}[1]{}" 
 # "\providecommand{\singleletter}[1]{#1}%" 
}

@article{a1976valeurs,
  title={Sur les valeurs propres de la transformation de Coxeter},
  author={A'Campo, Norbert},
  journal={Inventiones mathematicae},
  volume={33},
  number={1},
  pages={61--67},
  year={1976},
  publisher={Springer}
}

@article{ammann1992aperiodic,
  title={Aperiodic tiles},
  author={Ammann, Robert and Gr{\"u}nbaum, Branko and Shephard, Geoffrey C},
  journal={Discrete \& Computational Geometry},
  volume={8},
  number={1},
  pages={1--25},
  year={1992},
  publisher={Springer},
doi={10.1007/BF02293033}
}

@article{Aoki:1998bq,
    author = "Aoki, Hajime and Iso, Satoshi and Kawai, Hikaru and Kitazawa, Yoshihisa and Tsuchiya, Asato and Tada, Tsukasa",
    editor = "Iso, S. and Kawai, H. and Natsuume, M.",
    title = "{IIB matrix model}",
    eprint = "hep-th/9908038",
    archivePrefix = "arXiv",
    reportNumber = "KEK-TH-635",
    doi = "10.1143/PTPS.134.47",
    journal = "Prog. Theor. Phys. Suppl.",
    volume = "134",
    pages = "47--83",
    year = "1999"
}

@article{arkani1998hierarchy,
  title={The {H}ierarchy {P}roblem and {N}ew {D}imensions at a {M}illimeter},
  author={Arkani--Hamed, Nima and Dimopoulos, Savas and Dvali, Gia},
  journal={Physics Letters B},
  volume={429},
  number={3-4},
  pages={263--272},
  year={1998},
  publisher={Elsevier}
}

@article{baake1991quasiperiodic,
  title={Quasiperiodic tilings with tenfold symmetry and equivalence with respect to local derivability},
  author={Baake, Michael and Scholottmann, M and Jarvis, Peter D},
  journal={Journal of Physics A: Mathematical and General},
  volume={24},
  number={19},
  pages={4637},
  year={1991},
  publisher={IOP Publishing}
}

@article{baake1997torus,
  title={The torus parametrization of quasiperiodic LI-classes},
  author={Baake, Michael and Hermisson, Joachim and Pleasants, Peter AB},
  journal={Journal of Physics A: Mathematical and General},
  volume={30},
  number={9},
  pages={3029},
  year={1997},
  publisher={IOP Publishing}
}

@misc{baake1999guidemathematicalquasicrystals,
      title={A guide to mathematical quasicrystals}, 
      author={Michael Baake},
      year={1999},
      eprint={math-ph/9901014},
      archivePrefix={arXiv},
      primaryClass={math-ph},
      url={https://arxiv.org/abs/math-ph/9901014}, 
}

@book{baake2013aperiodic,
  title={Aperiodic order},
  author={Baake, Michael and Grimm, Uwe},
  volume={1},
  year={2013},
  publisher={Cambridge University Press}
}

@article{Baykara2024Quasicrystalline,
    author = "Baykara, Zihni Kaan and Tarazi, Houri-Christina and Vafa, Cumrun",
    title = "{Quasicrystalline string landscape}",
    eprint = "2406.00129",
    archivePrefix = "arXiv",
    primaryClass = "hep-th",
    doi = "10.1103/PhysRevD.111.086025",
    journal = "Phys. Rev. D",
    volume = "111",
    number = "8",
    pages = "086025",
    year = "2025"
}

@article{beenker1982algebraic,
  title={Algebraic theory of non-periodic tilings of the plane by two simple building blocks: a square and a rhombus, EUT report},
  author={Beenker, F. P. M.},
  journal={EUT report 82-WSK-04, Dept. of Mathematics and Computing Science, Eindhoven University of Technology},
  year={1982},
  url={https://api.semanticscholar.org/CorpusID:55737502}
}

@article{bindi2009natural,
  title={Natural quasicrystals},
  author={Bindi, Luca and Steinhardt, Paul J and Yao, Nan and Lu, Peter J},
  journal={Science},
  volume={324},
  number={5932},
  pages={1306--1309},
  year={2009},
  publisher={American Association for the Advancement of Science},
doi={10.1126/science.1170827}
}

@article{bindi2023electrical,
  title={Electrical discharge triggers quasicrystal formation in an eolian dune},
  author={Bindi, Luca and Pasek, Matthew A and Ma, Chi and Hu, Jinping and Cheng, Guangming and Yao, Nan and Asimow, Paul D and Steinhardt, Paul J},
  journal={Proceedings of the National Academy of Sciences},
  volume={120},
  number={1},
  pages={e2215484119},
  year={2023},
  publisher={National Acad Sciences},
doi={10.1073/pnas.2215484119}
}

@article{bindi2021accidental,
  title={Accidental synthesis of a previously unknown quasicrystal in the first atomic bomb test},
  author={Bindi, Luca and Kolb, William and Eby, G Nelson and Asimow, Paul D and Wallace, Terry C and Steinhardt, Paul J},
  journal={Proceedings of the National Academy of Sciences},
  volume={118},
  number={22},
  pages={e2101350118},
  year={2021},
  publisher={National Acad Sciences},
doi={10.1073/pnas.2101350118}
}

@article{Bombelli:1987aa,
    author = "Bombelli, Luca and Lee, Joohan and Meyer, David and Sorkin, Rafael",
    title = "{Space-Time as a Causal Set}",
    reportNumber = "IASSNS-HEP-87-23A",
    doi = "10.1103/PhysRevLett.59.521",
    journal = "Phys. Rev. Lett.",
    volume = "59",
    pages = "521--524",
    year = "1987"
}

@article{Boyle:2018uiv,
    author = "Boyle, Latham and Dickens, Madeline and Flicker, Felix",
    title = "{Conformal Quasicrystals and Holography}",
    eprint = "1805.02665",
    archivePrefix = "arXiv",
    primaryClass = "hep-th",
    doi = "10.1103/PhysRevX.10.011009",
    journal = "Phys. Rev. X",
    volume = "10",
    number = "1",
    pages = "011009",
    year = "2020"
}

@article{Boyle:2024qzn,
    author = "Boyle, Latham and Kulp, Justin",
    title = "{Holographic foliations: Self-similar quasicrystals from hyperbolic honeycombs}",
    eprint = "2408.15316",
    archivePrefix = "arXiv",
    primaryClass = "hep-th",
    doi = "10.1103/PhysRevD.111.046001",
    journal = "Phys. Rev. D",
    volume = "111",
    number = "4",
    pages = "046001",
    year = "2025"
}

@article{coxeterSimplices,
    author = {Coxeter, H. S. M.},
    title = "{Groups Whose Fundamental Regions Are Simplexes}",
    journal = {Journal of the London Mathematical Society},
    volume = {s1-6},
    number = {2},
    pages = {132-136},
    year = {1931},
    month = {04},
    issn = {0024-6107},
    doi = {10.1112/jlms/s1-6.2.132},
    url = {https://doi.org/10.1112/jlms/s1-6.2.132},
    eprint = {https://academic.oup.com/jlms/article-pdf/s1-6/2/132/2379423/s1-6-2-132.pdf},
}

@article{coxeterDiscreteGroups,
 ISSN = {0003486X},
 URL = {http://www.jstor.org/stable/1968753},
 author = {H. S. M. Coxeter},
 journal = {Annals of Mathematics},
 number = {3},
 pages = {588--621},
 publisher = {Annals of Mathematics},
 title = {Discrete Groups Generated by Reflections},
 volume = {35},
 year = {1934}
}

@article{coxeterCompleteEnumeration,
author = {Coxeter, H. S. M.},
title = {The Complete Enumeration of Finite Groups of the Form Ri2=(RiRj)kij=1},
journal = {Journal of the London Mathematical Society},
volume = {s1-10},
number = {1},
pages = {21-25},
doi = {https://doi.org/10.1112/jlms/s1-10.37.21},
url = {https://londmathsoc.onlinelibrary.wiley.com/doi/abs/10.1112/jlms/s1-10.37.21},
eprint = {https://londmathsoc.onlinelibrary.wiley.com/doi/pdf/10.1112/jlms/s1-10.37.21},
year = {1935}
}

@article{cassels1967algebraic,
  title={Algebraic number theory: Proceedings of an instructional conference organized by the london mathematical society (a nato advanced study institute) with the support of the international mathematical union},
  author={Cassels, John William Scott and Fr{\"o}hlich, Albrecht and others},
  journal={(No Title)},
  year={1967}
}

@article{conway1983automorphism,
  title={The automorphism group of the 26-dimensional even unimodular Lorentzian lattice},
  author={Conway, John H},
  journal={Journal of Algebra},
  volume={80},
  number={1},
  pages={159--163},
  year={1983},
  publisher={Elsevier}
}

@book{conway2013sphere,
  title={Sphere {P}ackings, {L}attices and {G}roups},
  author={Conway, John Horton and Sloane, Neil James Alexander},
  volume={290},
  year={2013},
  publisher={Springer Science \& Business Media}
}

@article{Damour:2000hv,
    author = "Damour, Thibault and Henneaux, Marc",
    title = "{E(10), BE(10) and arithmetical chaos in superstring cosmology}",
    eprint = "hep-th/0012172",
    archivePrefix = "arXiv",
    reportNumber = "IHES-P-00-83",
    doi = "10.1103/PhysRevLett.86.4749",
    journal = "Phys. Rev. Lett.",
    volume = "86",
    pages = "4749--4752",
    year = "2001"
}

@article{Damour:2002cu,
    author = "Damour, T. and Henneaux, M. and Nicolai, H.",
    title = "{E(10) and a 'small tension expansion' of M theory}",
    eprint = "hep-th/0207267",
    archivePrefix = "arXiv",
    reportNumber = "AEI-2002-054, IHES-P-02-48",
    doi = "10.1103/PhysRevLett.89.221601",
    journal = "Phys. Rev. Lett.",
    volume = "89",
    pages = "221601",
    year = "2002"
}

@article{Damour:2002et,
    author = "Damour, T. and Henneaux, M. and Nicolai, H.",
    title = "{Cosmological billiards}",
    eprint = "hep-th/0212256",
    archivePrefix = "arXiv",
    reportNumber = "IHES-P-02-80, AEI-2002-092, ULB-TH-02-33, IHES-P-02-08",
    doi = "10.1088/0264-9381/20/9/201",
    journal = "Class. Quant. Grav.",
    volume = "20",
    pages = "R145--R200",
    year = "2003"
}

@article{Dupuis:2020fhh,
    author = "Dupuis, N. and Canet, L. and Eichhorn, A. and Metzner, W. and Pawlowski, J. M. and Tissier, M. and Wschebor, N.",
    title = "{The nonperturbative functional renormalization group and its applications}",
    eprint = "2006.04853",
    archivePrefix = "arXiv",
    primaryClass = "cond-mat.stat-mech",
    doi = "10.1016/j.physrep.2021.01.001",
    journal = "Phys. Rept.",
    volume = "910",
    pages = "1--114",
    year = "2021"
}

@article{ElserSloane4D,
  title={A highly symmetric four-dimensional quasicrystal},
  author={Elser, Veit and Sloane, Neil JA},
  journal={Journal of Physics A: Mathematical and General},
  volume={20},
  number={18},
  pages={6161},
  year={1987},
  publisher={IOP Publishing}
}

@article{ESSELMANN1996103,
title = {Über die maximale Dimension von Lorentz–Gittern mit coendlicher Spiegelungsgruppe},
journal = {Journal of Number Theory},
volume = {61},
number = {1},
pages = {103-144},
year = {1996},
issn = {0022-314X},
doi = {https://doi.org/10.1006/jnth.1996.0141},
url = {https://www.sciencedirect.com/science/article/pii/S0022314X96901419},
author = {Frank Esselmann}
}

@article{Flicker:2018tkr,
    author = "Flicker, Felix",
    title = "{Time Quasilattices in Dissipative Dynamical Systems}",
    eprint = "1707.09371",
    archivePrefix = "arXiv",
    doi = "10.21468/SciPostPhys.5.1.001",
    journal = "SciPost Phys.",
    volume = "5",
    number = "1",
    pages = "001",
    year = "2018"
}

@article{friedland2024spacetime,
  title={Spacetime quasicrystals in Bose-Einstein condensates},
  author={Friedland, L and Shagalov, AG},
  journal={Physical Review Research},
  volume={6},
  number={2},
  pages={023054},
  year={2024},
  publisher={APS}
}

@article{gardner1977extraordinary,
  title={Extraordinary nonperiodic tiling that enriches the theory of tiles},
  author={Gardner, Martin},
  journal={Scientific American},
  volume={236},
  number={1},
  pages={110--121},
  year={1977},
  publisher={Munn \& Company}
}

@book{gardner1997penrose,
  title={Penrose Tiles to Trapdoor Ciphers: And the Return of Dr Matrix},
  author={Gardner, Martin},
  year={1997},
  publisher={Cambridge University Press}
}

@book{Green:1987sp,
    author = "Green, Michael B. and Schwarz, J. H. and Witten, Edward",
    title = "{Superstring Theory Vol. 1: Introduction}",
    isbn = "978-0-521-35752-4",
    series = "Cambridge Monographs on Mathematical Physics",
    month = "7",
    year = "1988",
    publisher={Cambridge University Press}
}

@book{Green:1987mn,
    author = "Green, Michael B. and Schwarz, J. H. and Witten, Edward",
    title = "{Superstring Theory Vol. 2: Loop Amplitudes, Anomalies and Phenomenology}",
    isbn = "978-0-521-35753-1",
    month = "7",
    year = "1988",
    publisher={Cambridge University Press}
}

@article{GROSS20091034,
title = {Cyclotomic {F}actors of {C}oxeter {P}olynomials},
journal = {Journal of Number Theory},
volume = {129},
number = {5},
pages = {1034-1043},
year = {2009},
issn = {0022-314X},
doi = {https://doi.org/10.1016/j.jnt.2008.09.021},
url = {https://www.sciencedirect.com/science/article/pii/S0022314X09000432},
author = {Benedict H. Gross and Eriko Hironaka and Curtis T. McMullen}
}

@book{grunbaum1987tilings,
  title={Tilings and patterns},
  author={Gr{\"u}nbaum, Branko and Shephard, Geoffrey Colin},
  year={1987},
  publisher={Courier Dover Publications}
}

@article{Harvey:1987da,
    author = "Harvey, Jeffrey A. and Moore, Gregory W. and Vafa, C.",
    title = "{QUASICRYSTALLINE COMPACTIFICATION}",
    reportNumber = "PUPT-1068, IASSNS/HEP-87/48, HUTP-87/A072",
    doi = "10.1016/0550-3213(88)90627-X",
    journal = "Nucl. Phys. B",
    volume = "304",
    pages = "269--290",
    year = "1988"
}

@book{Hawking2001-HAWTUI,
	author = {(Figure taken from) S. W. Hawking},
	year = {2001},
	title = {The Universe in a Nutshell}
}

@article{he2025experimental,
  title={Experimental realization of discrete time quasicrystals},
  author={He, Guanghui and Ye, Bingtian and Gong, Ruotian and Yao, Changyu and Liu, Zhongyuan and Murch, Kater W and Yao, Norman Y and Zu, Chong},
  journal={Physical Review X},
  volume={15},
  number={1},
  pages={011055},
  year={2025},
  publisher={APS}
}

@article{hollister2014impact,
  title={Impact-induced shock and the formation of natural quasicrystals in the early solar system},
  author={Hollister, Lincoln S and Bindi, Luca and Yao, Nan and Poirier, Gerald R and Andronicos, Christopher L and MacPherson, Glenn J and Lin, Chaney and Distler, Vadim V and Eddy, Michael P and Kostin, Alexander and others},
  journal={Nature Communications},
  volume={5},
  number={1},
  pages={4040},
  year={2014},
  publisher={Nature Publishing Group UK London},
doi={10.1038/ncomms5040}
}

@book{Humphreys1973,
  title={Introduction to Lie Algebras and Representation Theory},
  author={James E. Humphreys},
  series={Graduate Texts in Mathematics},
  volume={290},
  year={2013},
  publisher={Springer New York, NY}
}

@book{humphreys_1990, place={Cambridge}, series={Cambridge Studies in Advanced Mathematics}, title={Reflection Groups and Coxeter Groups}, DOI={10.1017/CBO9780511623646}, publisher={Cambridge University Press}, author={Humphreys, James E.}, year={1990}, collection={Cambridge Studies in Advanced Mathematics}}

@article{Ishibashi:1996xs,
    author = "Ishibashi, N. and Kawai, H. and Kitazawa, Y. and Tsuchiya, A.",
    title = "{A Large N reduced model as superstring}",
    eprint = "hep-th/9612115",
    archivePrefix = "arXiv",
    reportNumber = "KEK-TH-503",
    doi = "10.1016/S0550-3213(97)00290-3",
    journal = "Nucl. Phys. B",
    volume = "498",
    pages = "467--491",
    year = "1997"
}

@book{kac1990infinite,
  title={Infinite-dimensional Lie algebras},
  author={Kac, Victor G},
  year={1990},
  publisher={Cambridge university press}
}

@article{kaplinskaja1978groups,
  title={On the groups O18, 1 (Z) and O19, 1 (Z)},
  author={Kaplinskaja, IM and Vinberg, E},
  journal={Doklady Akademii Nauk SSSR},
  volume={238},
  pages={1273--1275},
  year={1978}
}

@article{Kleinschmidt:2015sfa,
    author = "Kleinschmidt, Axel and Nicolai, Hermann",
    title = "{Standard model fermions and K (E$_{10}$)}",
    eprint = "1504.01586",
    archivePrefix = "arXiv",
    primaryClass = "hep-th",
    doi = "10.1016/j.physletb.2015.06.005",
    journal = "Phys. Lett. B",
    volume = "747",
    pages = "251--254",
    year = "2015"
}

@book{lang1994algebraic,
  title={Algebraic Number Theory},
  author={Lang, Serge},
  volume={110},
  year={1994},
  publisher={Springer Graduate Texts In Mathematics}
}

@article{levine1986quasicrystals,
  title = {Quasicrystals. I. Definition and structure},
  author = {Levine, Dov and Steinhardt, Paul J.},
  journal = {Phys. Rev. B},
  volume = {34},
  issue = {2},
  pages = {596--616},
  numpages = {0},
  year = {1986},
  month = {Jul},
  publisher = {American Physical Society},
  doi = {10.1103/PhysRevB.34.596},
  url = {https://link.aps.org/doi/10.1103/PhysRevB.34.596}
}

@misc{li2024penrosetilingquantumerrorcorrecting,
      title={The Penrose Tiling is a Quantum Error-Correcting Code}, 
      author={Zhi Li and Latham Boyle},
      year={2024},
      eprint={2311.13040},
      archivePrefix={arXiv},
      primaryClass={quant-ph},
      url={https://arxiv.org/abs/2311.13040}, 
}

@article{mcmullen2002coxeter,
  title={Coxeter {G}roups, {S}alem {N}umbers and the {H}ilbert {M}etric},
  author={McMullen, Curtis T},
  journal={Publications Math{\'e}matiques de l'IH{\'E}S},
  volume={95},
  pages={151--183},
  year={2002}
}

@article{Meissner:2025qrx,
    author = "Meissner, Krzysztof A. and Nicolai, Hermann",
    title = "{Standard model symmetries and K(E$_{10}$)}",
    eprint = "2503.13155",
    archivePrefix = "arXiv",
    primaryClass = "hep-th",
    doi = "10.1007/JHEP08(2025)054",
    journal = "JHEP",
    volume = "08",
    pages = "054",
    year = "2025"
}

@article{meyer1977prasentation,
  title={Pr{\"a}sentation der Einheitengruppe der quadratischen FormF (X)=- X 0 2+ X 1 2+...+ X 18 2},
  author={Meyer, Josef},
  journal={Archiv der Mathematik},
  volume={29},
  number={1},
  pages={261--266},
  year={1977},
  publisher={Springer}
}

@misc{moore1993finitedirections,
      title={Finite in All Directions}, 
      author={G. Moore},
      year={1993},
      eprint={hep-th/9305139},
      archivePrefix={arXiv},
      primaryClass={hep-th},
      url={https://arxiv.org/abs/hep-th/9305139}, 
}

@book{mukhanov2005physical,
  title={Physical foundations of cosmology},
  author={Mukhanov, Viatcheslav},
  year={2005},
  publisher={Cambridge university press}
}

@phdthesis{mukunda2007pisot,
  title={{P}isot and {S}alem {N}umbers from {P}olynomials of {H}eight {O}ne},
  author={Mukunda, Keshav},
  year={2007},
  school={Dept. of Mathematics-Simon Fraser University}
}

@article{Newman:1961qr,
    author = "Newman, Ezra and Penrose, Roger",
    title = "{An Approach to gravitational radiation by a method of spin coefficients}",
    doi = "10.1063/1.1724257",
    journal = "J. Math. Phys.",
    volume = "3",
    pages = "566--578",
    year = "1962"
}

@article{Niedermaier:2006wt,
    author = "Niedermaier, Max and Reuter, Martin",
    title = "{The Asymptotic Safety Scenario in Quantum Gravity}",
    doi = "10.12942/lrr-2006-5",
    journal = "Living Rev. Rel.",
    volume = "9",
    pages = "5--173",
    year = "2006"
}

@article{Pastawski:2015qua,
    author = "Pastawski, Fernando and Yoshida, Beni and Harlow, Daniel and Preskill, John",
    title = "{Holographic quantum error-correcting codes: Toy models for the bulk/boundary correspondence}",
    eprint = "1503.06237",
    archivePrefix = "arXiv",
    primaryClass = "hep-th",
    doi = "10.1007/JHEP06(2015)149",
    journal = "JHEP",
    volume = "06",
    pages = "149",
    year = "2015"
}

@article{penrose1974role,
  title={The role of aesthetics in pure and applied mathematical research},
  author={Penrose, Roger},
  journal={Bull. Inst. Math. Appl.},
  volume={10},
  pages={266--271},
  year={1974}
}

@book{Polchinski:1998rr,
    author = "Polchinski, J.",
    title = "{String theory. Vol. 2: Superstring theory and beyond}",
    doi = "10.1017/CBO9780511618123",
    isbn = "978-0-511-25228-0, 978-0-521-63304-8, 978-0-521-67228-3",
    publisher = "Cambridge University Press",
    series = "Cambridge Monographs on Mathematical Physics",
    month = "12",
    year = "2007"
}

@book{Polchinski:1998rq,
    author = "Polchinski, J.",
    title = "{String theory. Vol. 1: An introduction to the bosonic string}",
    doi = "10.1017/CBO9780511816079",
    isbn = "978-0-511-25227-3, 978-0-521-67227-6, 978-0-521-63303-1",
    publisher = "Cambridge University Press",
    series = "Cambridge Monographs on Mathematical Physics",
    month = "12",
    year = "2007"
}

@article{Schechtman1984,
  title = {Metallic Phase with Long-Range Orientational Order and No Translational Symmetry},
  author = {Shechtman, D. and Blech, I. and Gratias, D. and Cahn, J. W.},
  journal = {Phys. Rev. Lett.},
  volume = {53},
  issue = {20},
  pages = {1951--1953},
  numpages = {0},
  year = {1984},
  month = {Nov},
  publisher = {American Physical Society},
  doi = {10.1103/PhysRevLett.53.1951},
  url = {https://link.aps.org/doi/10.1103/PhysRevLett.53.1951}
}

@book{serre2012arithmetic,
  title={A Course in Arithmetic},
  author={Serre, Jean-Pierre},
  volume={7},
  year={1973},
  publisher={Springer-Verlag}
}

@article{Surya:2019ndm,
    author = "Surya, Sumati",
    title = "{The causal set approach to quantum gravity}",
    eprint = "1903.11544",
    archivePrefix = "arXiv",
    primaryClass = "gr-qc",
    doi = "10.1007/s41114-019-0023-1",
    journal = "Living Rev. Rel.",
    volume = "22",
    number = "1",
    pages = "5",
    year = "2019"
}

@article{vacaru2018space,
  title={Space-time quasicrystal structures and inflationary and late time evolution dynamics in accelerating cosmology},
  author={Vacaru, Sergiu I},
  journal={Classical and Quantum Gravity},
  volume={35},
  number={24},
  pages={245009},
  year={2018},
  publisher={IOP Publishing}
}

@article{vinberg1975some,
  title={Some {A}rithmetical {D}iscrete {G}roups in {L}obachevsky {S}paces},
  author={Vinberg, {\'E} B},
  journal={Discrete subgroups of Lie groups and applications to moduli},
  pages={328--348},
  year={1975},
  publisher={Oxford}
}

@article{vinberg1967discrete,
  title={Discrete {G}roups {G}enerated by {R}eflections in {L}obachevsky {S}paces},
  author={Vinberg, {\'E} B},
  journal={Mathematics of the USSR-Sbornik},
  volume={1},
  number={3},
  pages={429},
  year={1967},
  publisher={IOP Publishing}
}

@article{vinberg1972groups,
  title={On {G}roups of {U}nit {E}lements of {C}ertain {Q}uadratic {F}orms},
  author={Vinberg, {\`E} B},
  journal={Mathematics of the USSR-Sbornik},
  volume={16},
  number={1},
  pages={17},
  year={1972},
  publisher={IOP Publishing}
}

@inproceedings{Weinberg:1976xy,
    author = "Weinberg, Steven",
    title = "{Critical Phenomena for Field Theorists}",
    booktitle = "{14th International School of Subnuclear Physics: Understanding the Fundamental Constitutents of Matter}",
    reportNumber = "HUTP-76-160",
    doi = "10.1007/978-1-4684-0931-4_1",
    month = "8",
    year = "1976"
}

@inbook{Weinberg:1980gg,
    author = "Weinberg, Steven",
    title = "{Ultraviolet Divergences in Quantum Theories of Gravitation}",
    booktitle = "{General Relativity}: {An Einstein Centenary Survey}",
    pages = "790--831",
    year = "1980"
}

@book{weinberg2008cosmology,
  title={Cosmology},
  author={Weinberg, Steven},
  year={2008},
  publisher={OUP Oxford}
}

@article{Dechant2016,
   title={The $E_8$ Geometry from a Clifford Perspective},
   volume={27},
   ISSN={1661-4909},
   url={http://dx.doi.org/10.1007/s00006-016-0675-9},
   DOI={10.1007/s00006-016-0675-9},
   number={1},
   journal={Advances in Applied Clifford Algebras},
   publisher={Springer Science and Business Media LLC},
   author={Dechant, Pierre-Philippe},
   year={2016},
   month=apr, pages={397–421} }

@article{HamCycles_Felix2024,
  title = {Hamiltonian Cycles on Ammann-Beenker Tilings},
  author = {Singh, Shobhna and Lloyd, Jerome and Flicker, Felix},
  journal = {Phys. Rev. X},
  volume = {14},
  issue = {3},
  pages = {031005},
  numpages = {24},
  year = {2024},
  month = {Jul},
  publisher = {American Physical Society},
  doi = {10.1103/PhysRevX.14.031005},
  url = {https://link.aps.org/doi/10.1103/PhysRevX.14.031005}
}

\end{document}